\documentclass{llncs}
\usepackage{latexsym,amssymb,upgreek,graphicx}
\usepackage{enumerate}
\usepackage{amsmath}
\usepackage{amsfonts}
\usepackage{xspace}
\usepackage{color}
\usepackage{url}
\usepackage{mathrsfs}
\usepackage{listings}

\newlength{\defbaselineskip}
\newcommand{\setlinespacing}[1]{\setlength{\baselineskip}{#1 \defbaselineskip}}

\newlength{\mylength}

\newcommand{\bgsP}{\settoheight{\mylength}{t}\setlength{\mylength}{0.94\mylength}
\overleftarrow{gs\rule{0cm}{\mylength}}_{_{\p}}}
\newcommand{\fgsP}{\settoheight{\mylength}{t}\setlength{\mylength}{0.94\mylength}
\overrightarrow{gs\rule{0cm}{\mylength}}_{_{\p}}}
\newcommand{\fgs}{\settoheight{\mylength}{t}\setlength{\mylength}{0.94\mylength}
\overrightarrow{gs\rule{0cm}{\mylength}}}
\newcommand{\bgs}{\settoheight{\mylength}{t}\setlength{\mylength}{0.94\mylength}
\overleftarrow{gs\rule{0cm}{\mylength}}}
\newcommand{\last}{\mathit{last}}
\newcommand{\slen}{\mathit{slen}}

\def\qed {{\unskip\nobreak\hfil\penalty50
\hskip2em\hbox{}\nobreak\hfil \rule{2mm}{2mm}
\parfillskip=0pt \finalhyphendemerits=0 \par \medskip}}

\newcommand{\defAs}
{\:\mbox{\hbox{$= \! \! \raisebox{-0.5 ex}[0 ex][0 ex]{\tiny Def}$}}\:}

\newcommand{\tb}[1]{\textbf{#1}}  			
\newcommand{\length}{\ensuremath{\mathrm{length}}}	

\newcommand{\kmp}{Knuth-Morris-Pratt\xspace}		
\newcommand{\hor}{Horspool\xspace}			
\newcommand{\quick}{Quick-Search\xspace}		
\newcommand{\berry}{Berry-Ravindran\xspace}		
\newcommand{\boyerm}{Boyer-Moore\xspace}		
\newcommand{\fs}{Fast-Search\xspace}			
\newcommand{\bfs}{Backward-Fast-Search\xspace}		
\newcommand{\ffs}{Forward-Fast-Search\xspace}		
\newcommand{\shiftA}{Shift-And\xspace}			
\newcommand{\tbm}{Tuned-Boyer-Moore\xspace}		
\newcommand{\skipS}{Skip-Search\xspace}			
\newcommand{\maximal}{Maximal-Shift\xspace}		
\newcommand{\reversef}{Reverse-Factor\xspace}		
\newcommand{\alphaS}{Alpha-Skip-Search\xspace}		
\newcommand{\bndm}{Backward-Nondeterministic-DAWG-Matching\xspace}
\newcommand{\tsw}{Two-Sliding-Windows\xspace}		
\newcommand{\fjs}{Franek-Jennings-Smyth\xspace}		
\newcommand{\raita}{Raita\xspace}			
\newcommand{\hashq}{Hash$q$\xspace}			
\newcommand{\bom}{Backward-Oracle-Matching\xspace}	
\newcommand{\bdm}{Backward-DAWG-Matching\xspace}	
\newcommand{\fdm}{Forward-DAWG-Matching\xspace}		
\newcommand{\ebom}{Extended-BOM\xspace}			
\newcommand{\fbom}{Forward-BOM\xspace}			
\newcommand{\dfdm}{Double-Forward-DAWG-Matching\xspace}	
\newcommand{\aoso}{Average-Optimal-Shift-Or\xspace}		
\newcommand{\faoso}{Fast-Average-Optimal-Shift-Or\xspace}	
\newcommand{\blim}{Bit-Parallel Length Invariant Matcher\xspace}
\newcommand{\nsn}{Not-So-Naive\xspace}				
\newcommand{\nn}{Not-Naive\xspace}				
\newcommand{\tail}{Tailed-Substring\xspace}			
\newcommand{\tndm}{Two-Way-Nondeterministic-DAWG-Matching\xspace}
\newcommand{\fndm}{Forward-Nondeterministic-DAWG-Matching\xspace}
\newcommand{\bpww}{Bit-Parallel Wide-Window\xspace}
\newcommand{\bpdww}{Bit-Parallel$^2$ Wide-Window\xspace}
\newcommand{\bpwwd}{Bit-Parallel Wide-Window$^2$\xspace}

\newcommand{\rev}[1]{\bar{#1}}  	  	

\newcommand{\SMA}{\emph{SMA}\xspace}    	
\newcommand{\NSMA}{\emph{NSMA}\xspace}   	
\newcommand{\DAWG}{\emph{Dawg}\xspace}    	
\newcommand{\NDAWG}{\emph{NDawg}\xspace}   	

\newcommand{\bAnd}{\mathrel{\&}}
\newcommand{\bOr}{\mathrel{|}}
\newcommand{\bNot}{\mathop{\sim}}
\newcommand{\bsf}{\textsf{bsf}\xspace}
\newcommand{\popcount}{\textsf{popcount}\xspace}
\newcommand{\reverse}{\textsf{reverse}\xspace}
\newcommand{\first}{\mathit{first}\xspace}
\newcommand{\mw}{\lceil m/w\rceil}
\newcommand{\kw}{\lceil k/w\rceil}
\newcommand{\UShiftAnd}{\textsf{F-Shift-And}\xspace}
\newcommand{\UBNDM}{\textsf{F-BNDM}\xspace}
\newcommand{\AND}{\mathrel{\&}}
\newcommand{\OR}{\mathrel{|}}
\newcommand{\NOT}{\mathop{\sim}}

\def\endpos{endpos}				
\newcommand{\supply}{Sp}			
\newcommand{\setP}{\mathcal{P}}			

\newcommand{\p}{p\xspace}			
\def\t{t\xspace}				

\def\pp{\mathinner{\ldotp\ldotp}}
\def\Fact{\textit{Fact}}
\def\Suff{\textit{Suff}}
\def\nextl{\textit{nextl}}
\def\nextr{\textit{nextr}}
\def\shiftl{\textit{shiftl}}
\def\shiftr{\textit{shiftr}}
\def\shift{\textit{shift}}
\def\wsize{\textit{wsize}}
\def\sa#1{\texttt{#1}}
\newcommand{\bigO}{\mathcal{O}}

\newcommand{\hs}{@{\hspace{0.3cm}}}

\begin{document}

\title{The Exact String Matching Problem:\\a Comprehensive Experimental Evaluation}

\titlerunning{Efficient Macthing of Biological Sequences Allowing for Non-Overlapping Inversions}

\author{Simone Faro$^\dag$ \and Thierry Lecroq$^\ddag$}

\institute{$^\dag$Universit\`{a} degli Studi di Catania, Dipartimento di Matematica e Informatica\\
Viale Andrea Doria 6, I-95125, Catania, Italy\\
\email{ faro@dmi.unict.it}\\[0.2cm]
$^\ddag$Universit\'e de Rouen, LITIS EA 4108, 76821 Mont-Saint-Aignan Cedex, France\\
\email{thierry.lecroq@univ-rouen.fr}
}

\maketitle

\begin{abstract}
This paper addresses the \emph{online exact string matching problem} which consists in finding \emph{all} occurrences of a given pattern $\p$ in a text $t$.  
It is an extensively studied problem in computer science, mainly due to its direct applications to such diverse areas as text, 
image and signal processing, speech analysis and recognition, data compression, information retrieval, computational biology and chemistry.
Since 1970 more than 80 string matching algorithms have been proposed, and more than 50\% of them in the last ten years. 
In this note we present a comprehensive list of all string matching algorithms and present experimental results in order to compare them from a practical point of view.
>From our experimental evaluation it turns out that the performance of the algorithms are quite different for different alphabet sizes and pattern length.
\end{abstract}

\markboth{Simone Faro and Thierry Lecroq}{Exact String Matching: a Comprehensive Experimental Evaluation}

\section{Introduction}\label{intro}
Given a text $t$ of length $n$ and a pattern $\p$ of length $m$ over some alphabet $\Sigma$ of size $\sigma$, 
the \emph{string matching problem} consists in finding \emph{all} occurrences of the pattern $\p$ in the text $t$.  
It is an extensively studied problem in computer science, mainly due to its direct applications to such diverse areas as text, 
image and signal processing, speech analysis and recognition, data compression, information retrieval, computational biology and chemistry.

String matching algorithms are also basic components used in implementations
of practical softwares existing under most operating systems. Moreover, they emphasize programming methods that serve as paradigms in other fields
of computer science. Finally they also play an important role in theoretical computer science by providing challenging problems.



Applications require two kinds of solutions depending on which string, the pattern or the text, is given first.
Algorithms based on the use of automata or combinatorial properties of strings are commonly implemented to preprocess the
pattern and solve the first kind of problem. This kind of problem is generally referred as \emph{online} string matching. 
The notion of indexes realized by trees or automata is used instead in the second kind of problem, generally referred as \emph{offline} string matching.
In this paper we are only interested in algorithms of the first kind.

The worst case lower bound of the online string matching problem is $\bigO(n)$ and has been firstly reached by the well known Morris-Pratt algorithm~\cite{MP70}.
An average lower bound in $\bigO(n \log m /m)$ (with equiprobability and independence of letters) has been proved by Yao in~\cite{Yao79}.


\begin{figure}
\begin{center}
\begin{scriptsize}
\begin{tabular*}{\textwidth}{@{\extracolsep{\fill}}|\hs l\hs l\hs l\hs l\hs |}
\hline
   &&&\\
   \multicolumn{4}{|l|}{\hspace{0.2cm}\tb{Algorithms based on characters comparison}}\\[.1cm]
\hline
    &&&\\
    \textsf{BF} 	& 	Brute-Force 				& ~\cite{CLRS01} 	& \\
    \textsf{MP} 	& 	Morris-Pratt 				& ~\cite{MP70} 		& 1970 \\
    \textsf{KMP} 	& 	Knuth-Morris-Pratt			& ~\cite{KMP77} 	& 1977 \\
    \textsf{BM} 	& 	Boyer-Moore				& ~\cite{BM77} 		& 1977 \\
    \textsf{HOR} 	& 	Horspool				& ~\cite{Hor80} 	& 1980 \\
    \textsf{GS} 	& 	Galil-Seiferas				& ~\cite{GS83} 		& 1983 \\
    \textsf{AG} 	& 	Apostolico-Giancarlo			& ~\cite{AG86} 		& 1986  \\
    \textsf{KR} 	& 	Karp-Rabin 				& ~\cite{KR87} 		& 1987  \\
    \textsf{ZT} 	& 	Zhu-Takaoka				& ~\cite{ZT87} 		& 1987 \\
    \textsf{OM} 	& 	Optimal-Mismatch			& ~\cite{Sun90} 	& 1990 \\
    \textsf{MS} 	& 	Maximal-Shift				& ~\cite{Sun90} 	& 1990 \\
    \textsf{QS} 	& 	Quick-Search				& ~\cite{Sun90} 	& 1990 \\
    \textsf{AC} 	& 	Apostolico-Crochemore			& ~\cite{AC91} 		& 1991\\
    \textsf{TW} 	& 	Two-Way					& ~\cite{CP91} 		& 1991 \\
    \textsf{TunBM} 	& 	Tuned-Boyer-Moore			& ~\cite{HS91} 		& 1991\\
    \textsf{COL} 	& 	Colussi					& ~\cite{Col91b} 	& 1991  \\
    \textsf{SMITH} 	& 	Smith					& ~\cite{Smi91} 	& 1991\\
    \textsf{GG} 	& 	Galil-Giancarlo				& ~\cite{GG92} 		& 1992\\
    \textsf{RAITA} 	& 	Raita					& ~\cite{Rai92} 	& 1992 \\
    \textsf{SMOA} 	& 	String-Matching on Ordered ALphabet	& ~\cite{Cro92} 	& 1992 \\
    \textsf{NSN} 	& 	Not-So-Naive				& ~\cite{Han93} 	& 1993\\
    \textsf{TBM} 	& 	Turbo-Boyer-Moore			& ~\cite{CCGJLPR94} 	& 1994 \\
    \textsf{RCOL} 	& 	Reverse-Colussi				& ~\cite{Col94b} 	& 1994 \\
    \textsf{SKIP} 	& 	Skip-Search				& ~\cite{CLP98} 	& 1998 \\
    \textsf{ASKIP} 	& 	Alpha-Skip-Search			& ~\cite{CLP98} 	& 1998 \\
    \textsf{KMPS} 	& 	KMP-Skip-Search				& ~\cite{CLP98} 	& 1998 \\
    \textsf{BR} 	& 	Berry-Ravindran				& ~\cite{BR99} 		& 1999 \\
    \textsf{AKC} 	& 	Ahmed-Kaykobad-Chowdhury 			& ~\cite{AKC2003} 	& 2003 \\
    \textsf{FS} 	& 	Fast-Search 					& ~\cite{CF03a} 	& 2003 \\
    \textsf{FFS} 	& 	Forward-Fast-Search 				& ~\cite{CF05} 		& 2004 \\
    \textsf{BFS}	&	Backward-Fast-Search, Fast-Boyer-Moore		& ~\cite{CF05,CL08}	& 2004 \\
    \textsf{TS} 	&	Tailed-Substring				& ~\cite{CF04} 		& 2004 \\
    \textsf{SSABS} 	&	Sheik-Sumit-Anindya-Balakrishnan-Sekar		& ~\cite{SABS04} 	& 2004 \\
    \textsf{TVSBS} 	&	Thathoo-Virmani-Sai-Balakrishnan-Sekar		& ~\cite{TVSBS} 	& 2006 \\
    \textsf{PBMH} 	&	Boyer-Moore-Horspool using Probabilities	& ~\cite{Nebel06} 	& 2006 \\
    \textsf{FJS} 	&	Franek-Jennings-Smyth 				& ~\cite{FJS07} 	& 2007 \\
    \textsf{2BLOCK}	&	2-Block	Boyer-Moore	 			& ~\cite{SM07} 		& 2007 \\
    \textsf{HASH$q$} 	&	Wu-Manber for Single Pattern Matching		& ~\cite{Lec07} 	& 2007 \\
    \textsf{TSW}	&	Two-Sliding-Window 				& ~\cite{HASIA08} 	& 2008 \\
    \textsf{BMH$q$} 	&	Boyer-Moore-Horspool with $q$-grams		& ~\cite{KPT08}		& 2008 \\
    \textsf{GRASPm}	&	Genomic Rapid Algo for String Pm		& ~\cite{DC09} 		& 2009 \\
    \textsf{SSEF} 	&	SSEF			 			& ~\cite{KO09} 		& 2009 \\
    &&&\\
\hline
\end{tabular*}
\end{scriptsize}
\caption{\label{fig:table_algos1}The list of all comparison based string matching algorithms (1970-2010).} 
\end{center}
\end{figure}

\begin{figure}
\begin{center}
\begin{scriptsize}
\begin{tabular*}{\textwidth}{@{\extracolsep{\fill}}|\hs l\hs l\hs l\hs l\hs |}
\hline
   &&&\\
   \multicolumn{4}{|l|}{\hspace{0.2cm}\tb{Algorithms based on automata}}\\[.1cm]
\hline
    &&&\\
    \textsf{DFA}	&	Deterministic-Finite-Automaton			& ~\cite{CLRS01}	& \\
    \textsf{RF} 	& 	Reverse-Factor					& ~\cite{Lec92} 	& 1992\\
    \textsf{SIM}	&	Simon						& ~\cite{Sim93}		& 1993 \\
    \textsf{TRF} 	& 	Turbo-Reverse-Factor				& ~\cite{CCGJLPR94} 	& 1994\\
    \textsf{FDM} 	& 	Forward-DAWG-Matching				& ~\cite{CR94} 		& 1994\\
    \textsf{BDM} 	& 	Backward-DAWG-Matching				& ~\cite{CR94} 		& 1994\\
    \textsf{BOM} 	& 	Backward-Oracle-Matching			& ~\cite{ACR99} 	& 1999\\
    \textsf{DFDM}	&	Double Forward DAWG Matching			& ~\cite{AR00}		& 2000 \\
    \textsf{WW}		&	Wide Window					& ~\cite{HFS05}		& 2005 \\
    \textsf{LDM}	&	Linear DAWG Matching				& ~\cite{HFS05}		& 2005 \\
    \textsf{ILDM1}	&	Improved Linear DAWG Matching			& ~\cite{LWLL06}	& 2006 \\
    \textsf{ILDM2} 	&	Improved Linear DAWG Matching 2			& ~\cite{LWLL06}	& 2006 \\
    \textsf{EBOM}	&	Extended Backward Oracle Matching		& ~\cite{FL09}		& 2009 \\
    \textsf{FBOM}	&	Forward Backward Oracle Matching 		& ~\cite{FL09}		& 2009 \\
    \textsf{SEBOM}	&	Simplified Extended Backward Oracle Matching	& ~\cite{FYM09}		& 2009 \\
    \textsf{SFBOM} 	&	Simplified Forward Backward Oracle Matching 	& ~\cite{FYM09}		& 2009 \\
    \textsf{SBDM}	&	Succint Backward DAWG Matching			& ~\cite{Fred09}	& 2009 \\
    &&&\\
\hline
\end{tabular*}
\end{scriptsize}
\caption{\label{fig:table_algos2}The list of the automata based string matching algorithms (1992-2009).} 
\end{center}
\end{figure}

\begin{figure}
\begin{center}
\begin{scriptsize}
\begin{tabular*}{\textwidth}{@{\extracolsep{\fill}}|\hs l\hs l\hs l\hs l\hs |}
\hline
   &&&\\
   \multicolumn{4}{|l|}{\hspace{0.2cm}\tb{Algorithms based on bit-parallelism}}\\[.1cm]
\hline
    &&&\\
    \textsf{SO}		&	Shift-Or					& ~\cite{BYR92} 	& 1992\\
    \textsf{SA}		&	Shift-And					& ~\cite{BYR92} 	& 1992\\
    \textsf{BNDM}	&	Backward-Nondeterministic-DAWG-Matching		& ~\cite{NR98a} 	& 1998\\
    \textsf{BNDM-L}	&	BNDM for Long patterns				& ~\cite{NR00} 		& 2000 \\
    \textsf{SBNDM}	&	Simplified BNDM 				& ~\cite{PT03,Nav01}	& 2003 \\
    \textsf{TNDM}	&	Two-Way Nondeterministic DAWG Matching		& ~\cite{PT03} 		& 2003 \\
    \textsf{LBNDM}	&	Long patterns BNDM				& ~\cite{PT03} 		& 2003 \\
    \textsf{SVM}	&	Shift Vector Matching				& ~\cite{PT03} 		& 2003 \\
    \textsf{BNDM2}	&	BNDM with loop-unrolling			& ~\cite{HD05} 		& 2005 \\
    \textsf{SBNDM2}	&	Simplified BNDM with loop-unrolling		& ~\cite{HD05} 		& 2005 \\
    \textsf{BNDM-BMH}	&	BNDM with Horspool Shift			& ~\cite{HD05} 		& 2005 \\
    \textsf{BMH-BNDM}	&	Horspool with BNDM test				& ~\cite{HD05} 		& 2005 \\
    \textsf{FNDM}	&	Forward Nondeterministic DAWG Matching		& ~\cite{HD05} 		& 2005 \\
    \textsf{BWW}	&	Bit parallel Wide Window			& ~\cite{HFS05}		& 2005 \\
    \textsf{FAOSO}	&	Fast Average Optimal Shift-Or 			& ~\cite{FG05}		& 2005 \\
    \textsf{AOSO}	&	Average Optimal Shift-Or 			& ~\cite{FG05}		& 2005 \\
    \textsf{BLIM}	&	Bit-Parallel Length Invariant Matcher		& ~\cite{OK08} 		& 2008 \\
    \textsf{FSBNDM}	&	Forward SBNDM 					& ~\cite{FL09}		& 2009 \\
    \textsf{BNDM$q$} 	&	BNDM with $q$-grams 				& ~\cite{DHPT09} 	& 2009 \\
    \textsf{SBNDM$q$} 	&	Simplified BNDM with $q$-grams			& ~\cite{DHPT09} 	& 2009 \\
    \textsf{UFNDM$q$} 	&	Shift-Or with $q$-grams 			& ~\cite{DHPT09} 	& 2009 \\
    \textsf{SABP}	&	Small Alphabet Bit-Parallel			& ~\cite{ZZMY2009}	& 2009 \\
    \textsf{BP2WW} 	&	\bpdww			 			& ~\cite{CFG10a} 	& 2010 \\
    \textsf{BPWW2} 	&	\bpwwd			 			& ~\cite{CFG10a} 	& 2010 \\
    \textsf{KBNDM} 	&	Factorized BNDM		 			& ~\cite{CFG10b}	& 2010 \\
    \textsf{KSA}	&	Factorized Shift-And	 			& ~\cite{CFG10b}	& 2010 \\
    &&&\\
\hline
\end{tabular*}
\end{scriptsize}
\caption{\label{fig:table_algos3}The list of all bit-parallel string matching algorithms (1992-2010).} 
\end{center}
\end{figure}

More than 80 online string matching algorithms (hereafter simply string matching algorithms) have been proposed over the years.
All solutions can be divided into two classes: 
algorithms which solve the problem by making use only of comparisons between characters, and algorithms which make use of automata
in order to locate all occurrences of the searched string. The latter class can be further divided into two classes:
algorithms which make use of deterministic automata and algorithms based on bit-parallelism which simulate the behavior of non-deterministic automata.

Fig.~\ref{fig:table_algos1}, Fig.~\ref{fig:table_algos2} and Fig.~\ref{fig:table_algos3} present the list of all string matching algorithms based on
comparison of characters, deterministic automata and bit-parallelism, respectively.

The class of algorithms based on comparison of characters is the wider class and consists of almost 50 per cent of all solutions.
Among the comparison based string matching algorithms the \boyerm algorithm~\cite{BM77} deserves a special mention, 
since it has been particularly successful and has inspired much work.

Also automata play a very important role in the design of efficient string matching algorithms. 
The first linear algorithm based on deterministic automata is the Automaton Matcher~\cite{CLRS01}.  

Over the years automata based solutions have been also developed to design algorithms which have optimal sublinear performance on average. 
This is done by using factor automata~\cite{BBEHMC83,Cro85,BBEHCS85,ACR99}, data structures which identify all factors of a word. 
Among the algorithms which make use of a factor automaton the BDM~\cite{CR94} and the \bom algorithm~\cite{ACR99} are among the most efficient solutions, 
especially for long patterns. 

In recent years, most of the work has been devoted to develop software techniques
to simulate efficiently the parallel computation of non-deterministic finite automata related to the search pattern. 
Such simulations can be done
efficiently using the bit-parallelism technique~\cite{BYG92}, which consists in exploiting the
intrinsic parallelism of the bit operations inside a computer word. In some
cases, bit-parallelism allows to reduce the overall number of operations up to a
factor equal to the number of bits in a computer word. Thus, although string
matching algorithms based on bit-parallelism are usually simple and have very
low memory requirements, they generally work well with patterns of moderate
length only.

The bit-parallelism technique has been
used to simulate efficiently the non-deterministic version of the Morris-Pratt
automaton.  The resulting algorithm, named Shift-Or~\cite{BYG92}, runs in
$\mathcal{O}(n\mw)$, where $w$ is the number of bits in a computer
word.  Later, a variant of the Shift-Or algorithm, called Shift-And,
and a very fast BDM-like algorithm (BNDM), based on the bit-parallel
simulation of the non-deterministic suffix automaton, were presented in
\cite{NR98}.

Bit-parallelism encoding requires one bit per pattern symbol, for a
total of $\mw$ computer words.  Thus, as long as a pattern fits in a
computer word, bit-parallel algorithms are extremely fast, otherwise
their performance degrades considerably as $\mw$ grows.  Though there
are a few techniques to maintain good performance in the case of long
patterns, such limitation is intrinsic.

\begin{figure}[t]
\begin{center}
\includegraphics[width=\textwidth]{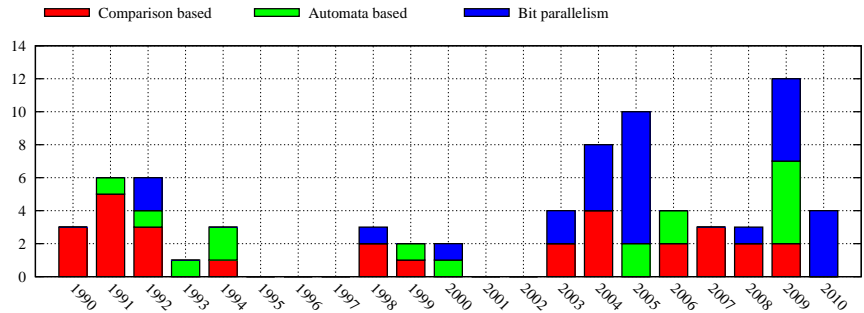}
\end{center}
\caption{\label{fig:number_algos}Number of algorithms proposed in the last 21 years (1990-2010)}
\end{figure}

Fig.~\ref{fig:number_algos} presents a plot of the number of algorithms (for each class) proposed in the last 21 years (1990-2010).
Observe that the number of proposed solutions have doubled in the last ten years, demonstrating the increasing interest in this issue.
It is interesting to observe also that almost 50 per cent of solutions in the last ten years are based on bit-parallelism. Moreover it seems 
that the number of bit-parallel solutions proposed in the years follows an increasing trend.

In the rest of the paper we present a comprehensive experimental evaluation of all string matching algorithms listed above 
in order to compare them from a practical point of view.

\newpage

\renewcommand{\hs}{@{\hspace{0.2cm}}}
\newcommand{\algo}[1]{\textsf{#1}}
\definecolor{red}{rgb}{1,0,0}
\definecolor{green}{rgb}{0,1,0}
\definecolor{blu}{rgb}{0,0,1}
\definecolor{best}{rgb}{0.9,0.9,0.9}
\definecolor{color0}{rgb}{0,0,1}
\definecolor{color1}{rgb}{0.05,0,0.95}
\definecolor{color2}{rgb}{0.1,0,0.90}
\definecolor{color3}{rgb}{0.15,0,0.85}
\definecolor{color4}{rgb}{0.20,0,0.80}
\definecolor{color5}{rgb}{0.25,0,0.75}
\definecolor{color6}{rgb}{0.30,0,0.70}
\definecolor{color7}{rgb}{0.35,0,0.65}
\definecolor{color8}{rgb}{0.40,0,0.60}
\definecolor{color9}{rgb}{0.45,0,0.55}
\definecolor{color10}{rgb}{0.50,0,0.50}
\definecolor{color11}{rgb}{0.55,0,0.45}
\definecolor{color12}{rgb}{0.60,0,0.40}
\definecolor{color13}{rgb}{0.65,0,0.35}
\definecolor{color14}{rgb}{0.70,0,0.30}
\definecolor{color15}{rgb}{0.75,0,0.25}
\definecolor{color16}{rgb}{0.80,0,0.20}
\definecolor{color17}{rgb}{0.85,0,0.15}
\definecolor{color18}{rgb}{0.90,0,0.10}
\definecolor{color19}{rgb}{0.95,0,0.05}
\definecolor{color20}{rgb}{1,0,0}

\section{Experimental Results}
We present next experimental data which allow to compare in terms of running time all the algorithms listed 
in Fig.~\ref{fig:table_algos1}, Fig.~\ref{fig:table_algos2} and Fig.~\ref{fig:table_algos3}.

In particular we tested the \textsf{Hash}$q$ algorithm with $q$ equal to $3$, $5$ and $8$. 
The \textsf{AOSO} and \textsf{BNDM$q$} algorithms have been tested with a value of $q$ equal to $2$, $4$ and $6$.
Finally the \textsf{SBNDM$q$} and \textsf{UFNDM$q$} have been tested with $q$ equal to $2$, $4$, $6$ and $8$.

All algorithms have been implemented in the \textbf{C} programming language and were used to 
search for the same strings in large fixed text buffers on a PC with Intel Core2 processor of 1.66GHz and running times have been measured with a
hardware cycle counter, available on modern CPUs. The codes have been compiled with the \textsf{GNU C} Compiler, using the optimization options
\textsf{-O2 -fno-guess-branch-probability}. 

In particular, the algorithms have been tested on the following 12 text buffers:

\begin{enumerate}[(i)]
\item eight Rand$\sigma$ text buffers, for $\sigma=2,4,8,16,32,64,128$ and $256$, 
      where each Rand$\sigma$ text buffer consists in a 5Mb random text over a common alphabet of size $\sigma$, with a uniform distribution of characters;
\item a genome sequence of $4,638,690$ base pairs of \textit{Escherichia coli} (with $\sigma = 4$);
\item a protein sequence (the \textsf{hs} file) from the \textit{Saccharomyces cerevisiae} genome, of length $3,295,751$ byte (with $\sigma = 20$);
\item the English King James version of the \textsf{Bible} composed of $4,047,392$ characters (with $\sigma = 63$);
\item the file \textsf{world192.txt} (The CIA World Fact Book) composed of $2,473,400$ characters (with $\sigma = 94$);
\end{enumerate}

Files (ii), (iv) and (v) are from the Large Canterbury Corpus (\textsf{http://www.data-compression.info/Corpora/CanterburyCorpus/}), while file (iii) is from the
Protein Corpus (\textsf{http://data-compression.info/Corpora/ProteinCorpus/}).

For each input file, we have generated sets of 400 patterns of fixed length $m$ randomly extracted from the text, for $m$ ranging over the values
$2$, $4$, $8$, $16$, $32$, $64$, $128$, $256$, $512$ and $1024$. 
For each set of patterns we reported in a table the mean over the running times of the 400 runs.
Running times are expressed in thousandths of seconds. 

Moreover we color each running time value with different shades of blue-red. 
In particular better results are presented in tones verging to red while worse results are presented in tones verging to blue.
In addition best results are highlighted with a light gray background.

Although we tested more than 85 different algorithms, for the sake of clearness we include in the following tables only the algorithms that obtain,
for each text buffer and each pattern length, the 25 best results. We add a red marker to comparison based algorithms, while a green and a blue marker is added
to automata and bit parallel algorithms, respectively.

Then, for each table, we briefly discuss the performance of the string matching algorithms by referring to the following
four classes of patterns:
\begin{itemize}
\item very short patterns (pattern with $m\leq 4$);
\item short patterns (pattern with $4 < m\leq 32$);
\item long patterns (pattern with $32 < m\leq 256$);
\item very long patterns (pattern with $m > 256$);
\end{itemize}
Finally we discuss the overall performance of the tested algorithms by considering those algorithms 
which maintain good performance for all classes of patterns.

\newpage
\subsection{Experimental Results on Rand$2$ Problem}\label{sec:exp2}
In this section we present experimental results on a random text buffer over a binary alphabet.
Matching binary data is an interesting problem in computer science, since
binary data are omnipresent in telecom and computer network applications.
Many formats for data exchange between nodes in distributed computer systems
as well as most network protocols use binary representations. 

\begin{scriptsize}
\begin{center}
\begin{tabular*}{\textwidth}{@{\extracolsep{\fill}}|l|cc|ccc|ccc|cc|}
\hline
$m$ & 2 & 4 & 8 & 16 & 32 & 64 & 128 & 256 & 512 & 1024 \\
\hline

 {\color{red}~\textbullet}\algo{BF}~&~{\color{color2}44.6}~~&~{\color{color0}46.4}~~&~{\color{color0}52.4}~~&~{\color{color0}52.6}~~&~{\color{color0}52.5}~~&~{\color{color0}52.5}~~&~{\color{color0}52.5}~~&~{\color{color0}52.5}~~&~{\color{color0}52.5}~~&~{\color{color0}52.5}~~\\
 {\color{red}~\textbullet}\algo{KR}~&~{\color{color0}48.2}~~&~{\color{color13}24.7}~~&~{\color{color12}16.9}~~&~{\color{color0}16.4}~~&~{\color{color0}16.4}~~&~{\color{color0}16.4}~~&~{\color{color0}16.4}~~&~{\color{color0}16.4}~~&~{\color{color0}16.4}~~&~{\color{color0}16.4}~~\\
 {\color{red}~\textbullet}\algo{QS}~&~{\color{color6}38.7}~~&~{\color{color0}41.2}~~&~{\color{color0}44.0}~~&~{\color{color0}45.3}~~&~{\color{color0}44.4}~~&~{\color{color0}45.2}~~&~{\color{color0}45.5}~~&~{\color{color0}45.5}~~&~{\color{color0}45.7}~~&~{\color{color0}44.7}~~\\
 {\color{red}~\textbullet}\algo{NSN}~&~{\color{color7}37.4}~~&~{\color{color0}43.1}~~&~{\color{color0}43.2}~~&~{\color{color0}43.4}~~&~{\color{color0}43.4}~~&~{\color{color0}43.3}~~&~{\color{color0}43.3}~~&~{\color{color0}43.3}~~&~{\color{color0}43.3}~~&~{\color{color0}43.3}~~\\
 {\color{red}~\textbullet}\algo{Smith}~&~{\color{color1}46.8}~~&~{\color{color0}41.2}~~&~{\color{color0}39.8}~~&~{\color{color0}39.2}~~&~{\color{color0}40.2}~~&~{\color{color0}39.7}~~&~{\color{color0}39.5}~~&~{\color{color0}40.5}~~&~{\color{color0}40.3}~~&~{\color{color0}40.0}~~\\
 {\color{red}~\textbullet}\algo{RCol}~&~{\color{color2}44.5}~~&~{\color{color0}37.5}~~&~{\color{color0}28.4}~~&~{\color{color0}20.4}~~&~{\color{color0}15.2}~~&~{\color{color0}11.9}~~&~{\color{color0}9.80}~~&~{\color{color0}8.59}~~&~{\color{color0}7.09}~~&~{\color{color0}6.15}~~\\
 {\color{red}~\textbullet}\algo{ASkip}~&~{\color{color0}77.3}~~&~{\color{color0}53.2}~~&~{\color{color0}28.5}~~&~{\color{color0}15.2}~~&~{\color{color0}8.27}~~&~{\color{color8}4.89}~~&~{\color{color8}5.09}~~&~{\color{color11}3.74}~~&~{\color{color12}3.20}~~&~{\color{color9}3.75}~~\\
 {\color{red}~\textbullet}\algo{BR}~&~{\color{color8}35.2}~~&~{\color{color2}35.6}~~&~{\color{color0}34.6}~~&~{\color{color0}33.4}~~&~{\color{color0}33.4}~~&~{\color{color0}32.6}~~&~{\color{color0}32.2}~~&~{\color{color0}32.5}~~&~{\color{color0}33.3}~~&~{\color{color0}33.0}~~\\
 {\color{red}~\textbullet}\algo{FS}~&~{\color{color2}44.5}~~&~{\color{color1}37.2}~~&~{\color{color0}28.4}~~&~{\color{color0}20.1}~~&~{\color{color0}15.4}~~&~{\color{color0}12.0}~~&~{\color{color0}10.0}~~&~{\color{color0}8.55}~~&~{\color{color0}7.25}~~&~{\color{color0}6.07}~~\\
 {\color{red}~\textbullet}\algo{FFS}~&~{\color{color5}39.6}~~&~{\color{color4}33.9}~~&~{\color{color0}25.2}~~&~{\color{color0}16.4}~~&~{\color{color0}11.6}~~&~{\color{color0}8.41}~~&~{\color{color0}7.05}~~&~{\color{color0}6.13}~~&~{\color{color4}5.03}~~&~{\color{color6}4.57}~~\\
 {\color{red}~\textbullet}\algo{BFS}~&~{\color{color3}43.6}~~&~{\color{color1}37.0}~~&~{\color{color0}29.1}~~&~{\color{color0}20.5}~~&~{\color{color0}15.5}~~&~{\color{color0}12.2}~~&~{\color{color0}10.2}~~&~{\color{color0}9.04}~~&~{\color{color0}7.88}~~&~{\color{color0}7.44}~~\\
 {\color{red}~\textbullet}\algo{TS}~&~{\color{color7}37.5}~~&~{\color{color4}34.1}~~&~{\color{color0}27.5}~~&~{\color{color0}22.9}~~&~{\color{color0}19.3}~~&~{\color{color0}17.1}~~&~{\color{color0}15.5}~~&~{\color{color0}13.9}~~&~{\color{color0}12.6}~~&~{\color{color0}11.5}~~\\
 {\color{red}~\textbullet}\algo{SSABS}~&~{\color{color10}32.1}~~&~{\color{color0}37.8}~~&~{\color{color0}43.4}~~&~{\color{color0}46.0}~~&~{\color{color0}43.8}~~&~{\color{color0}44.7}~~&~{\color{color0}45.6}~~&~{\color{color0}44.5}~~&~{\color{color0}44.8}~~&~{\color{color0}46.5}~~\\
 {\color{red}~\textbullet}\algo{TVSBS}~&~{\color{color12}29.8}~~&~{\color{color3}34.3}~~&~{\color{color0}36.9}~~&~{\color{color0}36.1}~~&~{\color{color0}34.9}~~&~{\color{color0}36.6}~~&~{\color{color0}35.6}~~&~{\color{color0}36.2}~~&~{\color{color0}36.3}~~&~{\color{color0}35.5}~~\\
 {\color{red}~\textbullet}\algo{FJS}~&~{\color{color5}39.7}~~&~{\color{color0}42.9}~~&~{\color{color0}50.2}~~&~{\color{color0}49.4}~~&~{\color{color0}49.1}~~&~{\color{color0}50.2}~~&~{\color{color0}50.6}~~&~{\color{color0}50.0}~~&~{\color{color0}50.2}~~&~{\color{color0}49.8}~~\\
 {\color{red}~\textbullet}\algo{HASH3}~ &~-~&~{\color{color9}28.2}~~&~{\color{color17}14.0}~~&~{\color{color10}9.78}~~&~{\color{color0}8.64}~~&~{\color{color0}8.71}~~&~{\color{color0}8.88}~~&~{\color{color0}8.71}~~&~{\color{color0}8.72}~~&~{\color{color0}8.65}~~\\
 {\color{red}~\textbullet}\algo{HASH5}~ &~-~ &~-~&~{\color{color16}14.4}~~&~\colorbox{best}{\color{color20}6.05}~~&~{\color{color19}3.72}~~&~{\color{color17}3.07}~~&~{\color{color19}3.15}~~&~{\color{color14}3.12}~~&~{\color{color12}3.12}~~&~{\color{color11}3.12}~~\\
 {\color{red}~\textbullet}\algo{HASH8}~ &~-~ &~-~ &~-~&~{\color{color16}7.67}~~&~\colorbox{best}{\color{color20}3.47}~~&~\colorbox{best}{\color{color20}2.47}~~&~\colorbox{best}{\color{color20}2.87}~~&~{\color{color20}1.97}~~&~{\color{color19}1.44}~~&~{\color{color18}1.30}~~\\
 {\color{red}~\textbullet}\algo{SSEF}~ &~-~ &~-~ &~-~ &~-~&~{\color{color11}5.38}~~&~{\color{color16}3.38}~~&~{\color{color17}3.44}~~&~\colorbox{best}{\color{color20}1.79}~~&~\colorbox{best}{\color{color20}0.99}~~&~\colorbox{best}{\color{color20}0.55}~~\\
 \hline
 {\color{green}~\textbullet}\algo{AUT}~&~{\color{color17}21.7}~~&~{\color{color16}21.7}~~&~{\color{color5}21.7}~~&~{\color{color0}21.7}~~&~{\color{color0}21.7}~~&~{\color{color0}21.8}~~&~{\color{color0}21.8}~~&~{\color{color0}21.9}~~&~{\color{color0}22.6}~~&~{\color{color0}23.9}~~\\
 {\color{green}~\textbullet}\algo{RF}~&~{\color{color0}68.3}~~&~{\color{color0}50.8}~~&~{\color{color0}31.6}~~&~{\color{color0}16.9}~~&~{\color{color0}9.48}~~&~{\color{color2}6.19}~~&~{\color{color3}5.89}~~&~{\color{color8}4.32}~~&~{\color{color5}4.93}~~&~{\color{color0}6.27}~~\\
 {\color{green}~\textbullet}\algo{BOM}~&~{\color{color0}94.1}~~&~{\color{color0}74.3}~~&~{\color{color0}47.4}~~&~{\color{color0}28.9}~~&~{\color{color0}17.1}~~&~{\color{color0}9.94}~~&~{\color{color0}7.52}~~&~{\color{color9}4.14}~~&~{\color{color15}2.27}~~&~{\color{color18}1.27}~~\\
 {\color{green}~\textbullet}\algo{BOM2}~&~{\color{color0}84.7}~~&~{\color{color0}61.1}~~&~{\color{color0}34.7}~~&~{\color{color0}17.9}~~&~{\color{color0}9.51}~~&~{\color{color6}5.30}~~&~{\color{color9}4.93}~~&~{\color{color15}2.91}~~&~{\color{color17}1.87}~~&~{\color{color13}2.70}~~\\
 {\color{green}~\textbullet}\algo{WW}~&~{\color{color0}70.0}~~&~{\color{color0}53.0}~~&~{\color{color0}35.1}~~&~{\color{color0}19.9}~~&~{\color{color0}11.8}~~&~{\color{color0}7.43}~~&~{\color{color0}6.90}~~&~{\color{color2}5.77}~~&~{\color{color0}7.07}~~&~{\color{color0}10.0}~~\\
 {\color{green}~\textbullet}\algo{ILDM1}~&~{\color{color5}40.5}~~&~{\color{color7}31.1}~~&~{\color{color2}23.9}~~&~{\color{color0}16.9}~~&~{\color{color0}11.2}~~&~{\color{color0}7.73}~~&~{\color{color0}7.16}~~&~{\color{color0}6.09}~~&~{\color{color0}7.39}~~&~{\color{color0}10.4}~~\\
 {\color{green}~\textbullet}\algo{ILDM2}~&~{\color{color0}54.7}~~&~{\color{color0}38.9}~~&~{\color{color2}23.6}~~&~{\color{color0}12.8}~~&~{\color{color2}7.42}~~&~{\color{color7}5.20}~~&~{\color{color5}5.56}~~&~{\color{color6}4.81}~~&~{\color{color0}6.46}~~&~{\color{color0}9.90}~~\\
 {\color{green}~\textbullet}\algo{EBOM}~&~{\color{color4}41.1}~~&~{\color{color1}37.2}~~&~{\color{color0}25.8}~~&~{\color{color0}14.4}~~&~{\color{color0}8.06}~~&~{\color{color9}4.77}~~&~{\color{color11}4.61}~~&~{\color{color16}2.79}~~&~{\color{color16}1.99}~~&~{\color{color12}2.92}~~\\
 {\color{green}~\textbullet}\algo{FBOM}~&~{\color{color0}55.6}~~&~{\color{color0}43.8}~~&~{\color{color0}28.3}~~&~{\color{color0}15.9}~~&~{\color{color0}8.71}~~&~{\color{color7}5.19}~~&~{\color{color9}4.91}~~&~{\color{color15}2.95}~~&~{\color{color16}2.05}~~&~{\color{color12}2.98}~~\\
 {\color{green}~\textbullet}\algo{SEBOM}~&~{\color{color4}41.4}~~&~{\color{color1}37.0}~~&~{\color{color0}25.3}~~&~{\color{color0}14.6}~~&~{\color{color0}8.17}~~&~{\color{color8}4.89}~~&~{\color{color10}4.79}~~&~{\color{color15}2.94}~~&~{\color{color16}2.07}~~&~{\color{color12}2.98}~~\\
 {\color{green}~\textbullet}\algo{SFBOM}~&~{\color{color0}52.0}~~&~{\color{color0}40.5}~~&~{\color{color0}26.2}~~&~{\color{color0}14.8}~~&~{\color{color0}8.24}~~&~{\color{color8}4.90}~~&~{\color{color10}4.75}~~&~{\color{color15}2.93}~~&~{\color{color16}2.06}~~&~{\color{color12}2.97}~~\\
 \hline
 {\color{blu}~\textbullet}\algo{SO}~&~{\color{color20}16.4}~~&~{\color{color20}16.4}~~&~{\color{color13}16.4}~~&~{\color{color0}16.4}~~&~{\color{color0}16.4}~~&~{\color{color0}21.7}~~&~{\color{color0}21.7}~~&~{\color{color0}21.8}~~&~{\color{color0}21.8}~~&~{\color{color0}21.7}~~\\
 {\color{blu}~\textbullet}\algo{SA}~&~\colorbox{best}{\color{color20}16.4}~~&~\colorbox{best}{\color{color20}16.4}~~&~{\color{color13}16.4}~~&~{\color{color0}16.4}~~&~{\color{color0}16.4}~~&~{\color{color0}19.1}~~&~{\color{color0}19.1}~~&~{\color{color0}19.1}~~&~{\color{color0}19.1}~~&~{\color{color0}19.1}~~\\
 {\color{blu}~\textbullet}\algo{BNDM}~&~{\color{color0}63.5}~~&~{\color{color0}47.9}~~&~{\color{color0}25.6}~~&~{\color{color1}12.6}~~&~{\color{color6}6.48}~~&~{\color{color0}8.53}~~&~{\color{color0}8.52}~~&~{\color{color0}8.52}~~&~{\color{color0}8.53}~~&~{\color{color0}8.50}~~\\
 {\color{blu}~\textbullet}\algo{BNDM-L}~&~{\color{color0}63.4}~~&~{\color{color0}46.6}~~&~{\color{color0}25.3}~~&~{\color{color2}12.5}~~&~{\color{color7}6.40}~~&~{\color{color0}13.7}~~&~{\color{color0}15.9}~~&~{\color{color0}16.3}~~&~{\color{color0}16.4}~~&~{\color{color0}17.0}~~\\
 {\color{blu}~\textbullet}\algo{SBNDM}~&~{\color{color0}56.1}~~&~{\color{color0}38.1}~~&~{\color{color3}23.4}~~&~{\color{color4}11.8}~~&~{\color{color8}6.17}~~&~{\color{color3}5.92}~~&~{\color{color3}5.91}~~&~{\color{color1}5.91}~~&~{\color{color1}5.91}~~&~{\color{color1}5.90}~~\\
 {\color{blu}~\textbullet}\algo{SBNDM2}~&~{\color{color0}52.5}~~&~{\color{color2}35.8}~~&~{\color{color6}21.0}~~&~{\color{color6}10.9}~~&~{\color{color9}5.93}~~&~{\color{color3}5.98}~~&~{\color{color3}5.98}~~&~{\color{color0}5.99}~~&~{\color{color0}5.98}~~&~{\color{color1}5.99}~~\\
 {\color{blu}~\textbullet}\algo{SBNDM-BMH}~&~{\color{color1}46.6}~~&~{\color{color0}37.6}~~&~{\color{color2}23.6}~~&~{\color{color4}11.8}~~&~{\color{color8}6.13}~~&~{\color{color3}5.92}~~&~{\color{color3}5.91}~~&~{\color{color1}5.91}~~&~{\color{color1}5.91}~~&~{\color{color1}5.90}~~\\
 {\color{blu}~\textbullet}\algo{FAOSOq2}~&~{\color{color0}150}~~&~{\color{color0}104}~~&~{\color{color0}39.7}~~&~{\color{color2}12.5}~~&~{\color{color0}9.97}~~&~{\color{color0}9.97}~~&~{\color{color0}9.97}~~&~{\color{color0}9.97}~~&~{\color{color0}9.96}~~&~{\color{color0}9.98}~~\\
 {\color{blu}~\textbullet}\algo{AOSO2}~&~{\color{color0}167}~~&~{\color{color1}36.7}~~&~\colorbox{best}{\color{color20}11.5}~~&~{\color{color10}9.66}~~&~{\color{color0}8.54}~~&~{\color{color0}8.53}~~&~{\color{color0}8.55}~~&~{\color{color0}8.54}~~&~{\color{color0}8.55}~~ &~-~\\
 {\color{blu}~\textbullet}\algo{AOSO4}~ &~-~&~{\color{color0}147}~~&~{\color{color0}90.9}~~&~{\color{color0}32.1}~~&~{\color{color4}6.92}~~&~{\color{color0}6.39}~~&~{\color{color0}6.40}~~&~{\color{color0}6.39}~~&~{\color{color0}6.40}~~&~{\color{color0}6.40}~~\\
 {\color{blu}~\textbullet}\algo{FSBNDM}~&~{\color{color0}56.6}~~&~{\color{color0}37.7}~~&~{\color{color8}20.0}~~&~{\color{color8}10.2}~~&~{\color{color10}5.69}~~&~{\color{color4}5.69}~~&~{\color{color4}5.70}~~&~{\color{color2}5.71}~~&~{\color{color2}5.70}~~&~{\color{color2}5.69}~~\\
 {\color{blu}~\textbullet}\algo{BNDMq2}~&~{\color{color0}51.8}~~&~{\color{color2}35.8}~~&~{\color{color6}21.1}~~&~{\color{color5}11.4}~~&~{\color{color6}6.45}~~&~{\color{color0}7.84}~~&~{\color{color0}7.83}~~&~{\color{color0}7.80}~~&~{\color{color0}7.83}~~&~{\color{color0}7.84}~~\\
 {\color{blu}~\textbullet}\algo{BNDMq4}~ &~-~&~{\color{color0}53.0}~~&~{\color{color10}18.4}~~&~{\color{color10}9.49}~~&~{\color{color13}5.10}~~&~{\color{color0}6.51}~~&~{\color{color0}6.49}~~&~{\color{color0}6.50}~~&~{\color{color0}6.51}~~&~{\color{color0}6.49}~~\\
 {\color{blu}~\textbullet}\algo{BNDMq6}~ &~-~ &~-~&~{\color{color0}26.3}~~&~{\color{color11}9.13}~~&~{\color{color13}5.08}~~&~{\color{color7}5.11}~~&~{\color{color8}5.13}~~&~{\color{color5}5.12}~~&~{\color{color4}5.14}~~&~{\color{color4}5.12}~~\\
 {\color{blu}~\textbullet}\algo{SBNDMq2}~&~{\color{color0}51.1}~~&~{\color{color3}35.0}~~&~{\color{color6}20.8}~~&~{\color{color6}10.9}~~&~{\color{color10}5.73}~~&~{\color{color3}5.99}~~&~{\color{color3}5.99}~~&~{\color{color0}5.99}~~&~{\color{color0}5.98}~~&~{\color{color1}6.00}~~\\
 {\color{blu}~\textbullet}\algo{SBNDMq4}~ &~-~&~{\color{color0}49.7}~~&~{\color{color11}17.9}~~&~{\color{color10}9.79}~~&~{\color{color11}5.49}~~&~{\color{color6}5.37}~~&~{\color{color6}5.39}~~&~{\color{color3}5.37}~~&~{\color{color3}5.39}~~&~{\color{color3}5.38}~~\\
 {\color{blu}~\textbullet}\algo{SBNDMq6}~ &~-~ &~-~&~{\color{color0}29.4}~~&~{\color{color9}9.80}~~&~{\color{color12}5.25}~~&~{\color{color8}4.86}~~&~{\color{color9}4.88}~~&~{\color{color6}4.88}~~&~{\color{color5}4.86}~~&~{\color{color5}4.88}~~\\
 {\color{blu}~\textbullet}\algo{SBNDMq8}~ &~-~ &~-~&~{\color{color0}97.0}~~&~{\color{color3}11.9}~~&~{\color{color13}5.02}~~&~{\color{color9}4.63}~~&~{\color{color11}4.63}~~&~{\color{color7}4.65}~~&~{\color{color6}4.66}~~&~{\color{color6}4.64}~~\\
 {\color{blu}~\textbullet}\algo{UFNDMq4}~&~{\color{color0}56.8}~~&~{\color{color6}31.9}~~&~{\color{color4}22.2}~~&~{\color{color0}13.4}~~&~{\color{color0}8.58}~~&~{\color{color0}8.63}~~&~{\color{color0}8.61}~~&~{\color{color0}8.58}~~&~{\color{color0}8.60}~~&~{\color{color0}8.57}~~\\
 {\color{blu}~\textbullet}\algo{UFNDMq6}~&~{\color{color0}57.6}~~&~{\color{color2}35.5}~~&~{\color{color11}17.9}~~&~{\color{color7}10.7}~~&~{\color{color1}7.57}~~&~{\color{color0}7.55}~~&~{\color{color0}7.59}~~&~{\color{color0}7.57}~~&~{\color{color0}7.56}~~&~{\color{color0}7.58}~~\\
 {\color{blu}~\textbullet}\algo{UFNDMq8}~&~{\color{color0}58.5}~~&~{\color{color0}38.7}~~&~{\color{color10}18.4}~~&~{\color{color9}10.1}~~&~{\color{color3}7.12}~~&~{\color{color0}7.12}~~&~{\color{color0}7.14}~~&~{\color{color0}7.12}~~&~{\color{color0}7.12}~~&~{\color{color0}7.14}~~\\
\hline\end{tabular*}\\
\end{center}
\end{scriptsize}

In the case of very short patterns the \textsf{SO} and \textsf{SA} algorithms obtain the best results.
The \textsf{AUT} algorithm obtains also good results. 
For short patterns the algorithms based on bit-parallelism achieves good results. The \textsf{AOSO2} algorithm is the best for patterns of length 8, while 
\textsf{HASH$q$} algorithms obtain best results for patterns of length $16$ and $32$.
In the case of long patterns the best results are obtained by the \textsf{HASH$q$} algorithms and by the \textsf{SSEF} algorithm (for patterns of length 256). 
For very long patterns the best results are obtained by the \textsf{SSEF} algorithm. 
Regarding the overall performance no algorithm maintains good performances for all patterns. However when the pattern is short the \textsf{SA} algorithm is a good 
choice while the \textsf{HASH$5$} and the \textsf{SSEF} algorithms are suggested for patterns with a length greater than $16$.

\vfill

\subsection{Experimental Results on Rand$4$ Problem}\label{sec:exp4}
Matching data over four characters alphabet is an interesting problem in computer science mostly related with computational biology.
It is the case, for instance, of DNA sequences which are constructed over an alphabet of four bases.

\begin{scriptsize}
\begin{center}
\begin{tabular*}{\textwidth}{@{\extracolsep{\fill}}|l|cc|ccc|ccc|cc|}
\hline
$m$ & 2 & 4 & 8 & 16 & 32 & 64 & 128 & 256 & 512 & 1024 \\
\hline

 {\color{red}~\textbullet}\algo{KR}~&~{\color{color0}29.7}~~&~{\color{color5}19.4}~~&~{\color{color0}16.5}~~&~{\color{color0}16.4}~~&~{\color{color0}16.3}~~&~{\color{color0}16.4}~~&~{\color{color0}16.4}~~&~{\color{color0}16.4}~~&~{\color{color0}16.4}~~&~{\color{color0}16.4}~~\\
 {\color{red}~\textbullet}\algo{QS}~&~{\color{color2}28.8}~~&~{\color{color0}21.8}~~&~{\color{color0}16.6}~~&~{\color{color0}15.4}~~&~{\color{color0}15.2}~~&~{\color{color0}15.4}~~&~{\color{color0}15.4}~~&~{\color{color0}15.4}~~&~{\color{color0}15.2}~~&~{\color{color0}15.4}~~\\
 {\color{red}~\textbullet}\algo{NSN}~&~{\color{color4}27.1}~~&~{\color{color0}30.2}~~&~{\color{color0}30.1}~~&~{\color{color0}29.9}~~&~{\color{color0}30.0}~~&~{\color{color0}30.4}~~&~{\color{color0}30.0}~~&~{\color{color0}29.8}~~&~{\color{color0}29.6}~~&~{\color{color0}30.2}~~\\
 {\color{red}~\textbullet}\algo{Raita}~&~{\color{color1}29.4}~~&~{\color{color9}18.3}~~&~{\color{color0}13.7}~~&~{\color{color0}12.9}~~&~{\color{color0}12.8}~~&~{\color{color0}13.2}~~&~{\color{color0}12.8}~~&~{\color{color0}12.6}~~&~{\color{color0}12.3}~~&~{\color{color0}12.8}~~\\
 {\color{red}~\textbullet}\algo{RCol}~&~{\color{color4}27.5}~~&~{\color{color7}18.8}~~&~{\color{color0}13.7}~~&~{\color{color0}11.5}~~&~{\color{color0}9.87}~~&~{\color{color0}8.56}~~&~{\color{color0}7.71}~~&~{\color{color0}7.03}~~&~{\color{color0}6.04}~~&~{\color{color0}5.51}~~\\
 {\color{red}~\textbullet}\algo{ASkip}~&~{\color{color0}55.4}~~&~{\color{color0}31.9}~~&~{\color{color0}15.1}~~&~{\color{color0}7.90}~~&~{\color{color0}4.68}~~&~{\color{color3}3.54}~~&~{\color{color0}4.34}~~&~{\color{color0}3.94}~~&~{\color{color0}5.08}~~&~{\color{color0}8.88}~~\\
 {\color{red}~\textbullet}\algo{BR}~&~{\color{color8}24.7}~~&~{\color{color9}18.2}~~&~{\color{color3}12.2}~~&~{\color{color0}8.30}~~&~{\color{color0}6.31}~~&~{\color{color0}5.74}~~&~{\color{color0}5.65}~~&~{\color{color0}5.68}~~&~{\color{color0}5.69}~~&~{\color{color0}5.64}~~\\
 {\color{red}~\textbullet}\algo{FS}~&~{\color{color3}27.5}~~&~{\color{color7}18.9}~~&~{\color{color0}14.0}~~&~{\color{color0}11.8}~~&~{\color{color0}9.81}~~&~{\color{color0}8.62}~~&~{\color{color0}7.58}~~&~{\color{color0}6.99}~~&~{\color{color0}6.15}~~&~{\color{color0}5.54}~~\\
 {\color{red}~\textbullet}\algo{FFS}~&~{\color{color5}26.6}~~&~{\color{color9}18.3}~~&~{\color{color0}12.8}~~&~{\color{color0}9.51}~~&~{\color{color0}7.41}~~&~{\color{color0}5.65}~~&~{\color{color0}5.04}~~&~{\color{color0}4.44}~~&~{\color{color0}3.84}~~&~{\color{color2}3.56}~~\\
 {\color{red}~\textbullet}\algo{BFS}~&~{\color{color3}27.6}~~&~{\color{color7}18.8}~~&~{\color{color1}12.8}~~&~{\color{color0}9.88}~~&~{\color{color0}7.68}~~&~{\color{color0}6.23}~~&~{\color{color0}5.47}~~&~{\color{color0}4.94}~~&~{\color{color0}4.28}~~&~{\color{color0}4.14}~~\\
 {\color{red}~\textbullet}\algo{TS}~&~{\color{color4}27.4}~~&~{\color{color0}22.3}~~&~{\color{color0}16.1}~~&~{\color{color0}11.4}~~&~{\color{color0}9.11}~~&~{\color{color0}7.78}~~&~{\color{color0}6.86}~~&~{\color{color0}6.19}~~&~{\color{color0}5.75}~~&~{\color{color0}5.28}~~\\
 {\color{red}~\textbullet}\algo{SSABS}~&~{\color{color7}25.1}~~&~{\color{color1}20.9}~~&~{\color{color0}17.4}~~&~{\color{color0}16.8}~~&~{\color{color0}16.9}~~&~{\color{color0}16.9}~~&~{\color{color0}16.8}~~&~{\color{color0}16.7}~~&~{\color{color0}16.6}~~&~{\color{color0}16.4}~~\\
 {\color{red}~\textbullet}\algo{TVSBS}~&~{\color{color12}22.2}~~&~{\color{color11}17.4}~~&~{\color{color4}12.1}~~&~{\color{color0}8.67}~~&~{\color{color0}6.90}~~&~{\color{color0}6.40}~~&~{\color{color0}6.40}~~&~{\color{color0}6.30}~~&~{\color{color0}6.29}~~&~{\color{color0}6.37}~~\\
 {\color{red}~\textbullet}\algo{HASH3}~ &~-~&~{\color{color0}21.1}~~&~\colorbox{best}{\color{color20}8.30}~~&~{\color{color20}4.74}~~&~{\color{color14}3.42}~~&~{\color{color11}3.07}~~&~{\color{color14}3.13}~~&~{\color{color7}3.09}~~&~{\color{color6}3.10}~~&~{\color{color5}3.08}~~\\
 {\color{red}~\textbullet}\algo{HASH5}~ &~-~ &~-~&~{\color{color2}12.5}~~&~{\color{color19}5.01}~~&~{\color{color20}2.93}~~&~{\color{color20}2.48}~~&~{\color{color18}2.85}~~&~{\color{color13}2.49}~~&~{\color{color12}2.24}~~&~{\color{color10}2.19}~~\\
 {\color{red}~\textbullet}\algo{HASH8}~ &~-~ &~-~ &~-~&~{\color{color2}7.62}~~&~{\color{color14}3.45}~~&~\colorbox{best}{\color{color20}2.46}~~&~{\color{color18}2.85}~~&~{\color{color19}1.96}~~&~{\color{color17}1.45}~~&~{\color{color16}1.30}~~\\
 {\color{red}~\textbullet}\algo{TSW}~&~{\color{color1}29.2}~~&~{\color{color0}21.6}~~&~{\color{color0}14.6}~~&~{\color{color0}10.0}~~&~{\color{color0}7.71}~~&~{\color{color0}6.89}~~&~{\color{color0}6.83}~~&~{\color{color0}6.79}~~&~{\color{color0}6.85}~~&~{\color{color0}6.85}~~\\
 {\color{red}~\textbullet}\algo{SSEF}~ &~-~ &~-~ &~-~ &~-~&~{\color{color0}5.39}~~&~{\color{color6}3.36}~~&~{\color{color9}3.43}~~&~\colorbox{best}{\color{color20}1.79}~~&~\colorbox{best}{\color{color20}0.99}~~&~\colorbox{best}{\color{color20}0.54}~~\\
 \hline
 {\color{green}~\textbullet}\algo{AUT}~&~{\color{color12}21.7}~~&~{\color{color0}21.7}~~&~{\color{color0}21.7}~~&~{\color{color0}21.7}~~&~{\color{color0}21.7}~~&~{\color{color0}21.7}~~&~{\color{color0}21.8}~~&~{\color{color0}21.9}~~&~{\color{color0}22.6}~~&~{\color{color0}23.9}~~\\
 {\color{green}~\textbullet}\algo{RF}~&~{\color{color0}49.4}~~&~{\color{color0}30.8}~~&~{\color{color0}17.0}~~&~{\color{color0}9.71}~~&~{\color{color0}5.69}~~&~{\color{color1}3.69}~~&~{\color{color0}3.98}~~&~{\color{color9}2.92}~~&~{\color{color5}3.21}~~&~{\color{color0}4.83}~~\\
 {\color{green}~\textbullet}\algo{BOM}~&~{\color{color0}65.7}~~&~{\color{color0}44.0}~~&~{\color{color0}27.7}~~&~{\color{color0}17.6}~~&~{\color{color0}11.2}~~&~{\color{color0}6.84}~~&~{\color{color0}5.79}~~&~{\color{color6}3.20}~~&~{\color{color15}1.78}~~&~{\color{color18}1.03}~~\\
 {\color{green}~\textbullet}\algo{BOM2}~&~{\color{color0}56.5}~~&~{\color{color0}32.3}~~&~{\color{color0}18.0}~~&~{\color{color0}10.2}~~&~{\color{color0}5.77}~~&~{\color{color4}3.51}~~&~{\color{color3}3.83}~~&~{\color{color16}2.24}~~&~{\color{color17}1.50}~~&~{\color{color9}2.49}~~\\
 {\color{green}~\textbullet}\algo{ILDM2}~&~{\color{color0}43.4}~~&~{\color{color0}24.3}~~&~{\color{color0}13.7}~~&~{\color{color0}8.12}~~&~{\color{color0}4.72}~~&~{\color{color7}3.29}~~&~{\color{color0}3.99}~~&~{\color{color1}3.62}~~&~{\color{color0}5.03}~~&~{\color{color0}8.65}~~\\
 {\color{green}~\textbullet}\algo{EBOM}~&~{\color{color9}24.0}~~&~\colorbox{best}{\color{color20}14.2}~~&~{\color{color12}10.2}~~&~{\color{color7}6.94}~~&~{\color{color2}4.43}~~&~{\color{color11}3.06}~~&~{\color{color7}3.55}~~&~{\color{color17}2.16}~~&~{\color{color16}1.61}~~&~{\color{color7}2.75}~~\\
 {\color{green}~\textbullet}\algo{FBOM}~&~{\color{color1}29.3}~~&~{\color{color9}18.1}~~&~{\color{color4}12.0}~~&~{\color{color1}7.82}~~&~{\color{color0}5.06}~~&~{\color{color6}3.39}~~&~{\color{color3}3.81}~~&~{\color{color15}2.28}~~&~{\color{color16}1.66}~~&~{\color{color7}2.77}~~\\
 {\color{green}~\textbullet}\algo{SEBOM}~&~{\color{color8}24.8}~~&~{\color{color20}14.3}~~&~{\color{color11}10.3}~~&~{\color{color6}7.08}~~&~{\color{color1}4.58}~~&~{\color{color8}3.23}~~&~{\color{color5}3.71}~~&~{\color{color15}2.27}~~&~{\color{color16}1.67}~~&~{\color{color7}2.79}~~\\
 {\color{green}~\textbullet}\algo{SFBOM}~&~{\color{color1}28.9}~~&~{\color{color9}18.1}~~&~{\color{color6}11.7}~~&~{\color{color3}7.57}~~&~{\color{color0}4.83}~~&~{\color{color8}3.23}~~&~{\color{color5}3.68}~~&~{\color{color15}2.27}~~&~{\color{color16}1.67}~~&~{\color{color7}2.78}~~\\
 \hline
 {\color{blu}~\textbullet}\algo{SO}~&~{\color{color20}16.4}~~&~{\color{color14}16.4}~~&~{\color{color0}16.4}~~&~{\color{color0}16.4}~~&~{\color{color0}16.4}~~&~{\color{color0}21.8}~~&~{\color{color0}21.8}~~&~{\color{color0}21.7}~~&~{\color{color0}21.7}~~&~{\color{color0}21.8}~~\\
 {\color{blu}~\textbullet}\algo{SA}~&~\colorbox{best}{\color{color20}16.4}~~&~{\color{color14}16.4}~~&~{\color{color0}16.4}~~&~{\color{color0}16.4}~~&~{\color{color0}16.4}~~&~{\color{color0}19.1}~~&~{\color{color0}19.1}~~&~{\color{color0}19.1}~~&~{\color{color0}19.1}~~&~{\color{color0}19.1}~~\\
 {\color{blu}~\textbullet}\algo{BNDM}~&~{\color{color0}49.0}~~&~{\color{color0}27.8}~~&~{\color{color0}14.7}~~&~{\color{color0}7.91}~~&~{\color{color3}4.40}~~&~{\color{color0}5.85}~~&~{\color{color0}5.88}~~&~{\color{color0}5.86}~~&~{\color{color0}5.87}~~&~{\color{color0}5.86}~~\\
 {\color{blu}~\textbullet}\algo{BNDM-L}~&~{\color{color0}49.4}~~&~{\color{color0}27.8}~~&~{\color{color0}14.7}~~&~{\color{color0}7.90}~~&~{\color{color3}4.40}~~&~{\color{color0}7.70}~~&~{\color{color0}9.55}~~&~{\color{color0}8.82}~~&~{\color{color0}8.57}~~&~{\color{color0}8.98}~~\\
 {\color{blu}~\textbullet}\algo{SBNDM}~&~{\color{color0}49.3}~~&~{\color{color0}22.5}~~&~{\color{color1}12.6}~~&~{\color{color5}7.21}~~&~{\color{color7}4.03}~~&~{\color{color0}3.84}~~&~{\color{color3}3.83}~~&~{\color{color0}3.82}~~&~{\color{color1}3.81}~~&~{\color{color0}3.82}~~\\
 {\color{blu}~\textbullet}\algo{SBNDM2}~&~{\color{color0}39.4}~~&~{\color{color8}18.4}~~&~{\color{color9}10.8}~~&~{\color{color10}6.34}~~&~{\color{color10}3.73}~~&~{\color{color2}3.65}~~&~{\color{color5}3.66}~~&~{\color{color1}3.65}~~&~{\color{color2}3.66}~~&~{\color{color1}3.66}~~\\
 {\color{blu}~\textbullet}\algo{SBNDM-BMH}~&~{\color{color0}32.6}~~&~{\color{color0}21.1}~~&~{\color{color1}12.7}~~&~{\color{color5}7.23}~~&~{\color{color7}4.05}~~&~{\color{color0}3.83}~~&~{\color{color3}3.81}~~&~{\color{color0}3.82}~~&~{\color{color0}3.84}~~&~{\color{color0}3.82}~~\\
 {\color{blu}~\textbullet}\algo{BMH-SBNDM}~&~{\color{color1}29.4}~~&~{\color{color5}19.4}~~&~{\color{color0}12.8}~~&~{\color{color0}8.82}~~&~{\color{color0}5.79}~~&~{\color{color0}5.74}~~&~{\color{color0}5.75}~~&~{\color{color0}5.76}~~&~{\color{color0}5.78}~~&~{\color{color0}5.81}~~\\
 {\color{blu}~\textbullet}\algo{FAOSOq2}~&~{\color{color0}97.6}~~&~{\color{color0}37.8}~~&~{\color{color3}12.3}~~&~{\color{color0}10.6}~~&~{\color{color0}9.95}~~&~{\color{color0}9.95}~~&~{\color{color0}9.95}~~&~{\color{color0}9.95}~~&~{\color{color0}9.95}~~&~{\color{color0}9.96}~~\\
 {\color{blu}~\textbullet}\algo{FAOSOq4}~ &~-~&~{\color{color0}79.7}~~&~{\color{color0}31.2}~~&~{\color{color4}7.34}~~&~{\color{color0}5.35}~~&~{\color{color0}5.36}~~&~{\color{color0}5.36}~~&~{\color{color0}5.36}~~&~{\color{color0}5.35}~~&~{\color{color0}5.36}~~\\
 {\color{blu}~\textbullet}\algo{AOSO2}~&~{\color{color0}102}~~&~{\color{color0}35.0}~~&~{\color{color7}11.3}~~&~{\color{color0}9.69}~~&~{\color{color0}9.69}~~&~{\color{color0}8.54}~~&~{\color{color0}8.55}~~&~{\color{color0}8.54}~~&~{\color{color0}8.54}~~&~{\color{color0}8.55}~~\\
 {\color{blu}~\textbullet}\algo{AOSO4}~ &~-~&~{\color{color0}84.7}~~&~{\color{color0}29.6}~~&~{\color{color8}6.71}~~&~{\color{color0}5.06}~~&~{\color{color0}4.54}~~&~{\color{color0}4.55}~~&~{\color{color0}4.55}~~&~{\color{color0}4.55}~~&~{\color{color0}4.54}~~\\
 {\color{blu}~\textbullet}\algo{AOSO6}~ &~-~ &~-~&~{\color{color0}74.8}~~&~{\color{color0}28.2}~~&~{\color{color8}3.96}~~&~{\color{color1}3.70}~~&~{\color{color5}3.69}~~&~{\color{color0}3.70}~~&~{\color{color1}3.71}~~&~{\color{color1}3.71}~~\\
 {\color{blu}~\textbullet}\algo{FSBNDM}~&~{\color{color0}39.7}~~&~{\color{color1}21.1}~~&~{\color{color7}11.4}~~&~{\color{color11}6.23}~~&~{\color{color15}3.36}~~&~{\color{color6}3.38}~~&~{\color{color10}3.37}~~&~{\color{color4}3.37}~~&~{\color{color4}3.37}~~&~{\color{color3}3.37}~~\\
 {\color{blu}~\textbullet}\algo{BNDMq2}~&~{\color{color0}37.5}~~&~{\color{color8}18.4}~~&~{\color{color9}10.8}~~&~{\color{color11}6.28}~~&~{\color{color11}3.66}~~&~{\color{color0}4.56}~~&~{\color{color0}4.56}~~&~{\color{color0}4.55}~~&~{\color{color0}4.55}~~&~{\color{color0}4.55}~~\\
 {\color{blu}~\textbullet}\algo{BNDMq4}~ &~-~&~{\color{color0}48.7}~~&~{\color{color9}10.8}~~&~{\color{color19}4.89}~~&~\colorbox{best}{\color{color20}2.86}~~&~{\color{color4}3.53}~~&~{\color{color7}3.54}~~&~{\color{color2}3.54}~~&~{\color{color3}3.54}~~&~{\color{color2}3.53}~~\\
 {\color{blu}~\textbullet}\algo{BNDMq6}~ &~-~ &~-~&~{\color{color0}24.0}~~&~{\color{color5}7.24}~~&~{\color{color13}3.53}~~&~{\color{color9}3.22}~~&~{\color{color12}3.20}~~&~{\color{color6}3.21}~~&~{\color{color5}3.21}~~&~{\color{color4}3.22}~~\\
 {\color{blu}~\textbullet}\algo{SBNDMq2}~&~{\color{color0}37.1}~~&~{\color{color10}17.8}~~&~{\color{color10}10.7}~~&~{\color{color11}6.23}~~&~{\color{color11}3.67}~~&~{\color{color2}3.66}~~&~{\color{color5}3.66}~~&~{\color{color1}3.66}~~&~{\color{color2}3.66}~~&~{\color{color1}3.66}~~\\
 {\color{blu}~\textbullet}\algo{SBNDMq4}~ &~-~&~{\color{color0}46.0}~~&~{\color{color12}10.2}~~&~\colorbox{best}{\color{color20}4.72}~~&~{\color{color20}2.87}~~&~{\color{color17}2.68}~~&~\colorbox{best}{\color{color20}2.68}~~&~{\color{color11}2.69}~~&~{\color{color9}2.68}~~&~{\color{color7}2.69}~~\\
 {\color{blu}~\textbullet}\algo{SBNDMq6}~ &~-~ &~-~&~{\color{color0}27.4}~~&~{\color{color0}8.04}~~&~{\color{color10}3.76}~~&~{\color{color9}3.22}~~&~{\color{color12}3.21}~~&~{\color{color6}3.22}~~&~{\color{color5}3.22}~~&~{\color{color4}3.22}~~\\
 {\color{blu}~\textbullet}\algo{SBNDMq8}~ &~-~ &~-~&~{\color{color0}97.0}~~&~{\color{color0}11.4}~~&~{\color{color1}4.55}~~&~{\color{color0}4.17}~~&~{\color{color0}4.17}~~&~{\color{color0}4.17}~~&~{\color{color0}4.17}~~&~{\color{color0}4.17}~~\\
 {\color{blu}~\textbullet}\algo{UFNDMq4}~&~{\color{color0}45.2}~~&~{\color{color0}21.8}~~&~{\color{color6}11.6}~~&~{\color{color9}6.52}~~&~{\color{color7}4.07}~~&~{\color{color0}4.06}~~&~{\color{color0}4.07}~~&~{\color{color0}4.06}~~&~{\color{color0}4.06}~~&~{\color{color0}4.07}~~\\
 {\color{blu}~\textbullet}\algo{UFNDMq6}~&~{\color{color0}52.5}~~&~{\color{color0}28.2}~~&~{\color{color0}14.2}~~&~{\color{color3}7.49}~~&~{\color{color0}4.89}~~&~{\color{color0}4.87}~~&~{\color{color0}4.89}~~&~{\color{color0}4.89}~~&~{\color{color0}4.90}~~&~{\color{color0}4.89}~~\\
 {\color{blu}~\textbullet}\algo{KBNDM}~&~{\color{color0}52.5}~~&~{\color{color0}28.2}~~&~{\color{color0}17.1}~~&~{\color{color0}10.5}~~&~{\color{color0}6.25}~~&~{\color{color0}3.92}~~&~{\color{color1}3.92}~~&~{\color{color0}3.92}~~&~{\color{color0}3.91}~~&~{\color{color0}3.90}~~\\
\hline\end{tabular*}\\
\end{center}
\end{scriptsize}

In the case of very short patterns the \textsf{SA} and \textsf{SO} algorithms obtain the best results. 
For short patterns the algorithms based on bit-parallelism achieve better results, in particular \textsf{BNDMq4} and \textsf{SBNDMq4}.
Other algorithms like \textsf{HASH5}, \textsf{HASH8}, \textsf{EBOM} and \textsf{SEBOM} are quite competitive. In particular the \textsf{HASH3} algorithm
obtains the best results for patterns of length 8.
In the case of long patterns the best results are obtained by the \textsf{SSEF} algorithm. However the algorithm in the \textsf{EBOM} family are good choices. 
Among the algorithm base on character comparisons the \textsf{HASH5} and \textsf{HASH8} algorithms achieve good results.
Among the algorithms based on bit-parallelism the \textsf{SBNDMq4} maintains quite competitive performance.
For very long patterns the best results are obtained by the \textsf{SSEF}, \textsf{HASH8} and \textsf{BOM} algorithms. 
Finally the algorithms \textsf{EBOM} maintains very good performance for all patterns.

\vfill

\subsection{Experimental Results on Rand$8$ Problem}\label{sec:exp8}
In this section we present experimental results on a random text buffer over an alphabet of eight characters.

\begin{scriptsize}
\begin{center}
\begin{tabular*}{\textwidth}{@{\extracolsep{\fill}}|l|cc|ccc|ccc|cc|}
\hline
$m$ & 2 & 4 & 8 & 16 & 32 & 64 & 128 & 256 & 512 & 1024 \\
\hline

 {\color{red}~\textbullet}\algo{KR}~&~{\color{color4}22.4}~~&~{\color{color0}17.8}~~&~{\color{color0}16.4}~~&~{\color{color0}16.4}~~&~{\color{color0}16.4}~~&~{\color{color0}16.4}~~&~{\color{color0}16.4}~~&~{\color{color0}16.4}~~&~{\color{color0}16.4}~~&~{\color{color0}16.4}~~\\
 {\color{red}~\textbullet}\algo{ZT}~&~{\color{color0}36.0}~~&~{\color{color0}18.3}~~&~{\color{color0}9.86}~~&~{\color{color0}5.84}~~&~{\color{color0}3.79}~~&~{\color{color7}2.96}~~&~{\color{color6}3.26}~~&~{\color{color4}3.02}~~&~{\color{color3}2.94}~~&~{\color{color3}2.94}~~\\
 {\color{red}~\textbullet}\algo{QS}~&~{\color{color10}19.6}~~&~{\color{color8}13.3}~~&~{\color{color4}8.91}~~&~{\color{color0}6.88}~~&~{\color{color0}6.16}~~&~{\color{color0}6.14}~~&~{\color{color0}6.10}~~&~{\color{color0}6.12}~~&~{\color{color0}6.17}~~&~{\color{color0}6.26}~~\\
 {\color{red}~\textbullet}\algo{TunBM}~&~{\color{color3}22.5}~~&~{\color{color8}13.1}~~&~{\color{color6}8.59}~~&~{\color{color0}6.65}~~&~{\color{color0}6.28}~~&~{\color{color0}6.21}~~&~{\color{color0}6.12}~~&~{\color{color0}6.19}~~&~{\color{color0}6.19}~~&~{\color{color0}6.12}~~\\
 {\color{red}~\textbullet}\algo{NSN}~&~{\color{color8}20.5}~~&~{\color{color0}22.5}~~&~{\color{color0}22.1}~~&~{\color{color0}22.1}~~&~{\color{color0}22.1}~~&~{\color{color0}22.1}~~&~{\color{color0}22.1}~~&~{\color{color0}22.1}~~&~{\color{color0}22.2}~~&~{\color{color0}22.1}~~\\
 {\color{red}~\textbullet}\algo{Raita}~&~{\color{color8}20.4}~~&~{\color{color13}11.3}~~&~{\color{color11}7.38}~~&~{\color{color2}5.65}~~&~{\color{color0}5.33}~~&~{\color{color0}5.25}~~&~{\color{color0}5.23}~~&~{\color{color0}5.29}~~&~{\color{color0}5.18}~~&~{\color{color0}5.17}~~\\
 {\color{red}~\textbullet}\algo{RCol}~&~{\color{color12}18.8}~~&~{\color{color13}11.1}~~&~{\color{color11}7.23}~~&~{\color{color4}5.47}~~&~{\color{color0}4.97}~~&~{\color{color0}4.67}~~&~{\color{color0}4.50}~~&~{\color{color0}4.44}~~&~{\color{color0}4.01}~~&~{\color{color0}3.81}~~\\
 {\color{red}~\textbullet}\algo{BR}~&~{\color{color16}16.8}~~&~{\color{color12}11.6}~~&~{\color{color10}7.47}~~&~{\color{color11}4.75}~~&~{\color{color9}3.24}~~&~{\color{color12}2.76}~~&~{\color{color3}3.39}~~&~{\color{color5}2.97}~~&~{\color{color3}2.93}~~&~{\color{color3}2.95}~~\\
 {\color{red}~\textbullet}\algo{FS}~&~{\color{color12}18.9}~~&~{\color{color13}11.1}~~&~{\color{color11}7.25}~~&~{\color{color3}5.54}~~&~{\color{color0}4.95}~~&~{\color{color0}4.62}~~&~{\color{color0}4.45}~~&~{\color{color0}4.35}~~&~{\color{color0}4.00}~~&~{\color{color0}3.79}~~\\
 {\color{red}~\textbullet}\algo{FFS}~&~{\color{color12}18.7}~~&~{\color{color13}11.2}~~&~{\color{color11}7.23}~~&~{\color{color6}5.24}~~&~{\color{color0}4.34}~~&~{\color{color0}3.61}~~&~{\color{color0}3.61}~~&~{\color{color0}3.36}~~&~{\color{color3}2.97}~~&~{\color{color3}2.85}~~\\
 {\color{red}~\textbullet}\algo{BFS}~&~{\color{color11}18.9}~~&~{\color{color13}11.1}~~&~{\color{color12}7.08}~~&~{\color{color7}5.15}~~&~{\color{color0}4.30}~~&~{\color{color0}3.61}~~&~{\color{color0}3.64}~~&~{\color{color0}3.44}~~&~{\color{color2}3.06}~~&~{\color{color2}3.02}~~\\
 {\color{red}~\textbullet}\algo{TS}~&~{\color{color12}18.6}~~&~{\color{color1}15.9}~~&~{\color{color0}12.2}~~&~{\color{color0}8.75}~~&~{\color{color0}6.13}~~&~{\color{color0}4.84}~~&~{\color{color0}4.22}~~&~{\color{color0}3.73}~~&~{\color{color0}3.48}~~&~{\color{color0}3.33}~~\\
 {\color{red}~\textbullet}\algo{SSABS}~&~{\color{color17}16.5}~~&~{\color{color12}11.6}~~&~{\color{color7}8.34}~~&~{\color{color0}6.74}~~&~{\color{color0}6.33}~~&~{\color{color0}6.32}~~&~{\color{color0}6.28}~~&~{\color{color0}6.29}~~&~{\color{color0}6.26}~~&~{\color{color0}6.17}~~\\
 {\color{red}~\textbullet}\algo{TVSBS}~&~\colorbox{best}{\color{color20}14.9}~~&~{\color{color15}10.5}~~&~{\color{color13}6.91}~~&~{\color{color13}4.50}~~&~{\color{color9}3.23}~~&~{\color{color10}2.82}~~&~{\color{color7}3.17}~~&~{\color{color5}2.96}~~&~{\color{color3}2.95}~~&~{\color{color3}2.92}~~\\
 {\color{red}~\textbullet}\algo{FJS}~&~{\color{color14}17.9}~~&~{\color{color9}12.8}~~&~{\color{color2}9.36}~~&~{\color{color0}7.61}~~&~{\color{color0}7.19}~~&~{\color{color0}7.13}~~&~{\color{color0}6.99}~~&~{\color{color0}7.09}~~&~{\color{color0}7.14}~~&~{\color{color0}7.15}~~\\
 {\color{red}~\textbullet}\algo{HASH3}~ &~-~&~{\color{color0}19.1}~~&~{\color{color11}7.25}~~&~{\color{color19}3.88}~~&~{\color{color19}2.66}~~&~{\color{color19}2.46}~~&~{\color{color15}2.75}~~&~{\color{color10}2.60}~~&~{\color{color7}2.45}~~&~{\color{color7}2.38}~~\\
 {\color{red}~\textbullet}\algo{HASH5}~ &~-~ &~-~&~{\color{color0}12.2}~~&~{\color{color10}4.79}~~&~{\color{color18}2.71}~~&~\colorbox{best}{\color{color20}2.41}~~&~{\color{color15}2.78}~~&~{\color{color17}2.06}~~&~{\color{color15}1.64}~~&~{\color{color14}1.47}~~\\
 {\color{red}~\textbullet}\algo{HASH8}~ &~-~ &~-~ &~-~&~{\color{color0}7.61}~~&~{\color{color6}3.45}~~&~{\color{color19}2.46}~~&~{\color{color13}2.85}~~&~{\color{color18}1.96}~~&~{\color{color16}1.45}~~&~{\color{color15}1.30}~~\\
 {\color{red}~\textbullet}\algo{TSW}~&~{\color{color11}19.3}~~&~{\color{color7}13.5}~~&~{\color{color5}8.80}~~&~{\color{color2}5.69}~~&~{\color{color0}3.91}~~&~{\color{color4}3.07}~~&~{\color{color0}3.84}~~&~{\color{color1}3.28}~~&~{\color{color0}3.24}~~&~{\color{color0}3.24}~~\\
 {\color{red}~\textbullet}\algo{GRASPm}~&~{\color{color6}21.5}~~&~{\color{color10}12.4}~~&~{\color{color8}7.94}~~&~{\color{color0}5.84}~~&~{\color{color0}4.76}~~&~{\color{color0}3.84}~~&~{\color{color0}4.06}~~&~{\color{color0}3.35}~~&~{\color{color5}2.70}~~&~{\color{color8}2.17}~~\\
 {\color{red}~\textbullet}\algo{SSEF}~ &~-~ &~-~ &~-~ &~-~&~{\color{color0}5.39}~~&~{\color{color0}3.36}~~&~{\color{color2}3.43}~~&~\colorbox{best}{\color{color20}1.79}~~&~\colorbox{best}{\color{color20}1.00}~~&~\colorbox{best}{\color{color20}0.55}~~\\
 \hline
 {\color{green}~\textbullet}\algo{AUT}~&~{\color{color4}22.3}~~&~{\color{color0}22.3}~~&~{\color{color0}21.7}~~&~{\color{color0}22.3}~~&~{\color{color0}22.4}~~&~{\color{color0}21.7}~~&~{\color{color0}21.8}~~&~{\color{color0}21.9}~~&~{\color{color0}22.6}~~&~{\color{color0}23.9}~~\\
 {\color{green}~\textbullet}\algo{RF}~&~{\color{color0}34.5}~~&~{\color{color0}22.0}~~&~{\color{color0}12.6}~~&~{\color{color0}7.02}~~&~{\color{color0}4.31}~~&~{\color{color8}2.89}~~&~{\color{color2}3.47}~~&~{\color{color10}2.59}~~&~{\color{color4}2.87}~~&~{\color{color0}4.38}~~\\
 {\color{green}~\textbullet}\algo{BOM}~&~{\color{color0}48.6}~~&~{\color{color0}33.3}~~&~{\color{color0}22.2}~~&~{\color{color0}15.1}~~&~{\color{color0}9.60}~~&~{\color{color0}5.98}~~&~{\color{color0}5.11}~~&~{\color{color7}2.82}~~&~{\color{color15}1.60}~~&~{\color{color18}0.94}~~\\
 {\color{green}~\textbullet}\algo{BOM2}~&~{\color{color0}36.8}~~&~{\color{color0}23.1}~~&~{\color{color0}13.2}~~&~{\color{color0}7.16}~~&~{\color{color0}4.37}~~&~{\color{color10}2.82}~~&~{\color{color3}3.40}~~&~{\color{color18}1.96}~~&~{\color{color17}1.36}~~&~{\color{color7}2.41}~~\\
 {\color{green}~\textbullet}\algo{ILDM1}~&~{\color{color0}30.3}~~&~{\color{color0}20.1}~~&~{\color{color0}11.6}~~&~{\color{color0}6.38}~~&~{\color{color0}4.01}~~&~{\color{color7}2.94}~~&~{\color{color0}3.54}~~&~{\color{color0}3.29}~~&~{\color{color0}4.67}~~&~{\color{color0}8.20}~~\\
 {\color{green}~\textbullet}\algo{ILDM2}~&~{\color{color0}31.9}~~&~{\color{color0}19.4}~~&~{\color{color0}10.8}~~&~{\color{color0}5.93}~~&~{\color{color0}3.77}~~&~{\color{color10}2.83}~~&~{\color{color0}3.53}~~&~{\color{color0}3.29}~~&~{\color{color0}4.65}~~&~{\color{color0}8.21}~~\\
 {\color{green}~\textbullet}\algo{EBOM}~&~{\color{color10}19.6}~~&~\colorbox{best}{\color{color20}8.37}~~&~\colorbox{best}{\color{color20}5.04}~~&~\colorbox{best}{\color{color20}3.70}~~&~{\color{color13}3.00}~~&~{\color{color15}2.63}~~&~{\color{color8}3.13}~~&~{\color{color19}1.90}~~&~{\color{color16}1.48}~~&~{\color{color5}2.65}~~\\
 {\color{green}~\textbullet}\algo{FBOM}~&~{\color{color15}17.4}~~&~{\color{color15}10.4}~~&~{\color{color13}6.72}~~&~{\color{color12}4.63}~~&~{\color{color6}3.45}~~&~{\color{color10}2.83}~~&~{\color{color5}3.30}~~&~{\color{color18}2.01}~~&~{\color{color16}1.52}~~&~{\color{color5}2.69}~~\\
 {\color{green}~\textbullet}\algo{SEBOM}~&~{\color{color8}20.6}~~&~{\color{color20}8.73}~~&~{\color{color20}5.22}~~&~{\color{color19}3.82}~~&~{\color{color11}3.12}~~&~{\color{color12}2.76}~~&~{\color{color6}3.25}~~&~{\color{color17}2.02}~~&~{\color{color15}1.56}~~&~{\color{color4}2.72}~~\\
 {\color{green}~\textbullet}\algo{SFBOM}~&~{\color{color15}17.2}~~&~{\color{color15}10.4}~~&~{\color{color13}6.77}~~&~{\color{color11}4.68}~~&~{\color{color5}3.49}~~&~{\color{color9}2.88}~~&~{\color{color4}3.33}~~&~{\color{color17}2.05}~~&~{\color{color15}1.56}~~&~{\color{color4}2.72}~~\\
 \hline
 {\color{blu}~\textbullet}\algo{SO}~&~{\color{color16}16.8}~~&~{\color{color0}16.8}~~&~{\color{color0}16.8}~~&~{\color{color0}16.8}~~&~{\color{color0}16.8}~~&~{\color{color0}21.8}~~&~{\color{color0}21.8}~~&~{\color{color0}21.7}~~&~{\color{color0}21.7}~~&~{\color{color0}21.8}~~\\
 {\color{blu}~\textbullet}\algo{SA}~&~{\color{color17}16.4}~~&~{\color{color0}16.4}~~&~{\color{color0}16.4}~~&~{\color{color0}16.4}~~&~{\color{color0}16.4}~~&~{\color{color0}18.9}~~&~{\color{color0}18.9}~~&~{\color{color0}18.9}~~&~{\color{color0}18.9}~~&~{\color{color0}18.9}~~\\
 {\color{blu}~\textbullet}\algo{BNDM}~&~{\color{color0}37.3}~~&~{\color{color0}22.0}~~&~{\color{color0}11.6}~~&~{\color{color0}6.10}~~&~{\color{color2}3.66}~~&~{\color{color0}4.51}~~&~{\color{color0}4.51}~~&~{\color{color0}4.52}~~&~{\color{color0}4.52}~~&~{\color{color0}4.51}~~\\
 {\color{blu}~\textbullet}\algo{BNDM-L}~&~{\color{color0}37.1}~~&~{\color{color0}21.9}~~&~{\color{color0}11.6}~~&~{\color{color0}6.08}~~&~{\color{color2}3.67}~~&~{\color{color0}5.48}~~&~{\color{color0}6.86}~~&~{\color{color0}6.33}~~&~{\color{color0}5.98}~~&~{\color{color0}6.26}~~\\
 {\color{blu}~\textbullet}\algo{SBNDM}~&~{\color{color0}48.2}~~&~{\color{color0}17.8}~~&~{\color{color5}8.61}~~&~{\color{color8}5.05}~~&~{\color{color9}3.24}~~&~{\color{color4}3.09}~~&~{\color{color9}3.08}~~&~{\color{color3}3.11}~~&~{\color{color1}3.10}~~&~{\color{color2}3.09}~~\\
 {\color{blu}~\textbullet}\algo{TNDM}~&~{\color{color0}29.9}~~&~{\color{color0}19.1}~~&~{\color{color0}10.9}~~&~{\color{color0}5.89}~~&~{\color{color4}3.57}~~&~{\color{color0}3.55}~~&~{\color{color0}3.55}~~&~{\color{color0}3.54}~~&~{\color{color0}3.53}~~&~{\color{color0}3.56}~~\\
 {\color{blu}~\textbullet}\algo{TNDMa}~&~{\color{color0}27.0}~~&~{\color{color0}18.2}~~&~{\color{color0}11.2}~~&~{\color{color0}5.93}~~&~{\color{color5}3.47}~~&~{\color{color0}3.40}~~&~{\color{color3}3.39}~~&~{\color{color0}3.39}~~&~{\color{color0}3.38}~~&~{\color{color0}3.38}~~\\
 {\color{blu}~\textbullet}\algo{LBNDM}~&~{\color{color0}39.7}~~&~{\color{color0}22.9}~~&~{\color{color0}12.8}~~&~{\color{color0}7.08}~~&~{\color{color0}4.27}~~&~{\color{color7}2.95}~~&~{\color{color0}4.25}~~&~{\color{color0}3.86}~~&~{\color{color0}7.02}~~&~{\color{color0}33.7}~~\\
 {\color{blu}~\textbullet}\algo{SBNDM2}~&~{\color{color0}36.0}~~&~{\color{color7}13.5}~~&~{\color{color12}6.98}~~&~{\color{color15}4.30}~~&~{\color{color13}3.01}~~&~{\color{color11}2.79}~~&~{\color{color14}2.79}~~&~{\color{color7}2.81}~~&~{\color{color4}2.80}~~&~{\color{color4}2.80}~~\\
 {\color{blu}~\textbullet}\algo{SBNDM-BMH}~&~{\color{color5}21.9}~~&~{\color{color6}14.0}~~&~{\color{color6}8.49}~~&~{\color{color8}5.03}~~&~{\color{color10}3.21}~~&~{\color{color4}3.08}~~&~{\color{color8}3.11}~~&~{\color{color3}3.08}~~&~{\color{color2}3.09}~~&~{\color{color2}3.09}~~\\
 {\color{blu}~\textbullet}\algo{BMH-SBNDM}~&~{\color{color10}19.4}~~&~{\color{color14}11.0}~~&~{\color{color12}6.96}~~&~{\color{color7}5.10}~~&~{\color{color0}4.08}~~&~{\color{color0}4.22}~~&~{\color{color0}4.25}~~&~{\color{color0}4.21}~~&~{\color{color0}4.19}~~&~{\color{color0}4.18}~~\\
 {\color{blu}~\textbullet}\algo{AOSO2}~&~{\color{color0}58.0}~~&~{\color{color1}15.9}~~&~{\color{color0}9.79}~~&~{\color{color0}9.72}~~&~{\color{color0}9.71}~~&~{\color{color0}8.56}~~&~{\color{color0}8.58}~~&~{\color{color0}8.56}~~&~{\color{color0}8.57}~~&~{\color{color0}8.56}~~\\
 {\color{blu}~\textbullet}\algo{AOSO4}~ &~-~&~{\color{color0}49.7}~~&~{\color{color0}11.1}~~&~{\color{color7}5.15}~~&~{\color{color0}5.05}~~&~{\color{color0}4.56}~~&~{\color{color0}4.55}~~&~{\color{color0}4.57}~~&~{\color{color0}4.57}~~&~{\color{color0}4.56}~~\\
 {\color{blu}~\textbullet}\algo{AOSO6}~ &~-~ &~-~&~{\color{color0}44.8}~~&~{\color{color0}9.79}~~&~{\color{color4}3.53}~~&~{\color{color0}3.31}~~&~{\color{color5}3.31}~~&~{\color{color0}3.31}~~&~{\color{color0}3.30}~~&~{\color{color0}3.30}~~\\
 {\color{blu}~\textbullet}\algo{FSBNDM}~&~{\color{color0}28.1}~~&~{\color{color5}14.2}~~&~{\color{color9}7.85}~~&~{\color{color11}4.71}~~&~{\color{color18}2.74}~~&~{\color{color12}2.75}~~&~{\color{color15}2.74}~~&~{\color{color8}2.74}~~&~{\color{color5}2.74}~~&~{\color{color4}2.76}~~\\
 {\color{blu}~\textbullet}\algo{BNDMq2}~&~{\color{color0}33.8}~~&~{\color{color9}12.8}~~&~{\color{color14}6.58}~~&~{\color{color17}4.06}~~&~{\color{color16}2.84}~~&~{\color{color0}3.41}~~&~{\color{color2}3.44}~~&~{\color{color0}3.45}~~&~{\color{color0}3.45}~~&~{\color{color0}3.44}~~\\
 {\color{blu}~\textbullet}\algo{BNDMq4}~ &~-~&~{\color{color0}48.4}~~&~{\color{color0}10.4}~~&~{\color{color12}4.59}~~&~{\color{color20}2.57}~~&~{\color{color2}3.16}~~&~{\color{color8}3.15}~~&~{\color{color2}3.17}~~&~{\color{color1}3.16}~~&~{\color{color1}3.16}~~\\
 {\color{blu}~\textbullet}\algo{BNDMq6}~ &~-~ &~-~&~{\color{color0}24.0}~~&~{\color{color0}7.22}~~&~{\color{color5}3.52}~~&~{\color{color1}3.19}~~&~{\color{color7}3.19}~~&~{\color{color2}3.18}~~&~{\color{color1}3.18}~~&~{\color{color1}3.20}~~\\
 {\color{blu}~\textbullet}\algo{SBNDMq2}~&~{\color{color0}33.5}~~&~{\color{color9}12.7}~~&~{\color{color13}6.72}~~&~{\color{color15}4.25}~~&~{\color{color14}2.97}~~&~{\color{color11}2.79}~~&~{\color{color14}2.79}~~&~{\color{color7}2.79}~~&~{\color{color4}2.81}~~&~{\color{color4}2.82}~~\\
 {\color{blu}~\textbullet}\algo{SBNDMq4}~ &~-~&~{\color{color0}45.8}~~&~{\color{color0}9.90}~~&~{\color{color14}4.39}~~&~\colorbox{best}{\color{color20}2.56}~~&~{\color{color19}2.46}~~&~\colorbox{best}{\color{color20}2.46}~~&~{\color{color12}2.46}~~&~{\color{color7}2.45}~~&~{\color{color6}2.46}~~\\
 {\color{blu}~\textbullet}\algo{SBNDMq6}~ &~-~ &~-~&~{\color{color0}27.4}~~&~{\color{color0}8.03}~~&~{\color{color1}3.75}~~&~{\color{color0}3.21}~~&~{\color{color6}3.22}~~&~{\color{color2}3.21}~~&~{\color{color0}3.21}~~&~{\color{color1}3.21}~~\\
 {\color{blu}~\textbullet}\algo{UFNDMq4}~&~{\color{color0}42.3}~~&~{\color{color0}21.1}~~&~{\color{color0}10.9}~~&~{\color{color0}6.04}~~&~{\color{color4}3.53}~~&~{\color{color0}3.54}~~&~{\color{color0}3.53}~~&~{\color{color0}3.53}~~&~{\color{color0}3.52}~~&~{\color{color0}3.54}~~\\
 {\color{blu}~\textbullet}\algo{KBNDM}~&~{\color{color0}43.7}~~&~{\color{color0}22.0}~~&~{\color{color0}12.1}~~&~{\color{color0}7.25}~~&~{\color{color0}4.72}~~&~{\color{color3}3.10}~~&~{\color{color0}3.58}~~&~{\color{color0}3.58}~~&~{\color{color0}3.59}~~&~{\color{color0}3.58}~~\\
\hline\end{tabular*}\\
\end{center}
\end{scriptsize}

In the case of very short the best performance is obtained by the \textsf{TVSBS} and \textsf{SSABS} algorithms.
Algorithms with very good performance are also \textsf{FBOM} and \textsf{SFBOM}.
For short patterns the algorithms based on bit-parallelism achieve good results, in particular \textsf{BNDMq2}, \textsf{FSBNDM} and \textsf{SBNDM2}.
However the algorithms in the \textsf{EBOM} family are also good choices.
In the case of long patterns the best results are obtained by the \textsf{EBOM}, \textsf{HASH5} and \textsf{SBNDMq4} algorithms. 
For very long patterns the best results are obtained by the \textsf{SSEF} algorithm. 
For the overall performance we notice that the algorithms int the \textsf{EBOM} family, and the \textsf{TVSBS} and \textsf{FSBNDM} algorithms  
maintain very good performance for all patterns. 

\vfill

\subsection{Experimental Results on Rand$16$ Problem}\label{sec:exp16}
In this section we present experimental results on a random text buffer over an alphabet of 16 characters.

\begin{scriptsize}
\begin{center}
\begin{tabular*}{\textwidth}{@{\extracolsep{\fill}}|l|cc|ccc|ccc|cc|}
\hline
$m$ & 2 & 4 & 8 & 16 & 32 & 64 & 128 & 256 & 512 & 1024 \\
\hline

 {\color{red}~\textbullet}\algo{KR}~&~{\color{color6}19.4}~~&~{\color{color0}17.1}~~&~{\color{color0}16.4}~~&~{\color{color0}16.4}~~&~{\color{color0}16.4}~~&~{\color{color0}16.4}~~&~{\color{color0}16.4}~~&~{\color{color0}16.4}~~&~{\color{color0}16.4}~~&~{\color{color0}16.4}~~\\
 {\color{red}~\textbullet}\algo{ZT}~&~{\color{color0}31.6}~~&~{\color{color0}16.2}~~&~{\color{color0}8.71}~~&~{\color{color0}5.10}~~&~{\color{color0}3.22}~~&~{\color{color7}2.60}~~&~{\color{color4}3.00}~~&~{\color{color15}2.05}~~&~{\color{color14}1.56}~~&~{\color{color12}1.45}~~\\
 {\color{red}~\textbullet}\algo{QS}~&~{\color{color13}15.4}~~&~{\color{color10}9.98}~~&~{\color{color6}6.38}~~&~{\color{color3}4.34}~~&~{\color{color0}3.46}~~&~{\color{color0}3.22}~~&~{\color{color0}3.26}~~&~{\color{color0}3.27}~~&~{\color{color0}3.28}~~&~{\color{color0}3.24}~~\\
 {\color{red}~\textbullet}\algo{TunBM}~&~{\color{color9}17.4}~~&~{\color{color11}9.59}~~&~{\color{color10}5.71}~~&~{\color{color8}3.87}~~&~{\color{color0}3.17}~~&~{\color{color0}3.04}~~&~{\color{color2}3.06}~~&~{\color{color0}3.04}~~&~{\color{color0}3.04}~~&~{\color{color0}3.06}~~\\
 {\color{red}~\textbullet}\algo{NSN}~&~{\color{color11}16.8}~~&~{\color{color0}17.9}~~&~{\color{color0}17.8}~~&~{\color{color0}17.8}~~&~{\color{color0}17.9}~~&~{\color{color0}17.7}~~&~{\color{color0}17.8}~~&~{\color{color0}17.9}~~&~{\color{color0}17.8}~~&~{\color{color0}17.8}~~\\
 {\color{red}~\textbullet}\algo{Raita}~&~{\color{color10}17.3}~~&~{\color{color13}9.29}~~&~{\color{color11}5.59}~~&~{\color{color9}3.77}~~&~{\color{color2}3.06}~~&~{\color{color0}2.91}~~&~{\color{color5}2.96}~~&~{\color{color0}2.94}~~&~{\color{color0}2.93}~~&~{\color{color0}2.96}~~\\
 {\color{red}~\textbullet}\algo{RCol}~&~{\color{color13}15.3}~~&~{\color{color16}8.47}~~&~{\color{color14}5.08}~~&~{\color{color13}3.48}~~&~{\color{color7}2.88}~~&~{\color{color0}2.77}~~&~{\color{color9}2.80}~~&~{\color{color0}2.77}~~&~{\color{color0}2.72}~~&~{\color{color2}2.63}~~\\
 {\color{red}~\textbullet}\algo{BR}~&~{\color{color17}13.3}~~&~{\color{color13}9.22}~~&~{\color{color9}5.96}~~&~{\color{color9}3.83}~~&~{\color{color10}2.76}~~&~{\color{color9}2.56}~~&~{\color{color2}3.08}~~&~{\color{color15}2.08}~~&~{\color{color13}1.64}~~&~{\color{color12}1.51}~~\\
 {\color{red}~\textbullet}\algo{FS}~&~{\color{color13}15.3}~~&~{\color{color16}8.47}~~&~{\color{color14}5.08}~~&~{\color{color13}3.47}~~&~{\color{color7}2.86}~~&~{\color{color0}2.77}~~&~{\color{color9}2.80}~~&~{\color{color0}2.78}~~&~{\color{color0}2.72}~~&~{\color{color1}2.65}~~\\
 {\color{red}~\textbullet}\algo{FFS}~&~{\color{color13}15.2}~~&~{\color{color15}8.54}~~&~{\color{color13}5.13}~~&~{\color{color13}3.46}~~&~{\color{color8}2.84}~~&~{\color{color5}2.65}~~&~{\color{color6}2.90}~~&~{\color{color0}2.77}~~&~{\color{color3}2.53}~~&~{\color{color2}2.62}~~\\
 {\color{red}~\textbullet}\algo{BFS}~&~{\color{color13}15.4}~~&~{\color{color15}8.51}~~&~{\color{color14}5.08}~~&~{\color{color13}3.43}~~&~{\color{color9}2.79}~~&~{\color{color5}2.65}~~&~{\color{color7}2.89}~~&~{\color{color0}2.77}~~&~{\color{color2}2.56}~~&~{\color{color1}2.65}~~\\
 {\color{red}~\textbullet}\algo{TS}~&~{\color{color15}14.7}~~&~{\color{color0}13.3}~~&~{\color{color0}11.1}~~&~{\color{color0}8.52}~~&~{\color{color0}6.08}~~&~{\color{color0}4.21}~~&~{\color{color0}3.61}~~&~{\color{color0}3.56}~~&~{\color{color0}3.56}~~&~{\color{color0}3.60}~~\\
 {\color{red}~\textbullet}\algo{SSABS}~&~{\color{color19}12.3}~~&~{\color{color16}8.25}~~&~{\color{color11}5.51}~~&~{\color{color7}3.94}~~&~{\color{color0}3.28}~~&~{\color{color0}3.12}~~&~{\color{color0}3.14}~~&~{\color{color0}3.16}~~&~{\color{color0}3.15}~~&~{\color{color0}3.16}~~\\
 {\color{red}~\textbullet}\algo{TVSBS}~&~\colorbox{best}{\color{color20}11.6}~~&~{\color{color17}8.09}~~&~{\color{color12}5.33}~~&~{\color{color12}3.51}~~&~{\color{color11}2.74}~~&~{\color{color10}2.55}~~&~{\color{color7}2.87}~~&~{\color{color17}1.96}~~&~{\color{color14}1.52}~~&~{\color{color13}1.40}~~\\
 {\color{red}~\textbullet}\algo{FJS}~&~{\color{color18}12.7}~~&~{\color{color15}8.58}~~&~{\color{color10}5.75}~~&~{\color{color5}4.16}~~&~{\color{color0}3.44}~~&~{\color{color0}3.26}~~&~{\color{color0}3.31}~~&~{\color{color0}3.30}~~&~{\color{color0}3.28}~~&~{\color{color0}3.30}~~\\
 {\color{red}~\textbullet}\algo{HASH3}~ &~-~&~{\color{color0}18.3}~~&~{\color{color4}6.85}~~&~{\color{color12}3.58}~~&~{\color{color17}2.49}~~&~\colorbox{best}{\color{color20}2.33}~~&~{\color{color12}2.69}~~&~{\color{color10}2.31}~~&~{\color{color8}2.07}~~&~{\color{color8}1.96}~~\\
 {\color{red}~\textbullet}\algo{HASH5}~ &~-~ &~-~&~{\color{color0}12.1}~~&~{\color{color0}4.72}~~&~{\color{color13}2.65}~~&~{\color{color18}2.39}~~&~{\color{color11}2.74}~~&~{\color{color19}1.84}~~&~{\color{color16}1.37}~~&~{\color{color14}1.22}~~\\
 {\color{red}~\textbullet}\algo{HASH8}~ &~-~ &~-~ &~-~&~{\color{color0}7.59}~~&~{\color{color0}3.45}~~&~{\color{color14}2.46}~~&~{\color{color8}2.85}~~&~{\color{color17}1.96}~~&~{\color{color15}1.44}~~&~{\color{color14}1.29}~~\\
 {\color{red}~\textbullet}\algo{TSW}~&~{\color{color14}15.2}~~&~{\color{color8}10.6}~~&~{\color{color3}6.94}~~&~{\color{color0}4.53}~~&~{\color{color0}3.19}~~&~{\color{color7}2.60}~~&~{\color{color0}3.60}~~&~{\color{color7}2.47}~~&~{\color{color9}2.00}~~&~{\color{color9}1.86}~~\\
 {\color{red}~\textbullet}\algo{GRASPm}~&~{\color{color10}17.2}~~&~{\color{color12}9.42}~~&~{\color{color11}5.57}~~&~{\color{color10}3.72}~~&~{\color{color4}2.99}~~&~{\color{color0}2.74}~~&~{\color{color4}2.98}~~&~{\color{color3}2.65}~~&~{\color{color10}1.94}~~&~{\color{color12}1.43}~~\\
 {\color{red}~\textbullet}\algo{SSEF}~ &~-~ &~-~ &~-~ &~-~&~{\color{color0}5.38}~~&~{\color{color0}3.37}~~&~{\color{color0}3.44}~~&~\colorbox{best}{\color{color20}1.79}~~&~\colorbox{best}{\color{color20}0.99}~~&~\colorbox{best}{\color{color20}0.55}~~\\
 \hline
 {\color{green}~\textbullet}\algo{RF}~&~{\color{color0}26.5}~~&~{\color{color0}16.2}~~&~{\color{color0}10.3}~~&~{\color{color0}6.11}~~&~{\color{color0}3.51}~~&~{\color{color4}2.66}~~&~{\color{color0}3.28}~~&~{\color{color6}2.49}~~&~{\color{color0}2.77}~~&~{\color{color0}4.21}~~\\
 {\color{green}~\textbullet}\algo{BOM}~&~{\color{color0}40.9}~~&~{\color{color0}29.0}~~&~{\color{color0}22.6}~~&~{\color{color0}15.6}~~&~{\color{color0}9.65}~~&~{\color{color0}5.86}~~&~{\color{color0}4.89}~~&~{\color{color1}2.75}~~&~{\color{color14}1.55}~~&~{\color{color17}0.97}~~\\
 {\color{green}~\textbullet}\algo{BOM2}~&~{\color{color0}27.7}~~&~{\color{color0}16.9}~~&~{\color{color0}10.9}~~&~{\color{color0}6.40}~~&~{\color{color0}3.56}~~&~{\color{color7}2.60}~~&~{\color{color0}3.21}~~&~{\color{color19}1.86}~~&~{\color{color17}1.29}~~&~{\color{color4}2.38}~~\\
 {\color{green}~\textbullet}\algo{ILDM1}~&~{\color{color0}24.5}~~&~{\color{color0}15.2}~~&~{\color{color0}9.65}~~&~{\color{color0}5.59}~~&~{\color{color0}3.18}~~&~{\color{color3}2.68}~~&~{\color{color0}3.38}~~&~{\color{color0}3.19}~~&~{\color{color0}4.54}~~&~{\color{color0}8.16}~~\\
 {\color{green}~\textbullet}\algo{ILDM2}~&~{\color{color0}25.4}~~&~{\color{color0}15.2}~~&~{\color{color0}9.59}~~&~{\color{color0}5.48}~~&~{\color{color0}3.15}~~&~{\color{color3}2.69}~~&~{\color{color0}3.37}~~&~{\color{color0}3.18}~~&~{\color{color0}4.53}~~&~{\color{color0}8.18}~~\\
 {\color{green}~\textbullet}\algo{EBOM}~&~{\color{color7}18.6}~~&~\colorbox{best}{\color{color20}7.15}~~&~\colorbox{best}{\color{color20}3.88}~~&~\colorbox{best}{\color{color20}2.81}~~&~{\color{color16}2.55}~~&~{\color{color15}2.44}~~&~{\color{color8}2.83}~~&~{\color{color20}1.81}~~&~{\color{color16}1.42}~~&~{\color{color1}2.68}~~\\
 {\color{green}~\textbullet}\algo{FBOM}~&~{\color{color17}13.3}~~&~{\color{color17}8.17}~~&~{\color{color14}5.10}~~&~{\color{color14}3.41}~~&~{\color{color9}2.79}~~&~{\color{color4}2.66}~~&~{\color{color0}3.20}~~&~{\color{color19}1.88}~~&~{\color{color15}1.45}~~&~{\color{color1}2.69}~~\\
 {\color{green}~\textbullet}\algo{SEBOM}~&~{\color{color5}19.6}~~&~{\color{color19}7.57}~~&~{\color{color19}4.11}~~&~{\color{color19}2.94}~~&~{\color{color12}2.68}~~&~{\color{color9}2.56}~~&~{\color{color5}2.95}~~&~{\color{color18}1.93}~~&~{\color{color15}1.49}~~&~{\color{color0}2.74}~~\\
 {\color{green}~\textbullet}\algo{SFBOM}~&~{\color{color17}13.3}~~&~{\color{color16}8.25}~~&~{\color{color13}5.18}~~&~{\color{color13}3.48}~~&~{\color{color7}2.87}~~&~{\color{color0}2.74}~~&~{\color{color0}3.28}~~&~{\color{color17}1.96}~~&~{\color{color15}1.50}~~&~{\color{color1}2.73}~~\\
 \hline
 {\color{blu}~\textbullet}\algo{SO}~&~{\color{color11}16.8}~~&~{\color{color0}16.8}~~&~{\color{color0}16.8}~~&~{\color{color0}16.8}~~&~{\color{color0}16.8}~~&~{\color{color0}22.1}~~&~{\color{color0}22.1}~~&~{\color{color0}22.1}~~&~{\color{color0}22.1}~~&~{\color{color0}22.1}~~\\
 {\color{blu}~\textbullet}\algo{SA}~&~{\color{color11}16.4}~~&~{\color{color0}16.4}~~&~{\color{color0}16.4}~~&~{\color{color0}16.4}~~&~{\color{color0}16.4}~~&~{\color{color0}19.1}~~&~{\color{color0}19.1}~~&~{\color{color0}19.1}~~&~{\color{color0}19.1}~~&~{\color{color0}19.1}~~\\
 {\color{blu}~\textbullet}\algo{SBNDM}~&~{\color{color0}48.1}~~&~{\color{color0}16.8}~~&~{\color{color0}7.71}~~&~{\color{color4}4.20}~~&~{\color{color14}2.61}~~&~{\color{color7}2.60}~~&~{\color{color15}2.59}~~&~{\color{color4}2.60}~~&~{\color{color2}2.60}~~&~{\color{color2}2.60}~~\\
 {\color{blu}~\textbullet}\algo{TNDM}~&~{\color{color0}25.0}~~&~{\color{color0}14.9}~~&~{\color{color0}9.34}~~&~{\color{color0}5.32}~~&~{\color{color7}2.89}~~&~{\color{color0}2.87}~~&~{\color{color7}2.88}~~&~{\color{color0}2.89}~~&~{\color{color0}2.89}~~&~{\color{color0}2.88}~~\\
 {\color{blu}~\textbullet}\algo{TNDMa}~&~{\color{color0}22.3}~~&~{\color{color0}13.6}~~&~{\color{color0}9.26}~~&~{\color{color0}5.65}~~&~{\color{color7}2.88}~~&~{\color{color0}2.83}~~&~{\color{color8}2.83}~~&~{\color{color0}2.82}~~&~{\color{color0}2.82}~~&~{\color{color0}2.84}~~\\
 {\color{blu}~\textbullet}\algo{LBNDM}~&~{\color{color0}34.0}~~&~{\color{color0}19.0}~~&~{\color{color0}11.2}~~&~{\color{color0}6.31}~~&~{\color{color0}3.57}~~&~{\color{color6}2.62}~~&~{\color{color0}3.54}~~&~{\color{color7}2.47}~~&~{\color{color3}2.53}~~&~{\color{color0}4.08}~~\\
 {\color{blu}~\textbullet}\algo{SVM1}~&~{\color{color7}18.5}~~&~{\color{color0}13.9}~~&~{\color{color0}16.4}~~&~{\color{color0}11.9}~~&~{\color{color0}9.20}~~&~{\color{color0}20.8}~~&~{\color{color0}20.8}~~&~{\color{color0}20.8}~~&~{\color{color0}20.8}~~&~{\color{color0}20.8}~~\\
 {\color{blu}~\textbullet}\algo{SBNDM2}~&~{\color{color0}35.2}~~&~{\color{color1}12.5}~~&~{\color{color8}6.10}~~&~{\color{color13}3.45}~~&~{\color{color16}2.55}~~&~{\color{color15}2.44}~~&~{\color{color19}2.45}~~&~{\color{color7}2.45}~~&~{\color{color4}2.45}~~&~{\color{color3}2.44}~~\\
 {\color{blu}~\textbullet}\algo{SBNDM-BMH}~&~{\color{color12}16.1}~~&~{\color{color10}9.90}~~&~{\color{color5}6.57}~~&~{\color{color5}4.17}~~&~{\color{color14}2.62}~~&~{\color{color8}2.59}~~&~{\color{color15}2.59}~~&~{\color{color4}2.60}~~&~{\color{color2}2.59}~~&~{\color{color2}2.60}~~\\
 {\color{blu}~\textbullet}\algo{BMH-SBNDM}~&~{\color{color13}15.5}~~&~{\color{color16}8.42}~~&~{\color{color14}5.00}~~&~{\color{color14}3.35}~~&~{\color{color10}2.75}~~&~{\color{color0}2.84}~~&~{\color{color8}2.84}~~&~{\color{color0}2.85}~~&~{\color{color0}2.81}~~&~{\color{color0}2.84}~~\\
 {\color{blu}~\textbullet}\algo{FAOSOq2}~&~{\color{color0}36.6}~~&~{\color{color1}12.4}~~&~{\color{color0}10.7}~~&~{\color{color0}10.7}~~&~{\color{color0}10.2}~~&~{\color{color0}10.2}~~&~{\color{color0}10.2}~~&~{\color{color0}10.2}~~&~{\color{color0}10.2}~~&~{\color{color0}10.2}~~\\
 {\color{blu}~\textbullet}\algo{AOSO2}~&~{\color{color0}34.1}~~&~{\color{color5}11.2}~~&~{\color{color0}9.73}~~&~{\color{color0}9.72}~~&~{\color{color0}9.74}~~&~{\color{color0}8.52}~~&~{\color{color0}8.52}~~&~{\color{color0}8.52}~~&~{\color{color0}8.53}~~&~{\color{color0}8.52}~~\\
 {\color{blu}~\textbullet}\algo{AOSO4}~ &~-~&~{\color{color0}28.5}~~&~{\color{color5}6.57}~~&~{\color{color0}5.09}~~&~{\color{color0}5.10}~~&~{\color{color0}4.55}~~&~{\color{color0}4.57}~~&~{\color{color0}4.55}~~&~{\color{color0}4.55}~~&~{\color{color0}4.55}~~\\
 {\color{blu}~\textbullet}\algo{FSBNDM}~&~{\color{color0}23.6}~~&~{\color{color2}12.1}~~&~{\color{color6}6.46}~~&~{\color{color10}3.73}~~&~\colorbox{best}{\color{color20}2.38}~~&~{\color{color18}2.38}~~&~\colorbox{best}{\color{color20}2.38}~~&~{\color{color8}2.39}~~&~{\color{color4}2.39}~~&~{\color{color4}2.37}~~\\
 {\color{blu}~\textbullet}\algo{BNDMq2}~&~{\color{color0}33.3}~~&~{\color{color3}11.8}~~&~{\color{color11}5.61}~~&~{\color{color16}3.16}~~&~{\color{color18}2.48}~~&~{\color{color3}2.68}~~&~{\color{color13}2.67}~~&~{\color{color3}2.67}~~&~{\color{color1}2.68}~~&~{\color{color1}2.70}~~\\
 {\color{blu}~\textbullet}\algo{BNDMq4}~ &~-~&~{\color{color0}48.4}~~&~{\color{color0}10.4}~~&~{\color{color0}4.57}~~&~{\color{color15}2.57}~~&~{\color{color0}3.14}~~&~{\color{color0}3.14}~~&~{\color{color0}3.16}~~&~{\color{color0}3.15}~~&~{\color{color0}3.14}~~\\
 {\color{blu}~\textbullet}\algo{SBNDMq2}~&~{\color{color0}32.6}~~&~{\color{color4}11.7}~~&~{\color{color10}5.70}~~&~{\color{color14}3.35}~~&~{\color{color17}2.50}~~&~{\color{color15}2.45}~~&~{\color{color19}2.44}~~&~{\color{color7}2.44}~~&~{\color{color4}2.44}~~&~{\color{color3}2.44}~~\\
 {\color{blu}~\textbullet}\algo{SBNDMq4}~ &~-~&~{\color{color0}45.7}~~&~{\color{color0}9.88}~~&~{\color{color2}4.36}~~&~{\color{color16}2.54}~~&~{\color{color15}2.44}~~&~{\color{color19}2.45}~~&~{\color{color7}2.45}~~&~{\color{color4}2.44}~~&~{\color{color3}2.44}~~\\
\hline\end{tabular*}\\
\end{center}
\end{scriptsize}

In the case of very short patterns the best results are obtained by the \textsf{TVSBS} and \textsf{EBOM} algorithms for patterns of length
2 and 4, respectively.
For short patterns the algorithms \textsf{EBOM} is the fastest. However it is outperformed by the \textsf{FSVBNDM} algorithm for patterns of length 32. 
The \textsf{FSBNDM} algorithm is very fast also for long patterns but is outperformed by the \textsf{HASH3} algorithm and by the \textsf{SSEF} algorithm
for patterns of length 64 and 256, respectively.
For very long patterns the best results are obtained by the \textsf{SSEF} algorithms. 
Regarding the overall performance the algorithm \textsf{TVSBS}, the algorithm \textsf{BR} and the algorithms in the \textsf{EBOM} family  
maintain very good performance for all patterns. 

\vfill

\subsection{Experimental Results on Rand$32$ Problem}\label{sec:exp32}
In this section we present experimental results on a random text buffer over an alphabet of 32 characters.

\begin{scriptsize}
\begin{center}
\begin{tabular*}{\textwidth}{@{\extracolsep{\fill}}|l|cc|ccc|ccc|cc|}
\hline
$m$ & 2 & 4 & 8 & 16 & 32 & 64 & 128 & 256 & 512 & 1024 \\
\hline

 {\color{red}~\textbullet}\algo{BM}~&~{\color{color0}21.1}~~&~{\color{color1}11.2}~~&~{\color{color0}6.34}~~&~{\color{color2}3.88}~~&~{\color{color0}2.79}~~&~{\color{color0}2.55}~~&~{\color{color4}2.76}~~&~{\color{color0}2.74}~~&~{\color{color0}2.72}~~&~{\color{color0}2.68}~~\\
 {\color{red}~\textbullet}\algo{KR}~&~{\color{color4}17.8}~~&~{\color{color0}16.7}~~&~{\color{color0}16.4}~~&~{\color{color0}16.4}~~&~{\color{color0}16.4}~~&~{\color{color0}16.4}~~&~{\color{color0}16.4}~~&~{\color{color0}16.4}~~&~{\color{color0}16.4}~~&~{\color{color0}16.4}~~\\
 {\color{red}~\textbullet}\algo{ZT}~&~{\color{color0}29.8}~~&~{\color{color0}15.4}~~&~{\color{color0}8.29}~~&~{\color{color0}4.86}~~&~{\color{color0}3.08}~~&~{\color{color0}2.55}~~&~{\color{color0}2.87}~~&~{\color{color19}1.67}~~&~{\color{color20}1.07}~~&~{\color{color18}0.77}~~\\
 {\color{red}~\textbullet}\algo{QS}~&~{\color{color13}13.5}~~&~{\color{color13}8.57}~~&~{\color{color8}5.29}~~&~{\color{color9}3.43}~~&~{\color{color2}2.64}~~&~{\color{color6}2.48}~~&~{\color{color2}2.82}~~&~{\color{color0}2.71}~~&~{\color{color0}2.70}~~&~{\color{color0}2.70}~~\\
 {\color{red}~\textbullet}\algo{TunBM}~&~{\color{color10}14.9}~~&~{\color{color15}8.01}~~&~{\color{color13}4.58}~~&~{\color{color17}2.92}~~&~{\color{color9}2.50}~~&~{\color{color15}2.36}~~&~{\color{color11}2.57}~~&~{\color{color0}2.58}~~&~{\color{color0}2.57}~~&~{\color{color1}2.58}~~\\
 {\color{red}~\textbullet}\algo{NSN}~&~{\color{color10}15.0}~~&~{\color{color0}15.7}~~&~{\color{color0}15.7}~~&~{\color{color0}15.6}~~&~{\color{color0}15.6}~~&~{\color{color0}15.5}~~&~{\color{color0}15.7}~~&~{\color{color0}15.7}~~&~{\color{color0}15.7}~~&~{\color{color0}15.5}~~\\
 {\color{red}~\textbullet}\algo{Raita}~&~{\color{color8}15.9}~~&~{\color{color13}8.48}~~&~{\color{color12}4.84}~~&~{\color{color14}3.07}~~&~{\color{color8}2.52}~~&~{\color{color12}2.40}~~&~{\color{color8}2.63}~~&~{\color{color0}2.60}~~&~{\color{color0}2.61}~~&~{\color{color0}2.60}~~\\
 {\color{red}~\textbullet}\algo{RCol}~&~{\color{color13}13.8}~~&~{\color{color18}7.43}~~&~{\color{color16}4.27}~~&~{\color{color19}2.76}~~&~{\color{color13}2.43}~~&~{\color{color17}2.33}~~&~{\color{color12}2.53}~~&~{\color{color2}2.52}~~&~{\color{color1}2.48}~~&~{\color{color2}2.47}~~\\
 {\color{red}~\textbullet}\algo{Skip}~&~{\color{color0}23.6}~~&~{\color{color0}13.4}~~&~{\color{color0}8.19}~~&~{\color{color0}5.51}~~&~{\color{color0}3.80}~~&~{\color{color0}2.85}~~&~{\color{color0}4.78}~~&~{\color{color0}3.44}~~&~{\color{color0}2.58}~~&~{\color{color3}2.32}~~\\
 {\color{red}~\textbullet}\algo{BR}~&~{\color{color17}11.8}~~&~{\color{color15}8.20}~~&~{\color{color8}5.34}~~&~{\color{color8}3.50}~~&~{\color{color2}2.64}~~&~{\color{color3}2.51}~~&~{\color{color0}2.85}~~&~{\color{color19}1.67}~~&~{\color{color19}1.08}~~&~{\color{color18}0.79}~~\\
 {\color{red}~\textbullet}\algo{FS}~&~{\color{color13}13.8}~~&~{\color{color18}7.44}~~&~{\color{color16}4.27}~~&~{\color{color19}2.76}~~&~{\color{color13}2.43}~~&~{\color{color17}2.33}~~&~{\color{color12}2.53}~~&~{\color{color2}2.51}~~&~{\color{color0}2.49}~~&~{\color{color2}2.47}~~\\
 {\color{red}~\textbullet}\algo{FFS}~&~{\color{color12}13.8}~~&~{\color{color18}7.52}~~&~{\color{color15}4.33}~~&~{\color{color19}2.79}~~&~{\color{color13}2.44}~~&~{\color{color15}2.35}~~&~{\color{color8}2.64}~~&~{\color{color0}2.62}~~&~{\color{color1}2.47}~~&~{\color{color1}2.53}~~\\
 {\color{red}~\textbullet}\algo{BFS}~&~{\color{color12}13.8}~~&~{\color{color18}7.47}~~&~{\color{color16}4.29}~~&~{\color{color19}2.75}~~&~{\color{color13}2.44}~~&~{\color{color15}2.35}~~&~{\color{color8}2.65}~~&~{\color{color0}2.61}~~&~{\color{color0}2.49}~~&~{\color{color1}2.58}~~\\
 {\color{red}~\textbullet}\algo{TS}~&~{\color{color15}12.8}~~&~{\color{color0}12.1}~~&~{\color{color0}11.0}~~&~{\color{color0}9.22}~~&~{\color{color0}7.07}~~&~{\color{color0}4.98}~~&~{\color{color0}4.35}~~&~{\color{color0}3.58}~~&~{\color{color0}3.19}~~&~{\color{color0}3.10}~~\\
 {\color{red}~\textbullet}\algo{SSABS}~&~{\color{color20}10.6}~~&~{\color{color20}6.92}~~&~{\color{color15}4.43}~~&~{\color{color15}3.06}~~&~{\color{color5}2.58}~~&~{\color{color7}2.46}~~&~{\color{color7}2.66}~~&~{\color{color0}2.64}~~&~{\color{color0}2.65}~~&~{\color{color0}2.66}~~\\
 {\color{red}~\textbullet}\algo{TVSBS}~&~\colorbox{best}{\color{color20}10.2}~~&~{\color{color19}7.19}~~&~{\color{color12}4.73}~~&~{\color{color13}3.17}~~&~{\color{color3}2.62}~~&~{\color{color4}2.50}~~&~{\color{color7}2.68}~~&~\colorbox{best}{\color{color20}1.58}~~&~{\color{color20}1.02}~~&~{\color{color19}0.74}~~\\
 {\color{red}~\textbullet}\algo{FJS}~&~{\color{color20}10.5}~~&~{\color{color20}6.97}~~&~{\color{color14}4.49}~~&~{\color{color14}3.09}~~&~{\color{color4}2.61}~~&~{\color{color5}2.49}~~&~{\color{color7}2.68}~~&~{\color{color0}2.66}~~&~{\color{color0}2.66}~~&~{\color{color0}2.66}~~\\
 {\color{red}~\textbullet}\algo{HASH3}~ &~-~&~{\color{color0}18.1}~~&~{\color{color0}6.68}~~&~{\color{color9}3.45}~~&~{\color{color12}2.45}~~&~{\color{color19}2.30}~~&~{\color{color8}2.63}~~&~{\color{color14}1.92}~~&~{\color{color13}1.55}~~&~{\color{color12}1.38}~~\\
 {\color{red}~\textbullet}\algo{HASH5}~ &~-~ &~-~&~{\color{color0}12.1}~~&~{\color{color0}4.72}~~&~{\color{color2}2.64}~~&~{\color{color13}2.38}~~&~{\color{color5}2.73}~~&~{\color{color15}1.85}~~&~{\color{color16}1.37}~~&~{\color{color14}1.22}~~\\
 {\color{red}~\textbullet}\algo{HASH8}~ &~-~ &~-~ &~-~&~{\color{color0}7.59}~~&~{\color{color0}3.44}~~&~{\color{color8}2.45}~~&~{\color{color0}2.85}~~&~{\color{color13}1.95}~~&~{\color{color14}1.45}~~&~{\color{color13}1.29}~~\\
 {\color{red}~\textbullet}\algo{TSW}~&~{\color{color13}13.6}~~&~{\color{color9}9.51}~~&~{\color{color1}6.29}~~&~{\color{color0}4.18}~~&~{\color{color0}3.05}~~&~{\color{color0}2.55}~~&~{\color{color0}3.36}~~&~{\color{color11}2.06}~~&~{\color{color15}1.40}~~&~{\color{color15}1.06}~~\\
 {\color{red}~\textbullet}\algo{GRASPm}~&~{\color{color9}15.4}~~&~{\color{color14}8.25}~~&~{\color{color13}4.69}~~&~{\color{color16}2.95}~~&~{\color{color10}2.48}~~&~{\color{color15}2.35}~~&~{\color{color9}2.61}~~&~{\color{color2}2.49}~~&~{\color{color6}2.09}~~&~{\color{color10}1.61}~~\\
 {\color{red}~\textbullet}\algo{SSEF}~ &~-~ &~-~ &~-~ &~-~&~{\color{color0}5.38}~~&~{\color{color0}3.38}~~&~{\color{color0}3.44}~~&~{\color{color17}1.78}~~&~\colorbox{best}{\color{color20}1.00}~~&~\colorbox{best}{\color{color20}0.54}~~\\
 \hline
 {\color{green}~\textbullet}\algo{RF}~&~{\color{color0}22.8}~~&~{\color{color0}13.0}~~&~{\color{color0}8.05}~~&~{\color{color0}5.29}~~&~{\color{color0}3.36}~~&~{\color{color0}2.57}~~&~{\color{color0}3.07}~~&~{\color{color4}2.42}~~&~{\color{color0}2.62}~~&~{\color{color0}4.16}~~\\
 {\color{green}~\textbullet}\algo{BOM}~&~{\color{color0}37.8}~~&~{\color{color0}27.4}~~&~{\color{color0}24.6}~~&~{\color{color0}17.4}~~&~{\color{color0}11.4}~~&~{\color{color0}6.97}~~&~{\color{color0}5.31}~~&~{\color{color0}2.96}~~&~{\color{color11}1.73}~~&~{\color{color15}1.15}~~\\
 {\color{green}~\textbullet}\algo{BOM2}~&~{\color{color0}24.0}~~&~{\color{color0}13.6}~~&~{\color{color0}8.47}~~&~{\color{color0}5.59}~~&~{\color{color0}3.48}~~&~{\color{color3}2.51}~~&~{\color{color0}3.00}~~&~{\color{color17}1.75}~~&~{\color{color17}1.27}~~&~{\color{color2}2.42}~~\\
 {\color{green}~\textbullet}\algo{EBOM}~&~{\color{color3}18.3}~~&~\colorbox{best}{\color{color20}6.87}~~&~\colorbox{best}{\color{color20}3.63}~~&~\colorbox{best}{\color{color20}2.67}~~&~{\color{color10}2.49}~~&~{\color{color11}2.41}~~&~{\color{color5}2.72}~~&~{\color{color18}1.71}~~&~{\color{color15}1.38}~~&~{\color{color0}2.69}~~\\
 {\color{green}~\textbullet}\algo{FBOM}~&~{\color{color17}11.8}~~&~{\color{color18}7.41}~~&~{\color{color13}4.61}~~&~{\color{color15}3.05}~~&~{\color{color0}2.67}~~&~{\color{color0}2.61}~~&~{\color{color0}2.91}~~&~{\color{color16}1.79}~~&~{\color{color14}1.46}~~&~{\color{color0}2.72}~~\\
 {\color{green}~\textbullet}\algo{SEBOM}~&~{\color{color0}19.4}~~&~{\color{color19}7.29}~~&~{\color{color19}3.85}~~&~{\color{color19}2.79}~~&~{\color{color4}2.61}~~&~{\color{color3}2.52}~~&~{\color{color1}2.83}~~&~{\color{color16}1.79}~~&~{\color{color14}1.45}~~&~{\color{color0}2.73}~~\\
 {\color{green}~\textbullet}\algo{SFBOM}~&~{\color{color17}11.8}~~&~{\color{color18}7.48}~~&~{\color{color13}4.68}~~&~{\color{color13}3.14}~~&~{\color{color0}2.74}~~&~{\color{color0}2.67}~~&~{\color{color0}2.98}~~&~{\color{color14}1.89}~~&~{\color{color14}1.51}~~&~{\color{color0}2.75}~~\\
 \hline
 {\color{blu}~\textbullet}\algo{SO}~&~{\color{color6}16.8}~~&~{\color{color0}16.8}~~&~{\color{color0}16.8}~~&~{\color{color0}16.8}~~&~{\color{color0}16.8}~~&~{\color{color0}22.0}~~&~{\color{color0}22.1}~~&~{\color{color0}22.1}~~&~{\color{color0}22.1}~~&~{\color{color0}22.1}~~\\
 {\color{blu}~\textbullet}\algo{SA}~&~{\color{color7}16.4}~~&~{\color{color0}16.4}~~&~{\color{color0}16.4}~~&~{\color{color0}16.4}~~&~{\color{color0}16.4}~~&~{\color{color0}19.1}~~&~{\color{color0}19.1}~~&~{\color{color0}19.1}~~&~{\color{color0}19.1}~~&~{\color{color0}19.1}~~\\
 {\color{blu}~\textbullet}\algo{SBNDM}~&~{\color{color0}48.1}~~&~{\color{color0}16.6}~~&~{\color{color0}7.52}~~&~{\color{color0}4.00}~~&~{\color{color12}2.45}~~&~{\color{color0}2.61}~~&~{\color{color9}2.61}~~&~{\color{color0}2.61}~~&~{\color{color0}2.60}~~&~{\color{color0}2.61}~~\\
 {\color{blu}~\textbullet}\algo{LBNDM}~&~{\color{color0}31.5}~~&~{\color{color0}16.9}~~&~{\color{color0}9.64}~~&~{\color{color0}5.81}~~&~{\color{color0}3.43}~~&~{\color{color7}2.46}~~&~{\color{color0}3.23}~~&~{\color{color13}1.98}~~&~{\color{color13}1.58}~~&~{\color{color11}1.52}~~\\
 {\color{blu}~\textbullet}\algo{SVM1}~&~{\color{color8}15.9}~~&~{\color{color0}11.6}~~&~{\color{color0}15.4}~~&~{\color{color0}11.1}~~&~{\color{color0}8.70}~~&~{\color{color0}20.7}~~&~{\color{color0}20.7}~~&~{\color{color0}20.7}~~&~{\color{color0}20.7}~~&~{\color{color0}20.7}~~\\
 {\color{blu}~\textbullet}\algo{SBNDM2}~&~{\color{color0}35.0}~~&~{\color{color0}12.3}~~&~{\color{color4}5.90}~~&~{\color{color11}3.27}~~&~{\color{color12}2.45}~~&~{\color{color12}2.39}~~&~{\color{color17}2.40}~~&~{\color{color4}2.39}~~&~{\color{color2}2.39}~~&~{\color{color3}2.39}~~\\
 {\color{blu}~\textbullet}\algo{SBNDM-BMH}~&~{\color{color13}13.6}~~&~{\color{color16}7.81}~~&~{\color{color11}4.92}~~&~{\color{color9}3.45}~~&~{\color{color8}2.52}~~&~{\color{color0}2.60}~~&~{\color{color9}2.60}~~&~{\color{color0}2.60}~~&~{\color{color0}2.60}~~&~{\color{color0}2.61}~~\\
 {\color{blu}~\textbullet}\algo{BMH-SBNDM}~&~{\color{color12}13.9}~~&~{\color{color18}7.42}~~&~{\color{color16}4.24}~~&~{\color{color20}2.73}~~&~{\color{color14}2.41}~~&~{\color{color9}2.44}~~&~{\color{color15}2.45}~~&~{\color{color3}2.44}~~&~{\color{color1}2.44}~~&~{\color{color2}2.44}~~\\
 {\color{blu}~\textbullet}\algo{FAOSOq2}~&~{\color{color0}23.8}~~&~{\color{color2}11.1}~~&~{\color{color0}10.7}~~&~{\color{color0}10.7}~~&~{\color{color0}10.2}~~&~{\color{color0}10.2}~~&~{\color{color0}10.2}~~&~{\color{color0}10.2}~~&~{\color{color0}10.2}~~&~{\color{color0}10.2}~~\\
 {\color{blu}~\textbullet}\algo{FAOSOq4}~ &~-~&~{\color{color0}18.1}~~&~{\color{color2}6.12}~~&~{\color{color0}5.72}~~&~{\color{color0}5.46}~~&~{\color{color0}5.45}~~&~{\color{color0}5.45}~~&~{\color{color0}5.45}~~&~{\color{color0}5.45}~~&~{\color{color0}5.44}~~\\
 {\color{blu}~\textbullet}\algo{AOSO2}~&~{\color{color0}21.9}~~&~{\color{color6}10.1}~~&~{\color{color0}9.73}~~&~{\color{color0}9.73}~~&~{\color{color0}9.72}~~&~{\color{color0}8.52}~~&~{\color{color0}8.53}~~&~{\color{color0}8.52}~~&~{\color{color0}8.54}~~&~{\color{color0}8.53}~~\\
 {\color{blu}~\textbullet}\algo{AOSO4}~ &~-~&~{\color{color0}16.9}~~&~{\color{color7}5.45}~~&~{\color{color0}5.09}~~&~{\color{color0}5.08}~~&~{\color{color0}4.55}~~&~{\color{color0}4.55}~~&~{\color{color0}4.55}~~&~{\color{color0}4.55}~~&~{\color{color0}4.54}~~\\
 {\color{blu}~\textbullet}\algo{AOSO6}~ &~-~ &~-~&~{\color{color0}15.1}~~&~{\color{color1}3.96}~~&~{\color{color0}3.57}~~&~{\color{color0}3.29}~~&~{\color{color0}3.29}~~&~{\color{color0}3.29}~~&~{\color{color0}3.29}~~&~{\color{color0}3.30}~~\\
 {\color{blu}~\textbullet}\algo{FSBNDM}~&~{\color{color0}21.9}~~&~{\color{color1}11.2}~~&~{\color{color3}5.97}~~&~{\color{color9}3.43}~~&~\colorbox{best}{\color{color20}2.29}~~&~\colorbox{best}{\color{color20}2.28}~~&~\colorbox{best}{\color{color20}2.29}~~&~{\color{color6}2.31}~~&~{\color{color3}2.29}~~&~{\color{color3}2.30}~~\\
 {\color{blu}~\textbullet}\algo{BNDMq2}~&~{\color{color0}33.1}~~&~{\color{color0}11.5}~~&~{\color{color8}5.37}~~&~{\color{color16}2.94}~~&~{\color{color15}2.40}~~&~{\color{color6}2.47}~~&~{\color{color14}2.48}~~&~{\color{color3}2.48}~~&~{\color{color1}2.48}~~&~{\color{color2}2.47}~~\\
 {\color{blu}~\textbullet}\algo{BNDMq4}~ &~-~&~{\color{color0}48.5}~~&~{\color{color0}10.5}~~&~{\color{color0}4.59}~~&~{\color{color5}2.58}~~&~{\color{color0}3.15}~~&~{\color{color0}3.15}~~&~{\color{color0}3.15}~~&~{\color{color0}3.14}~~&~{\color{color0}3.15}~~\\
 {\color{blu}~\textbullet}\algo{SBNDMq2}~&~{\color{color0}32.4}~~&~{\color{color0}11.4}~~&~{\color{color7}5.46}~~&~{\color{color14}3.12}~~&~{\color{color14}2.42}~~&~{\color{color13}2.38}~~&~{\color{color18}2.37}~~&~{\color{color4}2.39}~~&~{\color{color2}2.39}~~&~{\color{color3}2.39}~~\\
 {\color{blu}~\textbullet}\algo{SBNDMq4}~ &~-~&~{\color{color0}45.7}~~&~{\color{color0}9.90}~~&~{\color{color0}4.37}~~&~{\color{color8}2.53}~~&~{\color{color9}2.44}~~&~{\color{color15}2.45}~~&~{\color{color3}2.45}~~&~{\color{color1}2.44}~~&~{\color{color2}2.44}~~\\
 {\color{blu}~\textbullet}\algo{UFNDMq2}~&~{\color{color0}30.3}~~&~{\color{color0}15.5}~~&~{\color{color0}8.19}~~&~{\color{color0}4.52}~~&~{\color{color0}2.79}~~&~{\color{color0}2.79}~~&~{\color{color3}2.79}~~&~{\color{color0}2.79}~~&~{\color{color0}2.80}~~&~{\color{color0}2.80}~~\\
 {\color{blu}~\textbullet}\algo{DBWW}~&~{\color{color0}19.3}~~&~{\color{color2}11.1}~~&~{\color{color0}6.90}~~&~{\color{color0}4.07}~~&~{\color{color0}4.07}~~&~{\color{color0}4.06}~~&~{\color{color0}4.07}~~&~{\color{color0}4.06}~~&~{\color{color0}4.07}~~&~{\color{color0}4.07}~~\\
 {\color{blu}~\textbullet}\algo{DBWW2}~&~{\color{color1}19.1}~~&~{\color{color2}11.1}~~&~{\color{color0}6.85}~~&~{\color{color0}4.03}~~&~{\color{color0}4.03}~~&~{\color{color0}4.03}~~&~{\color{color0}4.04}~~&~{\color{color0}4.04}~~&~{\color{color0}4.05}~~&~{\color{color0}4.04}~~\\
 {\color{blu}~\textbullet}\algo{KBNDM}~&~{\color{color0}40.2}~~&~{\color{color0}20.3}~~&~{\color{color0}10.6}~~&~{\color{color0}5.82}~~&~{\color{color0}3.49}~~&~{\color{color0}2.62}~~&~{\color{color0}3.01}~~&~{\color{color12}2.00}~~&~{\color{color7}2.00}~~&~{\color{color6}2.02}~~\\
\hline\end{tabular*}\\
\end{center}
\end{scriptsize}

In the case of very short patterns the \textsf{TVSBS} and the \textsf{EBOM} algorithms obtain the best results.
For short patterns the algorithm \textsf{EBOM} is still the best algorithm. However it is outperformed by the \textsf{FSBNDM} algorithm for patterns of length 32.
In the case of long patterns 
the algorithm \textsf{FSBNDM} achieves the best results when then length of the pattern is less than 256. For patterns of length 256 the best results are obtained by the
\textsf{TVSBS} algorithm. 
For very long patterns the best results are obtained by the \textsf{SSEF} algorithm. 
For the overall performance the algorithms \textsf{TVSBS} and \textsf{BR} maintain very good performance for all patterns. 

\vfill

\subsection{Experimental Results on Rand$64$ Problem}\label{sec:exp64}
In this section we present experimental results on a random text buffer over an alphabet of 64 characters.

\begin{scriptsize}
\begin{center}
\begin{tabular*}{\textwidth}{@{\extracolsep{\fill}}|l|cc|ccc|ccc|cc|}
\hline
$m$ & 2 & 4 & 8 & 16 & 32 & 64 & 128 & 256 & 512 & 1024 \\
\hline

 {\color{red}~\textbullet}\algo{BM}~&~{\color{color0}20.3}~~&~{\color{color2}10.7}~~&~{\color{color0}5.87}~~&~{\color{color3}3.49}~~&~{\color{color9}2.52}~~&~{\color{color13}2.32}~~&~{\color{color5}2.72}~~&~{\color{color0}2.53}~~&~{\color{color0}2.49}~~&~{\color{color0}2.49}~~\\
 {\color{red}~\textbullet}\algo{HOR}~&~{\color{color0}24.9}~~&~{\color{color0}13.0}~~&~{\color{color0}7.05}~~&~{\color{color0}4.04}~~&~{\color{color0}2.69}~~&~{\color{color6}2.43}~~&~{\color{color0}3.14}~~&~{\color{color0}2.72}~~&~{\color{color0}2.68}~~&~{\color{color0}2.69}~~\\
 {\color{red}~\textbullet}\algo{KR}~&~{\color{color1}17.1}~~&~{\color{color0}16.5}~~&~{\color{color0}16.4}~~&~{\color{color0}16.4}~~&~{\color{color0}16.4}~~&~{\color{color0}16.4}~~&~{\color{color0}16.4}~~&~{\color{color0}16.4}~~&~{\color{color0}16.4}~~&~{\color{color0}16.4}~~\\
 {\color{red}~\textbullet}\algo{ZT}~&~{\color{color0}29.4}~~&~{\color{color0}15.4}~~&~{\color{color0}8.42}~~&~{\color{color0}5.01}~~&~{\color{color0}3.25}~~&~{\color{color0}2.61}~~&~{\color{color0}2.86}~~&~{\color{color19}1.61}~~&~{\color{color20}0.97}~~&~{\color{color20}0.61}~~\\
 {\color{red}~\textbullet}\algo{QS}~&~{\color{color13}12.6}~~&~{\color{color13}7.92}~~&~{\color{color10}4.80}~~&~{\color{color11}3.03}~~&~{\color{color12}2.45}~~&~{\color{color15}2.28}~~&~{\color{color0}2.97}~~&~{\color{color0}2.59}~~&~{\color{color0}2.55}~~&~{\color{color0}2.57}~~\\
 {\color{red}~\textbullet}\algo{TunBM}~&~{\color{color10}13.7}~~&~{\color{color16}7.29}~~&~{\color{color17}4.09}~~&~{\color{color19}2.60}~~&~{\color{color17}2.34}~~&~{\color{color20}2.20}~~&~{\color{color9}2.60}~~&~{\color{color1}2.42}~~&~{\color{color1}2.40}~~&~{\color{color1}2.38}~~\\
 {\color{red}~\textbullet}\algo{NSN}~&~{\color{color8}14.3}~~&~{\color{color0}14.5}~~&~{\color{color0}14.5}~~&~{\color{color0}14.6}~~&~{\color{color0}14.5}~~&~{\color{color0}14.5}~~&~{\color{color0}14.5}~~&~{\color{color0}14.5}~~&~{\color{color0}14.4}~~&~{\color{color0}14.5}~~\\
 {\color{red}~\textbullet}\algo{Raita}~&~{\color{color5}15.4}~~&~{\color{color13}8.12}~~&~{\color{color13}4.55}~~&~{\color{color16}2.77}~~&~{\color{color14}2.41}~~&~{\color{color16}2.26}~~&~{\color{color7}2.66}~~&~{\color{color0}2.47}~~&~{\color{color0}2.45}~~&~{\color{color0}2.44}~~\\
 {\color{red}~\textbullet}\algo{RCol}~&~{\color{color11}13.1}~~&~{\color{color17}7.00}~~&~{\color{color18}3.94}~~&~{\color{color20}2.52}~~&~{\color{color18}2.32}~~&~{\color{color20}2.20}~~&~{\color{color10}2.58}~~&~{\color{color1}2.41}~~&~{\color{color1}2.38}~~&~{\color{color1}2.36}~~\\
 {\color{red}~\textbullet}\algo{Skip}~&~{\color{color0}22.1}~~&~{\color{color0}12.0}~~&~{\color{color0}6.92}~~&~{\color{color0}4.37}~~&~{\color{color0}3.11}~~&~{\color{color0}2.57}~~&~{\color{color0}3.74}~~&~{\color{color0}2.76}~~&~{\color{color4}2.14}~~&~{\color{color7}1.83}~~\\
 {\color{red}~\textbullet}\algo{BR}~&~{\color{color16}11.4}~~&~{\color{color13}8.02}~~&~{\color{color5}5.32}~~&~{\color{color1}3.60}~~&~{\color{color0}2.78}~~&~{\color{color0}2.60}~~&~{\color{color2}2.82}~~&~{\color{color19}1.59}~~&~{\color{color20}0.96}~~&~{\color{color20}0.63}~~\\
 {\color{red}~\textbullet}\algo{FS}~&~{\color{color11}13.1}~~&~{\color{color17}7.00}~~&~{\color{color18}3.94}~~&~{\color{color20}2.52}~~&~{\color{color19}2.31}~~&~\colorbox{best}{\color{color20}2.19}~~&~{\color{color10}2.58}~~&~{\color{color1}2.42}~~&~{\color{color1}2.39}~~&~{\color{color1}2.37}~~\\
 {\color{red}~\textbullet}\algo{FFS}~&~{\color{color11}13.1}~~&~{\color{color17}7.03}~~&~{\color{color18}3.96}~~&~{\color{color20}2.54}~~&~{\color{color17}2.34}~~&~{\color{color18}2.23}~~&~{\color{color7}2.66}~~&~{\color{color0}2.54}~~&~{\color{color0}2.58}~~&~{\color{color0}2.77}~~\\
 {\color{red}~\textbullet}\algo{BFS}~&~{\color{color11}13.2}~~&~{\color{color17}7.01}~~&~{\color{color18}3.95}~~&~{\color{color20}2.53}~~&~{\color{color17}2.34}~~&~{\color{color18}2.23}~~&~{\color{color6}2.68}~~&~{\color{color0}2.55}~~&~{\color{color0}2.59}~~&~{\color{color0}2.78}~~\\
 {\color{red}~\textbullet}\algo{TS}~&~{\color{color14}12.0}~~&~{\color{color0}11.6}~~&~{\color{color0}11.0}~~&~{\color{color0}9.94}~~&~{\color{color0}8.37}~~&~{\color{color0}6.39}~~&~{\color{color0}5.20}~~&~{\color{color0}3.87}~~&~{\color{color0}2.90}~~&~{\color{color0}2.68}~~\\
 {\color{red}~\textbullet}\algo{SSABS}~&~{\color{color20}9.75}~~&~{\color{color20}6.38}~~&~{\color{color18}4.02}~~&~{\color{color17}2.69}~~&~{\color{color13}2.43}~~&~{\color{color14}2.29}~~&~{\color{color7}2.66}~~&~{\color{color0}2.48}~~&~{\color{color0}2.44}~~&~{\color{color0}2.44}~~\\
 {\color{red}~\textbullet}\algo{TVSBS}~&~{\color{color20}9.83}~~&~{\color{color17}7.04}~~&~{\color{color11}4.75}~~&~{\color{color6}3.30}~~&~{\color{color0}2.73}~~&~{\color{color0}2.56}~~&~{\color{color7}2.67}~~&~\colorbox{best}{\color{color20}1.52}~~&~\colorbox{best}{\color{color20}0.92}~~&~{\color{color20}0.59}~~\\
 {\color{red}~\textbullet}\algo{PBMH}~&~{\color{color0}21.0}~~&~{\color{color0}11.0}~~&~{\color{color0}6.05}~~&~{\color{color1}3.58}~~&~{\color{color8}2.54}~~&~{\color{color10}2.36}~~&~{\color{color0}3.19}~~&~{\color{color0}2.91}~~&~{\color{color0}3.64}~~&~{\color{color0}6.68}~~\\
 {\color{red}~\textbullet}\algo{FJS}~&~\colorbox{best}{\color{color20}9.59}~~&~\colorbox{best}{\color{color20}6.30}~~&~{\color{color18}3.98}~~&~{\color{color17}2.69}~~&~{\color{color12}2.44}~~&~{\color{color13}2.31}~~&~{\color{color7}2.66}~~&~{\color{color0}2.48}~~&~{\color{color0}2.43}~~&~{\color{color0}2.44}~~\\
 {\color{red}~\textbullet}\algo{HASH3}~ &~-~&~{\color{color0}17.9}~~&~{\color{color0}6.62}~~&~{\color{color5}3.40}~~&~{\color{color13}2.42}~~&~{\color{color15}2.28}~~&~{\color{color10}2.58}~~&~{\color{color16}1.72}~~&~{\color{color16}1.28}~~&~{\color{color14}1.13}~~\\
 {\color{red}~\textbullet}\algo{HASH5}~ &~-~ &~-~&~{\color{color0}12.1}~~&~{\color{color0}4.73}~~&~{\color{color3}2.63}~~&~{\color{color8}2.39}~~&~{\color{color5}2.72}~~&~{\color{color14}1.84}~~&~{\color{color15}1.37}~~&~{\color{color14}1.21}~~\\
 {\color{red}~\textbullet}\algo{HASH8}~ &~-~ &~-~ &~-~&~{\color{color0}7.60}~~&~{\color{color0}3.44}~~&~{\color{color4}2.46}~~&~{\color{color0}2.86}~~&~{\color{color11}1.96}~~&~{\color{color14}1.44}~~&~{\color{color13}1.29}~~\\
 {\color{red}~\textbullet}\algo{TSW}~&~{\color{color3}16.3}~~&~{\color{color0}11.4}~~&~{\color{color0}7.42}~~&~{\color{color0}4.80}~~&~{\color{color0}3.33}~~&~{\color{color0}2.68}~~&~{\color{color0}3.34}~~&~{\color{color11}1.98}~~&~{\color{color16}1.26}~~&~{\color{color17}0.88}~~\\
 {\color{red}~\textbullet}\algo{GRASPm}~&~{\color{color8}14.6}~~&~{\color{color14}7.72}~~&~{\color{color15}4.31}~~&~{\color{color18}2.63}~~&~{\color{color16}2.36}~~&~{\color{color19}2.21}~~&~{\color{color8}2.64}~~&~{\color{color1}2.41}~~&~{\color{color3}2.26}~~&~{\color{color5}2.05}~~\\
 {\color{red}~\textbullet}\algo{SSEF}~ &~-~ &~-~ &~-~ &~-~&~{\color{color0}5.39}~~&~{\color{color0}3.37}~~&~{\color{color0}3.43}~~&~{\color{color14}1.80}~~&~{\color{color20}0.99}~~&~\colorbox{best}{\color{color20}0.55}~~\\
 \hline
 {\color{green}~\textbullet}\algo{RF}~&~{\color{color0}21.0}~~&~{\color{color0}11.5}~~&~{\color{color0}6.70}~~&~{\color{color0}4.32}~~&~{\color{color0}3.07}~~&~{\color{color0}2.56}~~&~{\color{color0}3.05}~~&~{\color{color3}2.34}~~&~{\color{color0}2.51}~~&~{\color{color0}3.89}~~\\
 {\color{green}~\textbullet}\algo{BOM}~&~{\color{color0}36.2}~~&~{\color{color0}26.6}~~&~{\color{color0}25.6}~~&~{\color{color0}19.0}~~&~{\color{color0}13.9}~~&~{\color{color0}9.40}~~&~{\color{color0}6.85}~~&~{\color{color0}3.71}~~&~{\color{color5}2.10}~~&~{\color{color12}1.36}~~\\
 {\color{green}~\textbullet}\algo{BOM2}~&~{\color{color0}22.2}~~&~{\color{color0}12.1}~~&~{\color{color0}7.10}~~&~{\color{color0}4.54}~~&~{\color{color0}3.18}~~&~{\color{color0}2.52}~~&~{\color{color0}2.99}~~&~{\color{color16}1.71}~~&~{\color{color16}1.24}~~&~{\color{color1}2.40}~~\\
 {\color{green}~\textbullet}\algo{EBOM}~&~{\color{color0}18.4}~~&~{\color{color18}6.92}~~&~\colorbox{best}{\color{color20}3.72}~~&~{\color{color16}2.77}~~&~{\color{color5}2.59}~~&~{\color{color1}2.51}~~&~{\color{color4}2.76}~~&~{\color{color16}1.73}~~&~{\color{color14}1.41}~~&~{\color{color0}2.68}~~\\
 {\color{green}~\textbullet}\algo{FBOM}~&~{\color{color16}11.5}~~&~{\color{color16}7.41}~~&~{\color{color11}4.70}~~&~{\color{color8}3.23}~~&~{\color{color0}2.76}~~&~{\color{color0}2.67}~~&~{\color{color0}2.86}~~&~{\color{color15}1.79}~~&~{\color{color13}1.45}~~&~{\color{color0}2.71}~~\\
 {\color{green}~\textbullet}\algo{SEBOM}~&~{\color{color0}19.4}~~&~{\color{color16}7.32}~~&~{\color{color18}3.94}~~&~{\color{color14}2.88}~~&~{\color{color0}2.70}~~&~{\color{color0}2.62}~~&~{\color{color0}2.87}~~&~{\color{color14}1.83}~~&~{\color{color13}1.48}~~&~{\color{color0}2.74}~~\\
 {\color{green}~\textbullet}\algo{SFBOM}~&~{\color{color16}11.5}~~&~{\color{color15}7.47}~~&~{\color{color11}4.76}~~&~{\color{color7}3.29}~~&~{\color{color0}2.83}~~&~{\color{color0}2.74}~~&~{\color{color0}2.95}~~&~{\color{color13}1.86}~~&~{\color{color13}1.52}~~&~{\color{color0}2.78}~~\\
 {\color{green}~\textbullet}\algo{SBDM}~&~{\color{color0}24.9}~~&~{\color{color0}13.0}~~&~{\color{color0}7.03}~~&~{\color{color0}4.06}~~&~{\color{color0}2.69}~~&~{\color{color5}2.44}~~&~{\color{color0}3.11}~~&~{\color{color0}2.72}~~&~{\color{color0}2.68}~~&~{\color{color0}2.70}~~\\
 \hline
 {\color{blu}~\textbullet}\algo{SO}~&~{\color{color2}16.8}~~&~{\color{color0}16.8}~~&~{\color{color0}16.8}~~&~{\color{color0}16.8}~~&~{\color{color0}16.8}~~&~{\color{color0}22.1}~~&~{\color{color0}22.1}~~&~{\color{color0}22.1}~~&~{\color{color0}22.1}~~&~{\color{color0}22.1}~~\\
 {\color{blu}~\textbullet}\algo{SA}~&~{\color{color3}16.4}~~&~{\color{color0}16.4}~~&~{\color{color0}16.4}~~&~{\color{color0}16.4}~~&~{\color{color0}16.4}~~&~{\color{color0}19.1}~~&~{\color{color0}19.1}~~&~{\color{color0}19.1}~~&~{\color{color0}19.1}~~&~{\color{color0}19.1}~~\\
 {\color{blu}~\textbullet}\algo{SBNDM}~&~{\color{color0}48.0}~~&~{\color{color0}16.5}~~&~{\color{color0}7.47}~~&~{\color{color0}3.94}~~&~{\color{color13}2.42}~~&~{\color{color0}2.53}~~&~{\color{color12}2.52}~~&~{\color{color0}2.52}~~&~{\color{color0}2.52}~~&~{\color{color0}2.52}~~\\
 {\color{blu}~\textbullet}\algo{TNDM}~&~{\color{color0}21.7}~~&~{\color{color0}11.7}~~&~{\color{color0}6.75}~~&~{\color{color0}4.22}~~&~{\color{color0}2.87}~~&~{\color{color0}2.84}~~&~{\color{color1}2.84}~~&~{\color{color0}2.85}~~&~{\color{color0}2.84}~~&~{\color{color0}2.86}~~\\
 {\color{blu}~\textbullet}\algo{TNDMa}~&~{\color{color0}19.5}~~&~{\color{color3}10.4}~~&~{\color{color0}5.94}~~&~{\color{color0}3.89}~~&~{\color{color0}2.92}~~&~{\color{color0}2.84}~~&~{\color{color0}2.85}~~&~{\color{color0}2.84}~~&~{\color{color0}2.84}~~&~{\color{color0}2.86}~~\\
 {\color{blu}~\textbullet}\algo{LBNDM}~&~{\color{color0}30.4}~~&~{\color{color0}15.9}~~&~{\color{color0}8.75}~~&~{\color{color0}5.16}~~&~{\color{color0}3.28}~~&~{\color{color6}2.43}~~&~{\color{color0}3.00}~~&~{\color{color15}1.77}~~&~{\color{color17}1.21}~~&~{\color{color15}1.06}~~\\
 {\color{blu}~\textbullet}\algo{SVM1}~&~{\color{color7}14.7}~~&~{\color{color2}10.5}~~&~{\color{color0}14.9}~~&~{\color{color0}10.5}~~&~{\color{color0}8.34}~~&~{\color{color0}20.7}~~&~{\color{color0}20.7}~~&~{\color{color0}20.7}~~&~{\color{color0}20.7}~~&~{\color{color0}20.7}~~\\
 {\color{blu}~\textbullet}\algo{SBNDM2}~&~{\color{color0}34.9}~~&~{\color{color0}12.3}~~&~{\color{color0}5.85}~~&~{\color{color8}3.23}~~&~{\color{color13}2.42}~~&~{\color{color9}2.38}~~&~{\color{color17}2.38}~~&~{\color{color2}2.38}~~&~{\color{color1}2.38}~~&~{\color{color1}2.38}~~\\
 {\color{blu}~\textbullet}\algo{SBNDM-BMH}~&~{\color{color13}12.4}~~&~{\color{color18}6.90}~~&~{\color{color17}4.08}~~&~{\color{color15}2.83}~~&~{\color{color14}2.41}~~&~{\color{color0}2.52}~~&~{\color{color12}2.52}~~&~{\color{color0}2.51}~~&~{\color{color0}2.52}~~&~{\color{color0}2.52}~~\\
 {\color{blu}~\textbullet}\algo{BMH-SBNDM}~&~{\color{color11}13.2}~~&~{\color{color18}6.98}~~&~{\color{color19}3.92}~~&~\colorbox{best}{\color{color20}2.50}~~&~{\color{color19}2.30}~~&~{\color{color11}2.34}~~&~{\color{color18}2.35}~~&~{\color{color2}2.36}~~&~{\color{color1}2.36}~~&~{\color{color1}2.35}~~\\
 {\color{blu}~\textbullet}\algo{FAOSOq2}~&~{\color{color0}17.3}~~&~{\color{color1}10.8}~~&~{\color{color0}10.7}~~&~{\color{color0}10.7}~~&~{\color{color0}10.2}~~&~{\color{color0}10.2}~~&~{\color{color0}10.2}~~&~{\color{color0}10.2}~~&~{\color{color0}10.2}~~&~{\color{color0}10.2}~~\\
 {\color{blu}~\textbullet}\algo{FAOSOq4}~ &~-~&~{\color{color0}12.0}~~&~{\color{color1}5.82}~~&~{\color{color0}5.72}~~&~{\color{color0}5.45}~~&~{\color{color0}5.45}~~&~{\color{color0}5.45}~~&~{\color{color0}5.46}~~&~{\color{color0}5.45}~~&~{\color{color0}5.46}~~\\
 {\color{blu}~\textbullet}\algo{AOSO2}~&~{\color{color4}15.8}~~&~{\color{color5}9.80}~~&~{\color{color0}9.73}~~&~{\color{color0}9.72}~~&~{\color{color0}9.72}~~&~{\color{color0}8.53}~~&~{\color{color0}8.52}~~&~{\color{color0}8.52}~~&~{\color{color0}8.52}~~&~{\color{color0}8.53}~~\\
 {\color{blu}~\textbullet}\algo{AOSO4}~ &~-~&~{\color{color0}11.0}~~&~{\color{color7}5.18}~~&~{\color{color0}5.09}~~&~{\color{color0}5.08}~~&~{\color{color0}4.54}~~&~{\color{color0}4.55}~~&~{\color{color0}4.55}~~&~{\color{color0}4.55}~~&~{\color{color0}4.54}~~\\
 {\color{blu}~\textbullet}\algo{FSBNDM}~&~{\color{color0}21.2}~~&~{\color{color0}10.9}~~&~{\color{color1}5.80}~~&~{\color{color6}3.32}~~&~\colorbox{best}{\color{color20}2.27}~~&~{\color{color16}2.27}~~&~\colorbox{best}{\color{color20}2.27}~~&~{\color{color5}2.26}~~&~{\color{color3}2.27}~~&~{\color{color2}2.27}~~\\
 {\color{blu}~\textbullet}\algo{BNDMq2}~&~{\color{color0}33.0}~~&~{\color{color0}11.4}~~&~{\color{color6}5.31}~~&~{\color{color14}2.88}~~&~{\color{color15}2.39}~~&~{\color{color5}2.44}~~&~{\color{color15}2.44}~~&~{\color{color1}2.43}~~&~{\color{color0}2.43}~~&~{\color{color1}2.43}~~\\
 {\color{blu}~\textbullet}\algo{BNDMq4}~ &~-~&~{\color{color0}48.6}~~&~{\color{color0}10.5}~~&~{\color{color0}4.60}~~&~{\color{color6}2.57}~~&~{\color{color0}3.14}~~&~{\color{color0}3.14}~~&~{\color{color0}3.16}~~&~{\color{color0}3.15}~~&~{\color{color0}3.14}~~\\
 {\color{blu}~\textbullet}\algo{SBNDMq2}~&~{\color{color0}32.3}~~&~{\color{color0}11.4}~~&~{\color{color5}5.40}~~&~{\color{color10}3.07}~~&~{\color{color15}2.39}~~&~{\color{color9}2.38}~~&~{\color{color17}2.38}~~&~{\color{color2}2.38}~~&~{\color{color1}2.38}~~&~{\color{color1}2.38}~~\\
 {\color{blu}~\textbullet}\algo{SBNDMq4}~ &~-~&~{\color{color0}45.7}~~&~{\color{color0}9.90}~~&~{\color{color0}4.37}~~&~{\color{color8}2.53}~~&~{\color{color5}2.44}~~&~{\color{color15}2.44}~~&~{\color{color0}2.45}~~&~{\color{color0}2.44}~~&~{\color{color0}2.44}~~\\
 {\color{blu}~\textbullet}\algo{UFNDMq2}~&~{\color{color0}30.2}~~&~{\color{color0}15.4}~~&~{\color{color0}8.07}~~&~{\color{color0}4.43}~~&~{\color{color0}2.71}~~&~{\color{color0}2.70}~~&~{\color{color6}2.70}~~&~{\color{color0}2.70}~~&~{\color{color0}2.70}~~&~{\color{color0}2.70}~~\\
 {\color{blu}~\textbullet}\algo{DBWW}~&~{\color{color0}17.4}~~&~{\color{color6}9.58}~~&~{\color{color2}5.73}~~&~{\color{color0}3.66}~~&~{\color{color0}3.65}~~&~{\color{color0}3.65}~~&~{\color{color0}3.65}~~&~{\color{color0}3.66}~~&~{\color{color0}3.66}~~&~{\color{color0}3.65}~~\\
 {\color{blu}~\textbullet}\algo{DBWW2}~&~{\color{color0}17.3}~~&~{\color{color6}9.55}~~&~{\color{color2}5.73}~~&~{\color{color0}3.63}~~&~{\color{color0}3.64}~~&~{\color{color0}3.63}~~&~{\color{color0}3.63}~~&~{\color{color0}3.63}~~&~{\color{color0}3.63}~~&~{\color{color0}3.64}~~\\
 {\color{blu}~\textbullet}\algo{KBNDM}~&~{\color{color0}40.2}~~&~{\color{color0}20.5}~~&~{\color{color0}10.8}~~&~{\color{color0}5.94}~~&~{\color{color0}3.54}~~&~{\color{color0}2.64}~~&~{\color{color0}3.01}~~&~{\color{color18}1.63}~~&~{\color{color13}1.47}~~&~{\color{color11}1.46}~~\\
\hline\end{tabular*}\\
\end{center}
\end{scriptsize}

In the case of very short patterns the \textsf{FJS}  algorithm obtains the best performance.
For short patterns the algorithms \textsf{SBNDM-BMH} and \textsf{BMH-SBNDM} are very fast. 
However other algorithms based on bit-parallelism obtain good results. In particular the \textsf{FSBNDM} algorithm is the fastest for patterns of length 32.
For patterns of length 8 the \textsf{EBOM} algorithm obtains the best results.
In the case of long patterns the \textsf{FSBNDM} algorithm obtains very good results. 
In some cases it is outperformed by the \textsf{TVSBS} algorithm.
For very long patterns the best results are obtained by the \textsf{SSEF} and \textsf{TVSBS} algorithms. 
Regarding the overall performance the algorithms \textsf{BR}, and \textsf{TVSBS}  
maintain very good performance for all patterns. 

\vfill

\subsection{Experimental Results on Rand$128$ Problem}\label{sec:exp128}
In this section we present experimental results on a random text buffer over an alphabet of 128 characters.

\begin{scriptsize}
\begin{center}
\begin{tabular*}{\textwidth}{@{\extracolsep{\fill}}|l|cc|ccc|ccc|cc|}
\hline
$m$ & 2 & 4 & 8 & 16 & 32 & 64 & 128 & 256 & 512 & 1024 \\
\hline

 {\color{red}~\textbullet}\algo{BM}~&~{\color{color0}19.9}~~&~{\color{color2}10.4}~~&~{\color{color2}5.65}~~&~{\color{color5}3.32}~~&~{\color{color11}2.44}~~&~{\color{color13}2.24}~~&~{\color{color6}2.68}~~&~{\color{color4}2.08}~~&~{\color{color4}2.11}~~&~{\color{color4}2.08}~~\\
 {\color{red}~\textbullet}\algo{HOR}~&~{\color{color0}24.2}~~&~{\color{color0}12.6}~~&~{\color{color0}6.74}~~&~{\color{color0}3.81}~~&~{\color{color6}2.53}~~&~{\color{color7}2.34}~~&~{\color{color0}3.06}~~&~{\color{color0}2.30}~~&~{\color{color0}2.36}~~&~{\color{color1}2.36}~~\\
 {\color{red}~\textbullet}\algo{KR}~&~{\color{color1}16.7}~~&~{\color{color0}16.4}~~&~{\color{color0}16.3}~~&~{\color{color0}16.4}~~&~{\color{color0}16.4}~~&~{\color{color0}16.4}~~&~{\color{color0}16.4}~~&~{\color{color0}16.4}~~&~{\color{color0}18.7}~~&~{\color{color0}17.4}~~\\
 {\color{red}~\textbullet}\algo{ZT}~&~{\color{color0}39.2}~~&~{\color{color0}20.3}~~&~{\color{color0}10.9}~~&~{\color{color0}6.16}~~&~{\color{color0}3.83}~~&~{\color{color0}2.77}~~&~{\color{color0}2.98}~~&~{\color{color18}1.62}~~&~{\color{color20}1.07}~~&~{\color{color20}0.60}~~\\
 {\color{red}~\textbullet}\algo{OM}~&~{\color{color0}18.5}~~&~{\color{color0}11.6}~~&~{\color{color0}6.99}~~&~{\color{color0}4.30}~~&~{\color{color0}2.83}~~&~{\color{color0}2.45}~~&~{\color{color0}2.83}~~&~{\color{color0}2.26}~~&~{\color{color0}2.35}~~&~{\color{color2}2.22}~~\\
 {\color{red}~\textbullet}\algo{QS}~&~{\color{color13}12.1}~~&~{\color{color14}7.64}~~&~{\color{color12}4.57}~~&~{\color{color13}2.86}~~&~{\color{color13}2.40}~~&~{\color{color15}2.21}~~&~{\color{color0}2.91}~~&~{\color{color0}2.21}~~&~{\color{color2}2.28}~~&~{\color{color4}2.05}~~\\
 {\color{red}~\textbullet}\algo{TunBM}~&~{\color{color10}13.1}~~&~{\color{color16}6.99}~~&~{\color{color19}3.85}~~&~{\color{color20}2.48}~~&~{\color{color19}2.29}~~&~{\color{color20}2.14}~~&~{\color{color9}2.58}~~&~{\color{color6}2.03}~~&~{\color{color5}2.06}~~&~{\color{color7}1.84}~~\\
 {\color{red}~\textbullet}\algo{NSN}~&~{\color{color8}13.8}~~&~{\color{color0}14.0}~~&~{\color{color0}13.9}~~&~{\color{color0}13.9}~~&~{\color{color0}13.9}~~&~{\color{color0}13.9}~~&~{\color{color0}14.0}~~&~{\color{color0}13.9}~~&~{\color{color0}15.9}~~&~{\color{color0}14.4}~~\\
 {\color{red}~\textbullet}\algo{Raita}~&~{\color{color5}15.1}~~&~{\color{color12}8.01}~~&~{\color{color14}4.40}~~&~{\color{color17}2.64}~~&~{\color{color14}2.38}~~&~{\color{color17}2.19}~~&~{\color{color6}2.66}~~&~{\color{color4}2.09}~~&~{\color{color4}2.13}~~&~{\color{color6}1.89}~~\\
 {\color{red}~\textbullet}\algo{RCol}~&~{\color{color11}12.8}~~&~{\color{color17}6.82}~~&~{\color{color20}3.80}~~&~{\color{color20}2.46}~~&~{\color{color20}2.27}~~&~\colorbox{best}{\color{color20}2.13}~~&~{\color{color9}2.58}~~&~{\color{color6}2.03}~~&~{\color{color5}2.04}~~&~{\color{color7}1.83}~~\\
 {\color{red}~\textbullet}\algo{Skip}~&~{\color{color0}21.4}~~&~{\color{color0}11.3}~~&~{\color{color0}6.29}~~&~{\color{color0}3.79}~~&~{\color{color0}2.71}~~&~{\color{color1}2.43}~~&~{\color{color0}3.28}~~&~{\color{color1}2.18}~~&~{\color{color7}1.90}~~&~{\color{color12}1.37}~~\\
 {\color{red}~\textbullet}\algo{BR}~&~{\color{color5}15.2}~~&~{\color{color1}10.6}~~&~{\color{color0}6.87}~~&~{\color{color0}4.43}~~&~{\color{color0}3.06}~~&~{\color{color0}2.74}~~&~{\color{color0}2.92}~~&~{\color{color19}1.59}~~&~{\color{color20}1.08}~~&~{\color{color20}0.60}~~\\
 {\color{red}~\textbullet}\algo{FS}~&~{\color{color11}12.8}~~&~{\color{color17}6.82}~~&~{\color{color20}3.79}~~&~{\color{color20}2.46}~~&~{\color{color19}2.29}~~&~\colorbox{best}{\color{color20}2.13}~~&~{\color{color9}2.57}~~&~{\color{color6}2.02}~~&~{\color{color5}2.04}~~&~{\color{color7}1.84}~~\\
 {\color{red}~\textbullet}\algo{FFS}~&~{\color{color11}12.8}~~&~{\color{color17}6.83}~~&~{\color{color20}3.81}~~&~{\color{color20}2.47}~~&~{\color{color17}2.32}~~&~{\color{color18}2.17}~~&~{\color{color6}2.66}~~&~{\color{color0}2.20}~~&~{\color{color0}2.44}~~&~{\color{color0}2.49}~~\\
 {\color{red}~\textbullet}\algo{BFS}~&~{\color{color11}12.8}~~&~{\color{color17}6.85}~~&~{\color{color20}3.80}~~&~{\color{color20}2.47}~~&~{\color{color18}2.31}~~&~{\color{color17}2.19}~~&~{\color{color6}2.66}~~&~{\color{color0}2.20}~~&~{\color{color0}2.44}~~&~{\color{color0}2.50}~~\\
 {\color{red}~\textbullet}\algo{TS}~&~{\color{color14}11.6}~~&~{\color{color0}11.4}~~&~{\color{color0}11.1}~~&~{\color{color0}10.5}~~&~{\color{color0}9.47}~~&~{\color{color0}7.96}~~&~{\color{color0}6.43}~~&~{\color{color0}5.01}~~&~{\color{color0}3.67}~~&~{\color{color0}2.54}~~\\
 {\color{red}~\textbullet}\algo{SSABS}~&~{\color{color20}9.34}~~&~{\color{color20}6.14}~~&~{\color{color20}3.81}~~&~{\color{color19}2.54}~~&~{\color{color14}2.39}~~&~{\color{color13}2.25}~~&~{\color{color8}2.61}~~&~{\color{color4}2.07}~~&~{\color{color5}2.08}~~&~{\color{color6}1.86}~~\\
 {\color{red}~\textbullet}\algo{TVSBS}~&~{\color{color9}13.6}~~&~{\color{color5}9.60}~~&~{\color{color0}6.30}~~&~{\color{color0}4.15}~~&~{\color{color0}2.95}~~&~{\color{color0}2.66}~~&~{\color{color3}2.77}~~&~\colorbox{best}{\color{color20}1.54}~~&~\colorbox{best}{\color{color20}1.05}~~&~{\color{color20}0.59}~~\\
 {\color{red}~\textbullet}\algo{PBMH}~&~{\color{color0}20.3}~~&~{\color{color0}10.6}~~&~{\color{color1}5.76}~~&~{\color{color4}3.38}~~&~{\color{color9}2.47}~~&~{\color{color11}2.27}~~&~{\color{color0}3.20}~~&~{\color{color0}2.54}~~&~{\color{color0}3.52}~~&~{\color{color0}6.30}~~\\
 {\color{red}~\textbullet}\algo{FJS}~&~\colorbox{best}{\color{color20}9.18}~~&~\colorbox{best}{\color{color20}6.04}~~&~{\color{color20}3.76}~~&~{\color{color19}2.54}~~&~{\color{color14}2.38}~~&~{\color{color13}2.25}~~&~{\color{color8}2.61}~~&~{\color{color5}2.04}~~&~{\color{color4}2.10}~~&~{\color{color6}1.86}~~\\
 {\color{red}~\textbullet}\algo{HASH3}~ &~-~&~{\color{color0}18.0}~~&~{\color{color0}6.62}~~&~{\color{color4}3.41}~~&~{\color{color11}2.44}~~&~{\color{color10}2.30}~~&~{\color{color9}2.57}~~&~{\color{color15}1.72}~~&~{\color{color14}1.45}~~&~{\color{color14}1.18}~~\\
 {\color{red}~\textbullet}\algo{HASH5}~ &~-~ &~-~&~{\color{color0}12.1}~~&~{\color{color0}4.71}~~&~{\color{color0}2.65}~~&~{\color{color4}2.38}~~&~{\color{color4}2.73}~~&~{\color{color11}1.85}~~&~{\color{color12}1.57}~~&~{\color{color13}1.26}~~\\
 {\color{red}~\textbullet}\algo{HASH8}~ &~-~ &~-~ &~-~&~{\color{color0}7.59}~~&~{\color{color0}3.44}~~&~{\color{color0}2.46}~~&~{\color{color0}2.85}~~&~{\color{color7}1.97}~~&~{\color{color11}1.65}~~&~{\color{color12}1.34}~~\\
 {\color{red}~\textbullet}\algo{TSW}~&~{\color{color0}18.8}~~&~{\color{color0}13.1}~~&~{\color{color0}8.52}~~&~{\color{color0}5.57}~~&~{\color{color0}4.05}~~&~{\color{color0}3.44}~~&~{\color{color0}3.67}~~&~{\color{color3}2.11}~~&~{\color{color14}1.47}~~&~{\color{color17}0.88}~~\\
 {\color{red}~\textbullet}\algo{GRASPm}~&~{\color{color7}14.2}~~&~{\color{color14}7.51}~~&~{\color{color17}4.13}~~&~{\color{color20}2.48}~~&~{\color{color16}2.35}~~&~{\color{color19}2.15}~~&~{\color{color7}2.64}~~&~{\color{color5}2.06}~~&~{\color{color5}2.06}~~&~{\color{color7}1.80}~~\\
 {\color{red}~\textbullet}\algo{SSEF}~ &~-~ &~-~ &~-~ &~-~&~{\color{color0}5.40}~~&~{\color{color0}3.38}~~&~{\color{color0}3.43}~~&~{\color{color13}1.79}~~&~{\color{color19}1.14}~~&~\colorbox{best}{\color{color20}0.57}~~\\
 \hline
 {\color{green}~\textbullet}\algo{BOM}~&~{\color{color0}35.5}~~&~{\color{color0}26.4}~~&~{\color{color0}26.6}~~&~{\color{color0}20.3}~~&~{\color{color0}16.0}~~&~{\color{color0}12.2}~~&~{\color{color0}9.55}~~&~{\color{color0}5.49}~~&~{\color{color0}3.45}~~&~{\color{color6}1.92}~~\\
 {\color{green}~\textbullet}\algo{BOM2}~&~{\color{color0}21.5}~~&~{\color{color0}11.4}~~&~{\color{color0}6.43}~~&~{\color{color0}3.93}~~&~{\color{color0}2.83}~~&~{\color{color0}2.54}~~&~{\color{color0}3.03}~~&~{\color{color15}1.72}~~&~{\color{color15}1.41}~~&~{\color{color0}2.48}~~\\
 {\color{green}~\textbullet}\algo{ILDM1}~&~{\color{color0}19.6}~~&~{\color{color1}10.4}~~&~{\color{color1}5.77}~~&~{\color{color2}3.51}~~&~{\color{color0}2.75}~~&~{\color{color0}2.59}~~&~{\color{color0}3.18}~~&~{\color{color0}3.02}~~&~{\color{color0}5.01}~~&~{\color{color0}7.90}~~\\
 {\color{green}~\textbullet}\algo{EBOM}~&~{\color{color0}26.0}~~&~{\color{color9}8.61}~~&~{\color{color12}4.65}~~&~{\color{color9}3.11}~~&~{\color{color0}2.69}~~&~{\color{color0}2.56}~~&~{\color{color0}2.83}~~&~{\color{color14}1.75}~~&~{\color{color12}1.62}~~&~{\color{color0}2.77}~~\\
 {\color{green}~\textbullet}\algo{FBOM}~&~{\color{color1}16.8}~~&~{\color{color0}10.7}~~&~{\color{color0}6.52}~~&~{\color{color0}4.16}~~&~{\color{color0}2.97}~~&~{\color{color0}2.83}~~&~{\color{color0}2.96}~~&~{\color{color12}1.83}~~&~{\color{color11}1.68}~~&~{\color{color0}2.79}~~\\
 {\color{green}~\textbullet}\algo{SEBOM}~&~{\color{color0}26.4}~~&~{\color{color8}8.82}~~&~{\color{color10}4.78}~~&~{\color{color7}3.23}~~&~{\color{color0}2.81}~~&~{\color{color0}2.67}~~&~{\color{color0}2.93}~~&~{\color{color12}1.83}~~&~{\color{color11}1.67}~~&~{\color{color0}2.79}~~\\
 {\color{green}~\textbullet}\algo{SFBOM}~&~{\color{color0}16.8}~~&~{\color{color0}10.7}~~&~{\color{color0}6.60}~~&~{\color{color0}4.22}~~&~{\color{color0}3.04}~~&~{\color{color0}2.87}~~&~{\color{color0}3.06}~~&~{\color{color7}1.98}~~&~{\color{color9}1.77}~~&~{\color{color0}2.88}~~\\
 {\color{green}~\textbullet}\algo{SBDM}~&~{\color{color0}24.2}~~&~{\color{color0}12.6}~~&~{\color{color0}6.75}~~&~{\color{color0}3.81}~~&~{\color{color7}2.52}~~&~{\color{color8}2.33}~~&~{\color{color0}3.05}~~&~{\color{color0}2.42}~~&~{\color{color0}2.37}~~&~{\color{color3}2.13}~~\\
 \hline
 {\color{blu}~\textbullet}\algo{SO}~&~{\color{color1}16.8}~~&~{\color{color0}16.9}~~&~{\color{color0}16.8}~~&~{\color{color0}16.8}~~&~{\color{color0}16.8}~~&~{\color{color0}22.1}~~&~{\color{color0}22.1}~~&~{\color{color0}25.3}~~&~{\color{color0}25.1}~~&~{\color{color0}22.8}~~\\
 {\color{blu}~\textbullet}\algo{SA}~&~{\color{color2}16.4}~~&~{\color{color0}16.5}~~&~{\color{color0}16.4}~~&~{\color{color0}16.4}~~&~{\color{color0}16.4}~~&~{\color{color0}19.1}~~&~{\color{color0}19.1}~~&~{\color{color0}21.5}~~&~{\color{color0}21.8}~~&~{\color{color0}19.7}~~\\
 {\color{blu}~\textbullet}\algo{SBNDM}~&~{\color{color0}48.1}~~&~{\color{color0}16.6}~~&~{\color{color0}7.46}~~&~{\color{color0}3.94}~~&~{\color{color12}2.42}~~&~{\color{color6}2.36}~~&~{\color{color17}2.36}~~&~{\color{color0}2.66}~~&~{\color{color0}2.70}~~&~{\color{color0}2.43}~~\\
 {\color{blu}~\textbullet}\algo{TNDM}~&~{\color{color0}21.1}~~&~{\color{color0}11.2}~~&~{\color{color0}6.27}~~&~{\color{color0}3.74}~~&~{\color{color0}2.64}~~&~{\color{color0}2.62}~~&~{\color{color8}2.62}~~&~{\color{color0}2.94}~~&~{\color{color0}3.01}~~&~{\color{color0}2.71}~~\\
 {\color{blu}~\textbullet}\algo{TNDMa}~&~{\color{color0}19.1}~~&~{\color{color4}9.95}~~&~{\color{color5}5.39}~~&~{\color{color6}3.28}~~&~{\color{color4}2.58}~~&~{\color{color0}2.51}~~&~{\color{color11}2.52}~~&~{\color{color0}2.81}~~&~{\color{color0}2.87}~~&~{\color{color0}2.59}~~\\
 {\color{blu}~\textbullet}\algo{LBNDM}~&~{\color{color0}29.8}~~&~{\color{color0}15.5}~~&~{\color{color0}8.29}~~&~{\color{color0}4.73}~~&~{\color{color0}3.01}~~&~{\color{color2}2.41}~~&~{\color{color0}2.91}~~&~{\color{color12}1.81}~~&~{\color{color19}1.18}~~&~{\color{color18}0.81}~~\\
 {\color{blu}~\textbullet}\algo{SVM1}~&~{\color{color8}14.1}~~&~{\color{color3}9.97}~~&~{\color{color0}14.6}~~&~{\color{color0}10.2}~~&~{\color{color0}8.12}~~&~{\color{color0}20.7}~~&~{\color{color0}20.7}~~&~{\color{color0}23.3}~~&~{\color{color0}23.7}~~&~{\color{color0}21.4}~~\\
 {\color{blu}~\textbullet}\algo{SBNDM2}~&~{\color{color0}34.9}~~&~{\color{color0}12.4}~~&~{\color{color0}5.84}~~&~{\color{color7}3.23}~~&~{\color{color12}2.43}~~&~{\color{color4}2.38}~~&~{\color{color16}2.38}~~&~{\color{color0}2.66}~~&~{\color{color0}2.70}~~&~{\color{color0}2.45}~~\\
 {\color{blu}~\textbullet}\algo{SBNDM-BMH}~&~{\color{color14}11.8}~~&~{\color{color18}6.55}~~&~\colorbox{best}{\color{color20}3.71}~~&~{\color{color18}2.58}~~&~{\color{color19}2.29}~~&~{\color{color6}2.36}~~&~{\color{color17}2.36}~~&~{\color{color0}2.67}~~&~{\color{color0}2.69}~~&~{\color{color0}2.43}~~\\
 {\color{blu}~\textbullet}\algo{BMH-SBNDM}~&~{\color{color11}12.8}~~&~{\color{color17}6.86}~~&~{\color{color20}3.78}~~&~\colorbox{best}{\color{color20}2.45}~~&~{\color{color19}2.28}~~&~{\color{color7}2.34}~~&~{\color{color18}2.33}~~&~{\color{color0}2.61}~~&~{\color{color0}2.67}~~&~{\color{color0}2.40}~~\\
 {\color{blu}~\textbullet}\algo{FNDM}~&~{\color{color0}22.5}~~&~{\color{color0}12.1}~~&~{\color{color0}6.65}~~&~{\color{color0}3.93}~~&~{\color{color0}2.71}~~&~{\color{color0}2.63}~~&~{\color{color7}2.63}~~&~{\color{color0}2.95}~~&~{\color{color0}3.00}~~&~{\color{color0}2.72}~~\\
 {\color{blu}~\textbullet}\algo{FAOSOq2}~&~{\color{color8}14.0}~~&~{\color{color0}10.8}~~&~{\color{color0}10.7}~~&~{\color{color0}10.7}~~&~{\color{color0}10.2}~~&~{\color{color0}10.2}~~&~{\color{color0}10.2}~~&~{\color{color0}11.4}~~&~{\color{color0}11.7}~~&~{\color{color0}10.5}~~\\
 {\color{blu}~\textbullet}\algo{FAOSOq4}~ &~-~&~{\color{color8}8.87}~~&~{\color{color1}5.76}~~&~{\color{color0}5.72}~~&~{\color{color0}5.44}~~&~{\color{color0}5.45}~~&~{\color{color0}5.46}~~&~{\color{color0}6.10}~~&~{\color{color0}6.20}~~&~{\color{color0}5.63}~~\\
 {\color{blu}~\textbullet}\algo{AOSO2}~&~{\color{color11}12.7}~~&~{\color{color4}9.74}~~&~{\color{color0}9.75}~~&~{\color{color0}9.74}~~&~{\color{color0}9.74}~~&~{\color{color0}8.53}~~&~{\color{color0}8.51}~~&~{\color{color0}9.58}~~&~{\color{color0}9.74}~~&~{\color{color0}8.79}~~\\
 {\color{blu}~\textbullet}\algo{AOSO4}~ &~-~&~{\color{color12}8.02}~~&~{\color{color7}5.12}~~&~{\color{color0}5.10}~~&~{\color{color0}5.09}~~&~{\color{color0}4.55}~~&~{\color{color0}4.55}~~&~{\color{color0}5.09}~~&~{\color{color0}5.18}~~&~{\color{color0}4.70}~~\\
 {\color{blu}~\textbullet}\algo{FSBNDM}~&~{\color{color0}20.9}~~&~{\color{color0}10.8}~~&~{\color{color1}5.74}~~&~{\color{color6}3.29}~~&~\colorbox{best}{\color{color20}2.26}~~&~{\color{color11}2.27}~~&~\colorbox{best}{\color{color20}2.25}~~&~{\color{color0}2.54}~~&~{\color{color0}2.57}~~&~{\color{color1}2.33}~~\\
 {\color{blu}~\textbullet}\algo{BNDMq2}~&~{\color{color0}33.0}~~&~{\color{color0}11.4}~~&~{\color{color6}5.30}~~&~{\color{color13}2.88}~~&~{\color{color14}2.38}~~&~{\color{color2}2.42}~~&~{\color{color15}2.41}~~&~{\color{color0}2.72}~~&~{\color{color0}2.77}~~&~{\color{color0}2.50}~~\\
 {\color{blu}~\textbullet}\algo{BNDMq4}~ &~-~&~{\color{color0}48.7}~~&~{\color{color0}10.5}~~&~{\color{color0}4.58}~~&~{\color{color4}2.58}~~&~{\color{color0}3.15}~~&~{\color{color0}3.17}~~&~{\color{color0}3.52}~~&~{\color{color0}3.58}~~&~{\color{color0}3.25}~~\\
 {\color{blu}~\textbullet}\algo{SBNDMq2}~&~{\color{color0}32.3}~~&~{\color{color0}11.4}~~&~{\color{color5}5.39}~~&~{\color{color10}3.06}~~&~{\color{color14}2.39}~~&~{\color{color5}2.37}~~&~{\color{color16}2.38}~~&~{\color{color0}2.65}~~&~{\color{color0}2.71}~~&~{\color{color0}2.46}~~\\
 {\color{blu}~\textbullet}\algo{SBNDMq4}~ &~-~&~{\color{color0}45.8}~~&~{\color{color0}9.89}~~&~{\color{color0}4.37}~~&~{\color{color6}2.54}~~&~{\color{color0}2.44}~~&~{\color{color14}2.44}~~&~{\color{color0}2.76}~~&~{\color{color0}2.79}~~&~{\color{color0}2.52}~~\\
 {\color{blu}~\textbullet}\algo{UFNDMq2}~&~{\color{color0}30.2}~~&~{\color{color0}15.4}~~&~{\color{color0}8.07}~~&~{\color{color0}4.40}~~&~{\color{color0}2.69}~~&~{\color{color0}2.68}~~&~{\color{color6}2.67}~~&~{\color{color0}3.00}~~&~{\color{color0}3.06}~~&~{\color{color0}2.77}~~\\
 {\color{blu}~\textbullet}\algo{DBWW}~&~{\color{color1}16.5}~~&~{\color{color8}8.85}~~&~{\color{color8}5.07}~~&~{\color{color8}3.17}~~&~{\color{color0}3.17}~~&~{\color{color0}3.18}~~&~{\color{color0}3.17}~~&~{\color{color0}3.88}~~&~{\color{color0}3.62}~~&~{\color{color0}3.29}~~\\
 {\color{blu}~\textbullet}\algo{DBWW2}~&~{\color{color2}16.4}~~&~{\color{color8}8.82}~~&~{\color{color8}5.05}~~&~{\color{color8}3.16}~~&~{\color{color0}3.17}~~&~{\color{color0}3.17}~~&~{\color{color0}3.17}~~&~{\color{color0}3.84}~~&~{\color{color0}3.59}~~&~{\color{color0}3.27}~~\\
 {\color{blu}~\textbullet}\algo{KBNDM}~&~{\color{color0}44.7}~~&~{\color{color0}22.7}~~&~{\color{color0}11.8}~~&~{\color{color0}6.38}~~&~{\color{color0}3.64}~~&~{\color{color0}2.64}~~&~{\color{color0}3.09}~~&~{\color{color4}2.07}~~&~{\color{color18}1.24}~~&~{\color{color14}1.13}~~\\
\hline\end{tabular*}\\
\end{center}
\end{scriptsize}

In the case of very short patterns the \textsf{FJS} algorithm have the best performance. 
For short patterns the algorithms \textsf{SBNDM-BMH}, \textsf{BMH-SBNDM} and \textsf{FSBNDM} obtain the best results. 
In the case of long patterns the algorithm \textsf{FSBNDM} is still a good choice.
Very good results are obtained by the \textsf{TVSBS} and \textsf{FS} algorithms. 
For very long patterns the best results are obtained by the \textsf{SSEF} and \textsf{TVSBS} algorithms.
Regarding the overall performance the algorithm \textsf{FS} shows good results for all patterns.
Good results are also maintained over all patterns by the algorithm \textsf{FJS}.

\vfill

\subsection{Experimental Results on Rand$256$ Problem}\label{sec:exp256}
In this section we present experimental results on a random text buffer over an alphabet of 256 characters.

\begin{scriptsize}
\begin{center}
\begin{tabular*}{\textwidth}{@{\extracolsep{\fill}}|l|cc|ccc|ccc|cc|}
\hline
$m$ & 2 & 4 & 8 & 16 & 32 & 64 & 128 & 256 & 512 & 1024 \\
\hline

 {\color{red}~\textbullet}\algo{BM}~&~{\color{color0}20.3}~~&~{\color{color0}10.6}~~&~{\color{color0}5.70}~~&~{\color{color1}3.34}~~&~{\color{color3}2.49}~~&~{\color{color14}2.26}~~&~{\color{color2}2.67}~~&~{\color{color13}1.75}~~&~{\color{color7}1.33}~~&~{\color{color4}1.21}~~\\
 {\color{red}~\textbullet}\algo{HOR}~&~{\color{color0}24.7}~~&~{\color{color0}12.8}~~&~{\color{color0}6.83}~~&~{\color{color0}3.83}~~&~{\color{color0}2.54}~~&~{\color{color4}2.39}~~&~{\color{color0}2.98}~~&~{\color{color0}1.98}~~&~{\color{color0}1.54}~~&~{\color{color0}1.40}~~\\
 {\color{red}~\textbullet}\algo{ZT}~&~{\color{color0}47.7}~~&~{\color{color0}24.7}~~&~{\color{color0}13.3}~~&~{\color{color0}7.84}~~&~{\color{color0}5.44}~~&~{\color{color0}4.32}~~&~{\color{color0}3.87}~~&~{\color{color0}2.02}~~&~{\color{color16}1.08}~~&~{\color{color19}0.63}~~\\
 {\color{red}~\textbullet}\algo{OM}~&~{\color{color0}18.8}~~&~{\color{color0}11.8}~~&~{\color{color0}7.05}~~&~{\color{color0}4.30}~~&~{\color{color0}2.81}~~&~{\color{color0}2.48}~~&~{\color{color0}2.85}~~&~{\color{color2}1.94}~~&~{\color{color0}1.52}~~&~{\color{color0}1.49}~~\\
 {\color{red}~\textbullet}\algo{QS}~&~{\color{color13}12.3}~~&~{\color{color13}7.71}~~&~{\color{color11}4.60}~~&~{\color{color12}2.86}~~&~{\color{color6}2.46}~~&~{\color{color14}2.26}~~&~{\color{color0}2.83}~~&~{\color{color5}1.88}~~&~{\color{color1}1.47}~~&~{\color{color1}1.35}~~\\
 {\color{red}~\textbullet}\algo{TunBM}~&~{\color{color10}13.2}~~&~{\color{color16}6.98}~~&~{\color{color18}3.87}~~&~{\color{color20}2.51}~~&~{\color{color17}2.34}~~&~\colorbox{best}{\color{color20}2.17}~~&~{\color{color5}2.60}~~&~{\color{color14}1.73}~~&~{\color{color8}1.30}~~&~{\color{color6}1.16}~~\\
 {\color{red}~\textbullet}\algo{NSN}~&~{\color{color8}14.1}~~&~{\color{color0}14.1}~~&~{\color{color0}14.1}~~&~{\color{color0}14.1}~~&~{\color{color0}14.0}~~&~{\color{color0}14.1}~~&~{\color{color0}14.1}~~&~{\color{color0}14.1}~~&~{\color{color0}14.1}~~&~{\color{color0}14.1}~~\\
 {\color{red}~\textbullet}\algo{Raita}~&~{\color{color5}15.5}~~&~{\color{color11}8.13}~~&~{\color{color12}4.49}~~&~{\color{color17}2.67}~~&~{\color{color8}2.44}~~&~{\color{color16}2.23}~~&~{\color{color0}2.69}~~&~{\color{color10}1.79}~~&~{\color{color6}1.35}~~&~{\color{color5}1.20}~~\\
 {\color{red}~\textbullet}\algo{RCol}~&~{\color{color11}13.1}~~&~{\color{color17}6.91}~~&~{\color{color18}3.84}~~&~{\color{color20}2.53}~~&~{\color{color19}2.32}~~&~{\color{color20}2.18}~~&~{\color{color7}2.58}~~&~{\color{color14}1.73}~~&~{\color{color7}1.31}~~&~{\color{color5}1.18}~~\\
 {\color{red}~\textbullet}\algo{Skip}~&~{\color{color0}21.7}~~&~{\color{color0}11.3}~~&~{\color{color0}6.20}~~&~{\color{color0}3.63}~~&~{\color{color0}2.60}~~&~{\color{color5}2.38}~~&~{\color{color0}3.10}~~&~{\color{color6}1.87}~~&~{\color{color7}1.32}~~&~{\color{color9}1.02}~~\\
 {\color{red}~\textbullet}\algo{BR}~&~{\color{color0}19.9}~~&~{\color{color0}13.8}~~&~{\color{color0}9.06}~~&~{\color{color0}6.15}~~&~{\color{color0}4.75}~~&~{\color{color0}4.21}~~&~{\color{color0}3.78}~~&~{\color{color0}2.00}~~&~{\color{color15}1.10}~~&~{\color{color19}0.60}~~\\
 {\color{red}~\textbullet}\algo{FS}~&~{\color{color11}13.1}~~&~{\color{color17}6.90}~~&~{\color{color18}3.84}~~&~\colorbox{best}{\color{color20}2.50}~~&~{\color{color16}2.35}~~&~\colorbox{best}{\color{color20}2.17}~~&~{\color{color6}2.59}~~&~{\color{color14}1.72}~~&~{\color{color8}1.29}~~&~{\color{color5}1.18}~~\\
 {\color{red}~\textbullet}\algo{FFS}~&~{\color{color11}13.1}~~&~{\color{color17}6.92}~~&~{\color{color18}3.84}~~&~{\color{color20}2.54}~~&~{\color{color15}2.36}~~&~{\color{color16}2.23}~~&~{\color{color1}2.68}~~&~{\color{color3}1.91}~~&~{\color{color0}1.69}~~&~{\color{color0}1.91}~~\\
 {\color{red}~\textbullet}\algo{BFS}~&~{\color{color11}13.1}~~&~{\color{color17}6.90}~~&~{\color{color18}3.84}~~&~{\color{color20}2.52}~~&~{\color{color16}2.35}~~&~{\color{color16}2.23}~~&~{\color{color0}2.69}~~&~{\color{color2}1.94}~~&~{\color{color0}1.69}~~&~{\color{color0}1.92}~~\\
 {\color{red}~\textbullet}\algo{TS}~&~{\color{color14}11.7}~~&~{\color{color0}11.6}~~&~{\color{color0}11.5}~~&~{\color{color0}11.1}~~&~{\color{color0}10.5}~~&~{\color{color0}9.48}~~&~{\color{color0}8.08}~~&~{\color{color0}6.68}~~&~{\color{color0}4.51}~~&~{\color{color0}2.90}~~\\
 {\color{red}~\textbullet}\algo{SSABS}~&~{\color{color20}9.47}~~&~{\color{color20}6.19}~~&~{\color{color18}3.85}~~&~{\color{color19}2.57}~~&~{\color{color9}2.43}~~&~{\color{color12}2.29}~~&~{\color{color5}2.61}~~&~{\color{color13}1.75}~~&~{\color{color7}1.31}~~&~{\color{color5}1.18}~~\\
 {\color{red}~\textbullet}\algo{TVSBS}~&~{\color{color0}18.4}~~&~{\color{color0}12.8}~~&~{\color{color0}8.46}~~&~{\color{color0}5.88}~~&~{\color{color0}4.64}~~&~{\color{color0}4.15}~~&~{\color{color0}3.70}~~&~{\color{color0}1.98}~~&~{\color{color16}1.08}~~&~{\color{color19}0.60}~~\\
 {\color{red}~\textbullet}\algo{PBMH}~&~{\color{color0}20.7}~~&~{\color{color0}10.8}~~&~{\color{color0}5.80}~~&~{\color{color0}3.40}~~&~{\color{color0}2.52}~~&~{\color{color10}2.31}~~&~{\color{color0}3.13}~~&~{\color{color0}2.24}~~&~{\color{color0}2.58}~~&~{\color{color0}5.59}~~\\
 {\color{red}~\textbullet}\algo{FJS}~&~\colorbox{best}{\color{color20}9.26}~~&~\colorbox{best}{\color{color20}6.07}~~&~{\color{color19}3.76}~~&~{\color{color19}2.55}~~&~{\color{color8}2.44}~~&~{\color{color12}2.29}~~&~{\color{color4}2.62}~~&~{\color{color13}1.74}~~&~{\color{color7}1.32}~~&~{\color{color5}1.18}~~\\
 {\color{red}~\textbullet}\algo{HASH3}~ &~-~&~{\color{color0}18.5}~~&~{\color{color0}6.84}~~&~{\color{color0}3.50}~~&~{\color{color2}2.50}~~&~{\color{color7}2.35}~~&~{\color{color3}2.65}~~&~{\color{color11}1.77}~~&~{\color{color7}1.31}~~&~{\color{color5}1.17}~~\\
 {\color{red}~\textbullet}\algo{HASH5}~ &~-~ &~-~&~{\color{color0}12.4}~~&~{\color{color0}4.86}~~&~{\color{color0}2.72}~~&~{\color{color0}2.47}~~&~{\color{color0}2.81}~~&~{\color{color3}1.91}~~&~{\color{color3}1.42}~~&~{\color{color3}1.25}~~\\
 {\color{red}~\textbullet}\algo{HASH8}~ &~-~ &~-~ &~-~&~{\color{color0}7.83}~~&~{\color{color0}3.56}~~&~{\color{color0}2.53}~~&~{\color{color0}2.94}~~&~{\color{color0}2.03}~~&~{\color{color1}1.47}~~&~{\color{color1}1.33}~~\\
 {\color{red}~\textbullet}\algo{TSW}~&~{\color{color0}29.7}~~&~{\color{color0}22.2}~~&~{\color{color0}15.8}~~&~{\color{color0}11.1}~~&~{\color{color0}7.72}~~&~{\color{color0}5.47}~~&~{\color{color0}4.65}~~&~{\color{color0}2.56}~~&~{\color{color0}1.49}~~&~{\color{color11}0.92}~~\\
 {\color{red}~\textbullet}\algo{GRASPm}~&~{\color{color7}14.5}~~&~{\color{color13}7.60}~~&~{\color{color15}4.18}~~&~\colorbox{best}{\color{color20}2.50}~~&~{\color{color13}2.38}~~&~{\color{color19}2.19}~~&~{\color{color3}2.64}~~&~{\color{color13}1.75}~~&~{\color{color7}1.32}~~&~{\color{color6}1.16}~~\\
 {\color{red}~\textbullet}\algo{SSEF}~ &~-~ &~-~ &~-~ &~-~&~{\color{color0}5.58}~~&~{\color{color0}3.49}~~&~{\color{color0}3.55}~~&~{\color{color7}1.84}~~&~{\color{color18}1.03}~~&~\colorbox{best}{\color{color20}0.55}~~\\
 \hline
 {\color{green}~\textbullet}\algo{BOM2}~&~{\color{color0}21.8}~~&~{\color{color0}11.5}~~&~{\color{color0}6.32}~~&~{\color{color0}3.77}~~&~{\color{color0}2.76}~~&~{\color{color0}2.61}~~&~{\color{color0}3.11}~~&~{\color{color7}1.85}~~&~{\color{color7}1.31}~~&~{\color{color0}2.41}~~\\
 {\color{green}~\textbullet}\algo{ILDM1}~&~{\color{color0}19.9}~~&~{\color{color0}10.4}~~&~{\color{color0}5.65}~~&~{\color{color1}3.35}~~&~{\color{color0}2.73}~~&~{\color{color0}2.66}~~&~{\color{color0}3.22}~~&~{\color{color0}3.16}~~&~{\color{color0}4.65}~~&~{\color{color0}7.83}~~\\
 {\color{green}~\textbullet}\algo{EBOM}~&~{\color{color0}30.0}~~&~{\color{color1}10.4}~~&~{\color{color2}5.48}~~&~{\color{color0}3.72}~~&~{\color{color0}3.13}~~&~{\color{color0}2.99}~~&~{\color{color0}3.06}~~&~{\color{color4}1.90}~~&~{\color{color0}1.52}~~&~{\color{color0}2.77}~~\\
 {\color{green}~\textbullet}\algo{SBDM}~&~{\color{color0}24.7}~~&~{\color{color0}12.8}~~&~{\color{color0}6.81}~~&~{\color{color0}3.82}~~&~{\color{color0}2.54}~~&~{\color{color4}2.39}~~&~{\color{color0}3.00}~~&~{\color{color0}1.96}~~&~{\color{color0}1.53}~~&~{\color{color0}1.37}~~\\
 \hline
 {\color{blu}~\textbullet}\algo{SA}~&~{\color{color1}16.9}~~&~{\color{color0}16.9}~~&~{\color{color0}17.0}~~&~{\color{color0}17.0}~~&~{\color{color0}17.0}~~&~{\color{color0}19.7}~~&~{\color{color0}19.7}~~&~{\color{color0}19.7}~~&~{\color{color0}19.7}~~&~{\color{color0}19.7}~~\\
 {\color{blu}~\textbullet}\algo{SBNDM}~&~{\color{color0}49.6}~~&~{\color{color0}17.0}~~&~{\color{color0}7.69}~~&~{\color{color0}4.05}~~&~{\color{color3}2.49}~~&~{\color{color6}2.36}~~&~{\color{color19}2.36}~~&~{\color{color0}2.35}~~&~{\color{color0}2.35}~~&~{\color{color0}2.36}~~\\
 {\color{blu}~\textbullet}\algo{TNDM}~&~{\color{color0}21.6}~~&~{\color{color0}11.3}~~&~{\color{color0}6.17}~~&~{\color{color0}3.62}~~&~{\color{color0}2.59}~~&~{\color{color0}2.59}~~&~{\color{color6}2.59}~~&~{\color{color0}2.59}~~&~{\color{color0}2.58}~~&~{\color{color0}2.58}~~\\
 {\color{blu}~\textbullet}\algo{TNDMa}~&~{\color{color0}19.5}~~&~{\color{color2}10.00}~~&~{\color{color4}5.33}~~&~{\color{color7}3.07}~~&~{\color{color0}2.52}~~&~{\color{color0}2.44}~~&~{\color{color17}2.40}~~&~{\color{color0}2.43}~~&~{\color{color0}2.44}~~&~{\color{color0}2.44}~~\\
 {\color{blu}~\textbullet}\algo{LBNDM}~&~{\color{color0}30.4}~~&~{\color{color0}15.7}~~&~{\color{color0}8.33}~~&~{\color{color0}4.66}~~&~{\color{color0}2.87}~~&~{\color{color0}2.46}~~&~{\color{color0}2.91}~~&~\colorbox{best}{\color{color20}1.61}~~&~\colorbox{best}{\color{color20}0.96}~~&~{\color{color18}0.66}~~\\
 {\color{blu}~\textbullet}\algo{SVM1}~&~{\color{color8}14.2}~~&~{\color{color2}9.98}~~&~{\color{color0}13.6}~~&~{\color{color0}10.4}~~&~{\color{color0}8.26}~~&~{\color{color0}21.4}~~&~{\color{color0}21.4}~~&~{\color{color0}21.4}~~&~{\color{color0}21.4}~~&~{\color{color0}21.4}~~\\
 {\color{blu}~\textbullet}\algo{SBNDM2}~&~{\color{color0}36.0}~~&~{\color{color0}12.6}~~&~{\color{color0}6.03}~~&~{\color{color1}3.34}~~&~{\color{color3}2.49}~~&~{\color{color0}2.45}~~&~{\color{color14}2.45}~~&~{\color{color0}2.45}~~&~{\color{color0}2.45}~~&~{\color{color0}2.45}~~\\
 {\color{blu}~\textbullet}\algo{SBNDM-BMH}~&~{\color{color14}11.9}~~&~{\color{color19}6.50}~~&~\colorbox{best}{\color{color20}3.63}~~&~{\color{color19}2.55}~~&~\colorbox{best}{\color{color20}2.30}~~&~{\color{color7}2.35}~~&~{\color{color19}2.35}~~&~{\color{color0}2.34}~~&~{\color{color0}2.36}~~&~{\color{color0}2.36}~~\\
 {\color{blu}~\textbullet}\algo{BMH-SBNDM}~&~{\color{color11}13.1}~~&~{\color{color17}6.90}~~&~{\color{color19}3.83}~~&~{\color{color20}2.51}~~&~{\color{color19}2.32}~~&~{\color{color4}2.39}~~&~{\color{color18}2.38}~~&~{\color{color0}2.38}~~&~{\color{color0}2.39}~~&~{\color{color0}2.40}~~\\
 {\color{blu}~\textbullet}\algo{FNDM}~&~{\color{color0}22.9}~~&~{\color{color0}12.0}~~&~{\color{color0}6.57}~~&~{\color{color0}3.76}~~&~{\color{color0}2.65}~~&~{\color{color0}2.59}~~&~{\color{color5}2.60}~~&~{\color{color0}2.60}~~&~{\color{color0}2.61}~~&~{\color{color0}2.60}~~\\
 {\color{blu}~\textbullet}\algo{FAOSOq2}~&~{\color{color11}12.8}~~&~{\color{color0}11.1}~~&~{\color{color0}11.1}~~&~{\color{color0}11.1}~~&~{\color{color0}10.5}~~&~{\color{color0}10.5}~~&~{\color{color0}10.5}~~&~{\color{color0}10.5}~~&~{\color{color0}10.5}~~&~{\color{color0}10.5}~~\\
 {\color{blu}~\textbullet}\algo{FAOSOq4}~ &~-~&~{\color{color14}7.57}~~&~{\color{color0}5.92}~~&~{\color{color0}5.90}~~&~{\color{color0}5.62}~~&~{\color{color0}5.64}~~&~{\color{color0}5.62}~~&~{\color{color0}5.62}~~&~{\color{color0}5.62}~~&~{\color{color0}5.62}~~\\
 {\color{blu}~\textbullet}\algo{AOSO2}~&~{\color{color14}11.6}~~&~{\color{color2}10.0}~~&~{\color{color0}10.1}~~&~{\color{color0}10.0}~~&~{\color{color0}10.1}~~&~{\color{color0}8.80}~~&~{\color{color0}8.79}~~&~{\color{color0}8.80}~~&~{\color{color0}8.79}~~&~{\color{color0}8.80}~~\\
 {\color{blu}~\textbullet}\algo{AOSO4}~ &~-~&~{\color{color17}6.81}~~&~{\color{color4}5.27}~~&~{\color{color0}5.25}~~&~{\color{color0}5.26}~~&~{\color{color0}4.69}~~&~{\color{color0}4.69}~~&~{\color{color0}4.70}~~&~{\color{color0}4.70}~~&~{\color{color0}4.70}~~\\
 {\color{blu}~\textbullet}\algo{AOSO6}~ &~-~ &~-~&~{\color{color5}5.20}~~&~{\color{color0}3.70}~~&~{\color{color0}3.69}~~&~{\color{color0}3.40}~~&~{\color{color0}3.41}~~&~{\color{color0}3.40}~~&~{\color{color0}3.40}~~&~{\color{color0}3.38}~~\\
 {\color{blu}~\textbullet}\algo{FSBNDM}~&~{\color{color0}21.4}~~&~{\color{color0}11.1}~~&~{\color{color0}5.89}~~&~{\color{color0}3.37}~~&~{\color{color19}2.32}~~&~{\color{color9}2.32}~~&~\colorbox{best}{\color{color20}2.33}~~&~{\color{color0}2.33}~~&~{\color{color0}2.33}~~&~{\color{color0}2.34}~~\\
 {\color{blu}~\textbullet}\algo{BNDMq2}~&~{\color{color0}34.0}~~&~{\color{color0}11.8}~~&~{\color{color2}5.45}~~&~{\color{color9}2.98}~~&~{\color{color7}2.45}~~&~{\color{color0}2.49}~~&~{\color{color12}2.48}~~&~{\color{color0}2.49}~~&~{\color{color0}2.49}~~&~{\color{color0}2.49}~~\\
 {\color{blu}~\textbullet}\algo{SBNDMq2}~&~{\color{color0}33.3}~~&~{\color{color0}11.7}~~&~{\color{color1}5.57}~~&~{\color{color5}3.17}~~&~{\color{color6}2.46}~~&~{\color{color0}2.45}~~&~{\color{color13}2.46}~~&~{\color{color0}2.47}~~&~{\color{color0}2.45}~~&~{\color{color0}2.45}~~\\
 {\color{blu}~\textbullet}\algo{SBNDMq4}~ &~-~&~{\color{color0}47.2}~~&~{\color{color0}10.2}~~&~{\color{color0}4.51}~~&~{\color{color0}2.60}~~&~{\color{color0}2.52}~~&~{\color{color11}2.51}~~&~{\color{color0}2.51}~~&~{\color{color0}2.52}~~&~{\color{color0}2.51}~~\\
 {\color{blu}~\textbullet}\algo{DBWW}~&~{\color{color2}16.6}~~&~{\color{color8}8.77}~~&~{\color{color8}4.88}~~&~{\color{color10}2.96}~~&~{\color{color0}2.96}~~&~{\color{color0}2.97}~~&~{\color{color0}2.95}~~&~{\color{color0}2.96}~~&~{\color{color0}2.97}~~&~{\color{color0}2.97}~~\\
 {\color{blu}~\textbullet}\algo{DBWW2}~&~{\color{color2}16.6}~~&~{\color{color8}8.78}~~&~{\color{color8}4.90}~~&~{\color{color10}2.95}~~&~{\color{color0}2.96}~~&~{\color{color0}2.96}~~&~{\color{color0}2.96}~~&~{\color{color0}2.95}~~&~{\color{color0}2.96}~~&~{\color{color0}2.96}~~\\
 {\color{blu}~\textbullet}\algo{KBNDM}~&~{\color{color0}46.6}~~&~{\color{color0}23.7}~~&~{\color{color0}12.5}~~&~{\color{color0}7.16}~~&~{\color{color0}4.75}~~&~{\color{color0}3.61}~~&~{\color{color0}3.42}~~&~{\color{color11}1.78}~~&~\colorbox{best}{\color{color20}0.96}~~&~{\color{color13}0.87}~~\\
\hline\end{tabular*}\\
\end{center}
\end{scriptsize}

In the case of very short patterns the best performance are obtained by the \textsf{FJS} algorithm. 
For short patterns the algorithms \textsf{SBNDM-BMH}, \textsf{BMH-SBNDM} and \textsf{FS} obtain the best results. 
In the case of long patterns the algorithms based on characters comparison are good choices, among them \textsf{FS}, \textsf{GRASPm} and \textsf{TunBM}.
Very good results are obtained also by the \textsf{LBNDM}, \textsf{KBNDM} and \textsf{FSBNDM} algorithms. 
For very long patterns the best results are obtained by the \textsf{SSEF}, \textsf{LBNDM} and \textsf{KBNDM}  algorithms.
Regarding the overall performance the algorithm \textsf{FS} shows good results for all patterns.
Good results are also maintained over all patterns by the algorithms \textsf{GRASPm} and \textsf{FJS}.

\vfill

\subsection{Experimental Results on a Genome Sequence}\label{sec:exp:ecoli}
In this section we present experimental results on a genomic sequence which consists of 4
 different characters.

\begin{scriptsize}
\begin{center}
\begin{tabular*}{\textwidth}{@{\extracolsep{\fill}}|l|cc|ccc|ccc|cc|}
\hline
$m$ & 2 & 4 & 8 & 16 & 32 & 64 & 128 & 256 & 512 & 1024 \\
\hline

 {\color{red}~\textbullet}\algo{KR}~&~{\color{color12}51.1}~~&~{\color{color12}35.8}~~&~{\color{color0}33.8}~~&~{\color{color0}33.7}~~&~{\color{color0}33.7}~~&~{\color{color0}33.8}~~&~{\color{color0}33.8}~~&~{\color{color0}33.8}~~&~{\color{color0}33.8}~~&~{\color{color0}33.8}~~\\
 {\color{red}~\textbullet}\algo{QS}~&~{\color{color8}58.4}~~&~{\color{color0}44.6}~~&~{\color{color0}34.7}~~&~{\color{color0}31.7}~~&~{\color{color0}31.1}~~&~{\color{color0}31.2}~~&~{\color{color0}31.3}~~&~{\color{color0}31.7}~~&~{\color{color0}31.4}~~&~{\color{color0}31.1}~~\\
 {\color{red}~\textbullet}\algo{TunBM}~&~{\color{color4}66.5}~~&~{\color{color1}44.0}~~&~{\color{color0}35.0}~~&~{\color{color0}32.3}~~&~{\color{color0}31.7}~~&~{\color{color0}32.7}~~&~{\color{color0}32.6}~~&~{\color{color0}32.6}~~&~{\color{color0}32.1}~~&~{\color{color0}31.5}~~\\
 {\color{red}~\textbullet}\algo{NSN}~&~{\color{color9}55.1}~~&~{\color{color0}63.0}~~&~{\color{color0}62.0}~~&~{\color{color0}61.6}~~&~{\color{color0}61.7}~~&~{\color{color0}61.2}~~&~{\color{color0}61.1}~~&~{\color{color0}61.6}~~&~{\color{color0}61.5}~~&~{\color{color0}61.8}~~\\
 {\color{red}~\textbullet}\algo{Raita}~&~{\color{color7}60.6}~~&~{\color{color10}37.4}~~&~{\color{color1}28.4}~~&~{\color{color0}26.2}~~&~{\color{color0}25.7}~~&~{\color{color0}26.6}~~&~{\color{color0}26.2}~~&~{\color{color0}25.8}~~&~{\color{color0}26.1}~~&~{\color{color0}26.0}~~\\
 {\color{red}~\textbullet}\algo{RCol}~&~{\color{color8}57.4}~~&~{\color{color7}39.5}~~&~{\color{color0}28.6}~~&~{\color{color0}23.7}~~&~{\color{color0}20.7}~~&~{\color{color0}17.7}~~&~{\color{color0}15.6}~~&~{\color{color0}14.4}~~&~{\color{color0}12.6}~~&~{\color{color0}11.3}~~\\
 {\color{red}~\textbullet}\algo{ASkip}~&~{\color{color0}112}~~&~{\color{color0}65.5}~~&~{\color{color0}31.7}~~&~{\color{color1}16.3}~~&~{\color{color2}9.67}~~&~{\color{color8}7.07}~~&~{\color{color2}8.41}~~&~{\color{color7}6.59}~~&~{\color{color5}6.61}~~&~{\color{color0}9.32}~~\\
 {\color{red}~\textbullet}\algo{BR}~&~{\color{color12}50.5}~~&~{\color{color10}37.2}~~&~{\color{color5}25.2}~~&~{\color{color0}17.0}~~&~{\color{color0}12.8}~~&~{\color{color0}11.3}~~&~{\color{color0}11.3}~~&~{\color{color0}11.4}~~&~{\color{color0}11.4}~~&~{\color{color0}11.4}~~\\
 {\color{red}~\textbullet}\algo{FS}~&~{\color{color8}57.4}~~&~{\color{color7}39.7}~~&~{\color{color0}28.9}~~&~{\color{color0}23.5}~~&~{\color{color0}20.4}~~&~{\color{color0}17.7}~~&~{\color{color0}15.8}~~&~{\color{color0}14.5}~~&~{\color{color0}12.9}~~&~{\color{color0}11.5}~~\\
 {\color{red}~\textbullet}\algo{FFS}~&~{\color{color9}56.1}~~&~{\color{color9}37.9}~~&~{\color{color3}27.1}~~&~{\color{color0}19.9}~~&~{\color{color0}14.9}~~&~{\color{color0}11.8}~~&~{\color{color0}10.4}~~&~{\color{color0}9.06}~~&~{\color{color2}7.47}~~&~{\color{color6}6.70}~~\\
 {\color{red}~\textbullet}\algo{BFS}~&~{\color{color8}57.2}~~&~{\color{color8}38.8}~~&~{\color{color3}26.9}~~&~{\color{color0}20.1}~~&~{\color{color0}15.7}~~&~{\color{color0}12.6}~~&~{\color{color0}11.2}~~&~{\color{color0}9.89}~~&~{\color{color0}8.43}~~&~{\color{color3}7.62}~~\\
 {\color{red}~\textbullet}\algo{TS}~&~{\color{color9}56.0}~~&~{\color{color0}45.9}~~&~{\color{color0}33.1}~~&~{\color{color0}23.7}~~&~{\color{color0}18.5}~~&~{\color{color0}15.5}~~&~{\color{color0}13.7}~~&~{\color{color0}12.4}~~&~{\color{color0}10.9}~~&~{\color{color0}10.2}~~\\
 {\color{red}~\textbullet}\algo{SSABS}~&~{\color{color11}51.4}~~&~{\color{color3}42.5}~~&~{\color{color0}36.2}~~&~{\color{color0}34.2}~~&~{\color{color0}33.6}~~&~{\color{color0}33.7}~~&~{\color{color0}33.8}~~&~{\color{color0}33.9}~~&~{\color{color0}33.6}~~&~{\color{color0}34.4}~~\\
 {\color{red}~\textbullet}\algo{TVSBS}~&~{\color{color15}44.6}~~&~{\color{color12}35.3}~~&~{\color{color6}24.6}~~&~{\color{color0}17.5}~~&~{\color{color0}13.8}~~&~{\color{color0}12.7}~~&~{\color{color0}12.6}~~&~{\color{color0}12.6}~~&~{\color{color0}12.8}~~&~{\color{color0}12.5}~~\\
 {\color{red}~\textbullet}\algo{FJS}~&~{\color{color7}60.3}~~&~{\color{color0}50.6}~~&~{\color{color0}43.7}~~&~{\color{color0}41.8}~~&~{\color{color0}42.2}~~&~{\color{color0}42.0}~~&~{\color{color0}41.7}~~&~{\color{color0}42.4}~~&~{\color{color0}41.7}~~&~{\color{color0}41.3}~~\\
 {\color{red}~\textbullet}\algo{HASH3}~ &~-~&~{\color{color7}39.8}~~&~\colorbox{best}{\color{color20}15.0}~~&~\colorbox{best}{\color{color20}7.93}~~&~\colorbox{best}{\color{color20}5.38}~~&~{\color{color20}4.96}~~&~{\color{color20}5.59}~~&~{\color{color14}5.22}~~&~{\color{color11}5.06}~~&~{\color{color11}5.03}~~\\
 {\color{red}~\textbullet}\algo{HASH5}~ &~-~ &~-~&~{\color{color6}25.1}~~&~{\color{color16}9.84}~~&~{\color{color20}5.54}~~&~\colorbox{best}{\color{color20}4.91}~~&~{\color{color20}5.66}~~&~{\color{color19}4.00}~~&~{\color{color18}3.04}~~&~{\color{color18}2.58}~~\\
 {\color{red}~\textbullet}\algo{HASH8}~ &~-~ &~-~ &~-~&~{\color{color2}15.8}~~&~{\color{color13}7.18}~~&~{\color{color19}5.09}~~&~{\color{color19}5.86}~~&~{\color{color18}4.19}~~&~{\color{color17}3.13}~~&~{\color{color17}2.70}~~\\
 {\color{red}~\textbullet}\algo{TSW}~&~{\color{color10}55.0}~~&~{\color{color5}40.7}~~&~{\color{color2}27.6}~~&~{\color{color0}19.0}~~&~{\color{color0}14.6}~~&~{\color{color0}12.8}~~&~{\color{color0}12.9}~~&~{\color{color0}12.7}~~&~{\color{color0}12.6}~~&~{\color{color0}12.9}~~\\
 {\color{red}~\textbullet}\algo{GRASPm}~&~{\color{color3}66.7}~~&~{\color{color2}43.8}~~&~{\color{color0}30.1}~~&~{\color{color0}23.1}~~&~{\color{color0}17.9}~~&~{\color{color0}14.2}~~&~{\color{color0}16.1}~~&~{\color{color0}12.5}~~&~{\color{color0}9.97}~~&~{\color{color0}8.90}~~\\
 {\color{red}~\textbullet}\algo{SSEF}~ &~-~ &~-~ &~-~ &~-~&~{\color{color0}11.5}~~&~{\color{color13}6.22}~~&~{\color{color13}6.68}~~&~\colorbox{best}{\color{color20}3.75}~~&~\colorbox{best}{\color{color20}2.26}~~&~\colorbox{best}{\color{color20}1.59}~~\\
 \hline
 {\color{green}~\textbullet}\algo{AUT}~&~{\color{color15}44.8}~~&~{\color{color0}46.0}~~&~{\color{color0}44.8}~~&~{\color{color0}46.1}~~&~{\color{color0}46.0}~~&~{\color{color0}44.8}~~&~{\color{color0}46.1}~~&~{\color{color0}45.0}~~&~{\color{color0}45.7}~~&~{\color{color0}47.1}~~\\
 {\color{green}~\textbullet}\algo{RF}~&~{\color{color0}102}~~&~{\color{color0}62.5}~~&~{\color{color0}35.5}~~&~{\color{color0}20.1}~~&~{\color{color0}11.8}~~&~{\color{color6}7.53}~~&~{\color{color4}8.09}~~&~{\color{color14}5.22}~~&~{\color{color12}4.79}~~&~{\color{color8}6.02}~~\\
 {\color{green}~\textbullet}\algo{TRF}~&~{\color{color0}112}~~&~{\color{color0}70.2}~~&~{\color{color0}41.2}~~&~{\color{color0}24.8}~~&~{\color{color0}15.1}~~&~{\color{color0}9.96}~~&~{\color{color0}9.64}~~&~{\color{color3}7.42}~~&~{\color{color3}7.39}~~&~{\color{color0}8.88}~~\\
 {\color{green}~\textbullet}\algo{BOM}~&~{\color{color0}136}~~&~{\color{color0}90.2}~~&~{\color{color0}56.5}~~&~{\color{color0}36.6}~~&~{\color{color0}23.0}~~&~{\color{color0}14.1}~~&~{\color{color0}11.8}~~&~{\color{color8}6.52}~~&~{\color{color16}3.59}~~&~{\color{color19}2.05}~~\\
 {\color{green}~\textbullet}\algo{BOM2}~&~{\color{color0}117}~~&~{\color{color0}66.7}~~&~{\color{color0}37.7}~~&~{\color{color0}21.2}~~&~{\color{color0}12.0}~~&~{\color{color8}7.20}~~&~{\color{color6}7.84}~~&~{\color{color18}4.37}~~&~{\color{color19}2.73}~~&~{\color{color16}3.20}~~\\
 {\color{green}~\textbullet}\algo{WW}~&~{\color{color0}108}~~&~{\color{color0}67.9}~~&~{\color{color0}40.3}~~&~{\color{color0}23.5}~~&~{\color{color0}13.9}~~&~{\color{color0}8.71}~~&~{\color{color0}9.43}~~&~{\color{color7}6.54}~~&~{\color{color5}6.69}~~&~{\color{color0}9.86}~~\\
 {\color{green}~\textbullet}\algo{ILDM1}~&~{\color{color0}80.4}~~&~{\color{color0}53.9}~~&~{\color{color0}33.3}~~&~{\color{color0}19.9}~~&~{\color{color0}11.8}~~&~{\color{color5}7.73}~~&~{\color{color3}8.33}~~&~{\color{color10}6.10}~~&~{\color{color5}6.65}~~&~{\color{color0}9.94}~~\\
 {\color{green}~\textbullet}\algo{ILDM2}~&~{\color{color0}89.7}~~&~{\color{color0}50.5}~~&~{\color{color1}28.5}~~&~{\color{color1}16.5}~~&~{\color{color1}9.90}~~&~{\color{color10}6.74}~~&~{\color{color5}7.95}~~&~{\color{color11}5.88}~~&~{\color{color6}6.51}~~&~{\color{color0}9.81}~~\\
 {\color{green}~\textbullet}\algo{EBOM}~&~{\color{color12}49.8}~~&~\colorbox{best}{\color{color20}29.1}~~&~{\color{color11}21.3}~~&~{\color{color6}14.3}~~&~{\color{color5}9.03}~~&~{\color{color14}6.11}~~&~{\color{color11}7.03}~~&~{\color{color19}4.08}~~&~{\color{color19}2.74}~~&~{\color{color15}3.42}~~\\
 {\color{green}~\textbullet}\algo{FBOM}~&~{\color{color7}60.2}~~&~{\color{color9}38.0}~~&~{\color{color6}24.8}~~&~{\color{color2}16.0}~~&~{\color{color0}10.2}~~&~{\color{color10}6.78}~~&~{\color{color7}7.59}~~&~{\color{color18}4.35}~~&~{\color{color18}2.84}~~&~{\color{color15}3.49}~~\\
 {\color{green}~\textbullet}\algo{SEBOM}~&~{\color{color11}51.4}~~&~{\color{color20}29.4}~~&~{\color{color11}21.4}~~&~{\color{color5}14.5}~~&~{\color{color4}9.14}~~&~{\color{color13}6.32}~~&~{\color{color10}7.24}~~&~{\color{color18}4.19}~~&~{\color{color19}2.79}~~&~{\color{color15}3.47}~~\\
 {\color{green}~\textbullet}\algo{SFBOM}~&~{\color{color8}57.5}~~&~{\color{color10}37.5}~~&~{\color{color7}24.1}~~&~{\color{color3}15.4}~~&~{\color{color2}9.68}~~&~{\color{color13}6.30}~~&~{\color{color10}7.22}~~&~{\color{color18}4.18}~~&~{\color{color19}2.80}~~&~{\color{color15}3.47}~~\\
 \hline
 {\color{blu}~\textbullet}\algo{SO}~&~{\color{color20}35.3}~~&~{\color{color12}35.3}~~&~{\color{color0}35.3}~~&~{\color{color0}35.3}~~&~{\color{color0}35.3}~~&~{\color{color0}44.8}~~&~{\color{color0}44.8}~~&~{\color{color0}44.8}~~&~{\color{color0}44.8}~~&~{\color{color0}44.9}~~\\
 {\color{blu}~\textbullet}\algo{SA}~&~\colorbox{best}{\color{color20}33.8}~~&~{\color{color14}33.9}~~&~{\color{color0}33.8}~~&~{\color{color0}33.8}~~&~{\color{color0}33.8}~~&~{\color{color0}39.0}~~&~{\color{color0}39.0}~~&~{\color{color0}39.0}~~&~{\color{color0}39.0}~~&~{\color{color0}38.9}~~\\
 {\color{blu}~\textbullet}\algo{BNDM}~&~{\color{color0}102}~~&~{\color{color0}57.1}~~&~{\color{color0}30.5}~~&~{\color{color1}16.5}~~&~{\color{color4}9.12}~~&~{\color{color0}12.1}~~&~{\color{color0}12.1}~~&~{\color{color0}12.0}~~&~{\color{color0}12.1}~~&~{\color{color0}12.1}~~\\
 {\color{blu}~\textbullet}\algo{BNDM-L}~&~{\color{color0}101}~~&~{\color{color0}57.2}~~&~{\color{color0}30.6}~~&~{\color{color1}16.4}~~&~{\color{color4}9.13}~~&~{\color{color0}15.6}~~&~{\color{color0}19.9}~~&~{\color{color0}18.1}~~&~{\color{color0}17.8}~~&~{\color{color0}18.9}~~\\
 {\color{blu}~\textbullet}\algo{SBNDM}~&~{\color{color0}102}~~&~{\color{color0}46.5}~~&~{\color{color4}26.2}~~&~{\color{color4}15.0}~~&~{\color{color7}8.44}~~&~{\color{color3}8.03}~~&~{\color{color5}8.02}~~&~{\color{color0}8.04}~~&~{\color{color0}8.03}~~&~{\color{color2}8.02}~~\\
 {\color{blu}~\textbullet}\algo{TNDM}~&~{\color{color0}82.3}~~&~{\color{color0}52.8}~~&~{\color{color0}29.7}~~&~{\color{color1}16.6}~~&~{\color{color3}9.54}~~&~{\color{color0}9.32}~~&~{\color{color0}9.32}~~&~{\color{color0}9.29}~~&~{\color{color0}9.30}~~&~{\color{color0}9.31}~~\\
 {\color{blu}~\textbullet}\algo{TNDMa}~&~{\color{color0}79.4}~~&~{\color{color0}52.7}~~&~{\color{color0}30.1}~~&~{\color{color0}16.9}~~&~{\color{color2}9.73}~~&~{\color{color0}9.47}~~&~{\color{color0}9.48}~~&~{\color{color0}9.44}~~&~{\color{color0}9.46}~~&~{\color{color0}9.43}~~\\
 {\color{blu}~\textbullet}\algo{LBNDM}~&~{\color{color0}108}~~&~{\color{color0}62.5}~~&~{\color{color0}34.2}~~&~{\color{color0}19.1}~~&~{\color{color0}10.8}~~&~{\color{color2}8.18}~~&~{\color{color0}12.4}~~&~{\color{color0}26.4}~~&~{\color{color0}113}~~&~{\color{color0}111}~~\\
 {\color{blu}~\textbullet}\algo{SBNDM2}~&~{\color{color0}81.5}~~&~{\color{color10}37.6}~~&~{\color{color10}22.3}~~&~{\color{color9}13.0}~~&~{\color{color10}7.75}~~&~{\color{color5}7.73}~~&~{\color{color6}7.73}~~&~{\color{color2}7.74}~~&~{\color{color1}7.74}~~&~{\color{color3}7.73}~~\\
 {\color{blu}~\textbullet}\algo{SBNDM-BMH}~&~{\color{color3}67.4}~~&~{\color{color2}43.5}~~&~{\color{color4}26.2}~~&~{\color{color4}14.9}~~&~{\color{color7}8.44}~~&~{\color{color3}8.04}~~&~{\color{color5}8.01}~~&~{\color{color0}8.04}~~&~{\color{color0}8.03}~~&~{\color{color2}8.02}~~\\
 {\color{blu}~\textbullet}\algo{BMH-SBNDM}~&~{\color{color6}61.3}~~&~{\color{color6}40.4}~~&~{\color{color4}26.2}~~&~{\color{color0}18.2}~~&~{\color{color0}11.9}~~&~{\color{color0}11.9}~~&~{\color{color0}11.8}~~&~{\color{color0}11.8}~~&~{\color{color0}11.8}~~&~{\color{color0}11.8}~~\\
 {\color{blu}~\textbullet}\algo{FAOSOq2}~&~{\color{color0}199}~~&~{\color{color0}78.2}~~&~{\color{color4}26.1}~~&~{\color{color0}22.4}~~&~{\color{color0}20.5}~~&~{\color{color0}20.5}~~&~{\color{color0}20.5}~~&~{\color{color0}20.5}~~&~{\color{color0}20.5}~~&~{\color{color0}20.5}~~\\
 {\color{blu}~\textbullet}\algo{FAOSOq4}~ &~-~&~{\color{color0}163}~~&~{\color{color0}64.7}~~&~{\color{color3}15.5}~~&~{\color{color0}11.0}~~&~{\color{color0}11.0}~~&~{\color{color0}11.0}~~&~{\color{color0}11.0}~~&~{\color{color0}11.0}~~&~{\color{color0}11.0}~~\\
 {\color{blu}~\textbullet}\algo{AOSO2}~&~{\color{color0}211}~~&~{\color{color0}73.0}~~&~{\color{color8}23.4}~~&~{\color{color0}20.0}~~&~{\color{color0}20.0}~~&~{\color{color0}17.6}~~&~{\color{color0}17.6}~~&~{\color{color0}17.6}~~&~{\color{color0}17.6}~~&~{\color{color0}17.6}~~\\
 {\color{blu}~\textbullet}\algo{AOSO4}~ &~-~&~{\color{color0}174}~~&~{\color{color0}61.1}~~&~{\color{color7}13.8}~~&~{\color{color0}10.4}~~&~{\color{color0}9.34}~~&~{\color{color0}9.34}~~&~{\color{color0}9.34}~~&~{\color{color0}9.33}~~&~{\color{color0}9.33}~~\\
 {\color{blu}~\textbullet}\algo{AOSO6}~ &~-~ &~-~&~{\color{color0}155}~~&~{\color{color0}57.5}~~&~{\color{color9}8.15}~~&~{\color{color5}7.62}~~&~{\color{color7}7.61}~~&~{\color{color2}7.62}~~&~{\color{color2}7.60}~~&~{\color{color3}7.61}~~\\
 {\color{blu}~\textbullet}\algo{FSBNDM}~&~{\color{color0}80.9}~~&~{\color{color2}43.0}~~&~{\color{color8}23.2}~~&~{\color{color9}12.9}~~&~{\color{color13}7.06}~~&~{\color{color8}7.07}~~&~{\color{color11}7.05}~~&~{\color{color5}7.07}~~&~{\color{color4}7.06}~~&~{\color{color5}7.07}~~\\
 {\color{blu}~\textbullet}\algo{BNDMq2}~&~{\color{color0}77.6}~~&~{\color{color10}37.6}~~&~{\color{color9}22.5}~~&~{\color{color9}13.0}~~&~{\color{color11}7.63}~~&~{\color{color0}9.49}~~&~{\color{color0}9.50}~~&~{\color{color0}9.48}~~&~{\color{color0}9.47}~~&~{\color{color0}9.48}~~\\
 {\color{blu}~\textbullet}\algo{BNDMq4}~ &~-~&~{\color{color0}101}~~&~{\color{color9}22.4}~~&~{\color{color15}10.2}~~&~{\color{color18}5.95}~~&~{\color{color7}7.34}~~&~{\color{color9}7.34}~~&~{\color{color4}7.35}~~&~{\color{color3}7.35}~~&~{\color{color4}7.33}~~\\
 {\color{blu}~\textbullet}\algo{BNDMq6}~ &~-~ &~-~&~{\color{color0}49.8}~~&~{\color{color4}14.9}~~&~{\color{color12}7.28}~~&~{\color{color11}6.59}~~&~{\color{color14}6.58}~~&~{\color{color7}6.58}~~&~{\color{color5}6.58}~~&~{\color{color6}6.58}~~\\
 {\color{blu}~\textbullet}\algo{SBNDMq2}~&~{\color{color0}76.5}~~&~{\color{color11}36.5}~~&~{\color{color10}22.0}~~&~{\color{color9}12.9}~~&~{\color{color11}7.63}~~&~{\color{color5}7.73}~~&~{\color{color6}7.73}~~&~{\color{color2}7.72}~~&~{\color{color2}7.72}~~&~{\color{color3}7.72}~~\\
 {\color{blu}~\textbullet}\algo{SBNDMq4}~ &~-~&~{\color{color0}94.9}~~&~{\color{color11}21.2}~~&~{\color{color16}9.76}~~&~{\color{color18}5.93}~~&~{\color{color17}5.55}~~&~\colorbox{best}{\color{color20}5.57}~~&~{\color{color12}5.56}~~&~{\color{color9}5.58}~~&~{\color{color9}5.58}~~\\
 {\color{blu}~\textbullet}\algo{SBNDMq6}~ &~-~ &~-~&~{\color{color0}56.7}~~&~{\color{color0}16.6}~~&~{\color{color10}7.73}~~&~{\color{color11}6.64}~~&~{\color{color13}6.65}~~&~{\color{color7}6.66}~~&~{\color{color5}6.64}~~&~{\color{color6}6.64}~~\\
 {\color{blu}~\textbullet}\algo{SBNDMq8}~ &~-~ &~-~&~{\color{color0}201}~~&~{\color{color0}23.6}~~&~{\color{color3}9.37}~~&~{\color{color0}8.64}~~&~{\color{color0}8.64}~~&~{\color{color0}8.63}~~&~{\color{color0}8.64}~~&~{\color{color0}8.65}~~\\
 {\color{blu}~\textbullet}\algo{UFNDMq4}~&~{\color{color0}93.5}~~&~{\color{color0}45.1}~~&~{\color{color7}24.1}~~&~{\color{color8}13.6}~~&~{\color{color7}8.47}~~&~{\color{color0}8.45}~~&~{\color{color2}8.45}~~&~{\color{color0}8.48}~~&~{\color{color0}8.48}~~&~{\color{color1}8.46}~~\\
 {\color{blu}~\textbullet}\algo{UFNDMq6}~&~{\color{color0}109}~~&~{\color{color0}58.2}~~&~{\color{color0}29.4}~~&~{\color{color3}15.4}~~&~{\color{color0}10.0}~~&~{\color{color0}10.1}~~&~{\color{color0}10.0}~~&~{\color{color0}10.0}~~&~{\color{color0}10.0}~~&~{\color{color0}10.1}~~\\
 {\color{blu}~\textbullet}\algo{KBNDM}~&~{\color{color0}109}~~&~{\color{color0}58.1}~~&~{\color{color0}35.5}~~&~{\color{color0}21.6}~~&~{\color{color0}12.9}~~&~{\color{color3}8.06}~~&~{\color{color5}8.02}~~&~{\color{color1}8.00}~~&~{\color{color0}8.02}~~&~{\color{color2}7.99}~~\\
\hline\end{tabular*}\\
\end{center}
\end{scriptsize}

In the case of very short patterns the \textsf{SA} and \textsf{EBOM} algorithms obtain the best performance for patterns of length 2 and 4, respectively. 
In the case of short patterns the \textsf{HASH3} algorithm achieves the best results.
In the case of long patterns the algorithms in the \textsf{HASHq} family are still very good choices. They are sporadically outperformed
by algorithms based on bit-parallelism.
For very long patterns the best results are obtained by the \textsf{SSEF}.
Regarding the overall performance the \textsf{EBOM} family of algorithms maintain good performance for all patterns.

\vfill

\subsection{Experimental Results on a Protein Sequence}\label{sec:exp:hs}
In this section we present experimental results on a protein sequence which consists of 20
 different characters.

\begin{scriptsize}
\begin{center}
\begin{tabular*}{\textwidth}{@{\extracolsep{\fill}}|l|cc|ccc|ccc|cc|}
\hline
$m$ & 2 & 4 & 8 & 16 & 32 & 64 & 128 & 256 & 512 & 1024 \\
\hline

 {\color{red}~\textbullet}\algo{KR}~&~{\color{color2}12.0}~~&~{\color{color0}10.9}~~&~{\color{color0}10.6}~~&~{\color{color0}10.6}~~&~{\color{color0}10.6}~~&~{\color{color0}10.7}~~&~{\color{color0}10.6}~~&~{\color{color0}10.7}~~&~{\color{color0}10.6}~~&~{\color{color0}10.7}~~\\
 {\color{red}~\textbullet}\algo{ZT}~&~{\color{color0}20.5}~~&~{\color{color0}10.6}~~&~{\color{color0}5.69}~~&~{\color{color0}3.35}~~&~{\color{color0}2.14}~~&~{\color{color5}1.76}~~&~{\color{color1}2.02}~~&~{\color{color16}1.34}~~&~{\color{color14}0.99}~~&~{\color{color14}0.84}~~\\
 {\color{red}~\textbullet}\algo{QS}~&~{\color{color6}9.88}~~&~{\color{color7}6.32}~~&~{\color{color7}4.00}~~&~{\color{color5}2.71}~~&~{\color{color0}2.08}~~&~{\color{color0}1.92}~~&~{\color{color0}2.06}~~&~{\color{color0}1.97}~~&~{\color{color0}1.96}~~&~{\color{color0}1.94}~~\\
 {\color{red}~\textbullet}\algo{TunBM}~&~{\color{color4}11.1}~~&~{\color{color7}6.17}~~&~{\color{color10}3.70}~~&~{\color{color10}2.44}~~&~{\color{color2}1.97}~~&~{\color{color0}1.83}~~&~{\color{color6}1.90}~~&~{\color{color0}1.89}~~&~{\color{color0}1.88}~~&~{\color{color1}1.87}~~\\
 {\color{red}~\textbullet}\algo{NSN}~&~{\color{color4}10.9}~~&~{\color{color0}11.4}~~&~{\color{color0}11.3}~~&~{\color{color0}11.3}~~&~{\color{color0}11.4}~~&~{\color{color0}11.5}~~&~{\color{color0}11.5}~~&~{\color{color0}11.4}~~&~{\color{color0}11.4}~~&~{\color{color0}11.2}~~\\
 {\color{red}~\textbullet}\algo{Raita}~&~{\color{color4}11.1}~~&~{\color{color8}5.97}~~&~{\color{color12}3.56}~~&~{\color{color11}2.38}~~&~{\color{color6}1.88}~~&~{\color{color5}1.76}~~&~{\color{color7}1.88}~~&~{\color{color0}1.87}~~&~{\color{color0}1.84}~~&~{\color{color2}1.82}~~\\
 {\color{red}~\textbullet}\algo{RCol}~&~{\color{color6}9.86}~~&~{\color{color10}5.46}~~&~{\color{color14}3.25}~~&~{\color{color15}2.19}~~&~{\color{color9}1.82}~~&~{\color{color9}1.70}~~&~{\color{color10}1.81}~~&~{\color{color3}1.75}~~&~{\color{color1}1.72}~~&~{\color{color4}1.68}~~\\
 {\color{red}~\textbullet}\algo{Skip}~&~{\color{color0}17.0}~~&~{\color{color0}10.4}~~&~{\color{color0}6.73}~~&~{\color{color0}4.59}~~&~{\color{color0}3.21}~~&~{\color{color0}2.49}~~&~{\color{color0}3.91}~~&~{\color{color0}2.79}~~&~{\color{color0}2.16}~~&~{\color{color1}1.93}~~\\
 {\color{red}~\textbullet}\algo{BR}~&~{\color{color9}8.61}~~&~{\color{color8}5.98}~~&~{\color{color8}3.90}~~&~{\color{color9}2.52}~~&~{\color{color7}1.85}~~&~{\color{color6}1.74}~~&~{\color{color0}2.05}~~&~{\color{color16}1.36}~~&~{\color{color14}1.02}~~&~{\color{color14}0.87}~~\\
 {\color{red}~\textbullet}\algo{FS}~&~{\color{color6}9.88}~~&~{\color{color10}5.49}~~&~{\color{color14}3.26}~~&~{\color{color16}2.17}~~&~{\color{color9}1.82}~~&~{\color{color9}1.70}~~&~{\color{color10}1.80}~~&~{\color{color3}1.77}~~&~{\color{color1}1.72}~~&~{\color{color4}1.69}~~\\
 {\color{red}~\textbullet}\algo{FFS}~&~{\color{color6}9.84}~~&~{\color{color10}5.52}~~&~{\color{color14}3.30}~~&~{\color{color15}2.18}~~&~{\color{color10}1.79}~~&~{\color{color10}1.69}~~&~{\color{color7}1.88}~~&~{\color{color0}1.84}~~&~{\color{color0}1.77}~~&~{\color{color0}1.98}~~\\
 {\color{red}~\textbullet}\algo{BFS}~&~{\color{color6}9.94}~~&~{\color{color10}5.48}~~&~{\color{color14}3.25}~~&~{\color{color16}2.15}~~&~{\color{color10}1.78}~~&~{\color{color10}1.68}~~&~{\color{color7}1.87}~~&~{\color{color0}1.85}~~&~{\color{color0}1.80}~~&~{\color{color0}2.01}~~\\
 {\color{red}~\textbullet}\algo{TS}~&~{\color{color7}9.41}~~&~{\color{color0}8.58}~~&~{\color{color0}7.31}~~&~{\color{color0}5.80}~~&~{\color{color0}4.42}~~&~{\color{color0}3.36}~~&~{\color{color0}3.03}~~&~{\color{color0}2.72}~~&~{\color{color0}2.43}~~&~{\color{color0}2.20}~~\\
 {\color{red}~\textbullet}\algo{SSABS}~&~{\color{color10}8.01}~~&~{\color{color10}5.28}~~&~{\color{color11}3.57}~~&~{\color{color10}2.48}~~&~{\color{color0}2.02}~~&~{\color{color0}1.87}~~&~{\color{color3}1.97}~~&~{\color{color0}1.93}~~&~{\color{color0}1.93}~~&~{\color{color1}1.93}~~\\
 {\color{red}~\textbullet}\algo{TVSBS}~&~{\color{color11}7.50}~~&~{\color{color10}5.28}~~&~{\color{color12}3.50}~~&~{\color{color12}2.34}~~&~{\color{color7}1.85}~~&~{\color{color7}1.73}~~&~{\color{color5}1.92}~~&~{\color{color17}1.31}~~&~{\color{color14}0.98}~~&~{\color{color15}0.82}~~\\
 {\color{red}~\textbullet}\algo{FJS}~&~{\color{color9}8.21}~~&~{\color{color10}5.52}~~&~{\color{color10}3.70}~~&~{\color{color6}2.65}~~&~{\color{color0}2.13}~~&~{\color{color0}1.96}~~&~{\color{color0}2.03}~~&~{\color{color0}1.99}~~&~{\color{color0}1.98}~~&~{\color{color0}1.98}~~\\
 {\color{red}~\textbullet}\algo{HASH3}~ &~-~&~{\color{color0}11.8}~~&~{\color{color4}4.39}~~&~{\color{color13}2.29}~~&~{\color{color18}1.61}~~&~\colorbox{best}{\color{color20}1.52}~~&~{\color{color13}1.74}~~&~{\color{color15}1.37}~~&~{\color{color11}1.17}~~&~{\color{color12}1.07}~~\\
 {\color{red}~\textbullet}\algo{HASH5}~ &~-~ &~-~&~{\color{color0}7.85}~~&~{\color{color0}3.07}~~&~{\color{color12}1.74}~~&~{\color{color18}1.56}~~&~{\color{color10}1.79}~~&~{\color{color20}1.22}~~&~{\color{color16}0.90}~~&~{\color{color15}0.80}~~\\
 {\color{red}~\textbullet}\algo{HASH8}~ &~-~ &~-~ &~-~&~{\color{color0}4.96}~~&~{\color{color0}2.24}~~&~{\color{color15}1.60}~~&~{\color{color7}1.87}~~&~{\color{color18}1.29}~~&~{\color{color15}0.96}~~&~{\color{color14}0.85}~~\\
 {\color{red}~\textbullet}\algo{TSW}~&~{\color{color6}9.82}~~&~{\color{color5}6.86}~~&~{\color{color2}4.55}~~&~{\color{color0}3.01}~~&~{\color{color0}2.18}~~&~{\color{color0}1.84}~~&~{\color{color0}2.48}~~&~{\color{color5}1.71}~~&~{\color{color8}1.33}~~&~{\color{color10}1.16}~~\\
 {\color{red}~\textbullet}\algo{GRASPm}~&~{\color{color4}11.2}~~&~{\color{color8}6.10}~~&~{\color{color11}3.60}~~&~{\color{color12}2.35}~~&~{\color{color7}1.86}~~&~{\color{color8}1.72}~~&~{\color{color6}1.90}~~&~{\color{color5}1.70}~~&~{\color{color8}1.33}~~&~{\color{color13}0.99}~~\\
 {\color{red}~\textbullet}\algo{SSEF}~ &~-~ &~-~ &~-~ &~-~&~{\color{color0}3.57}~~&~{\color{color0}2.24}~~&~{\color{color0}2.29}~~&~\colorbox{best}{\color{color20}1.21}~~&~\colorbox{best}{\color{color20}0.63}~~&~\colorbox{best}{\color{color20}0.36}~~\\
 \hline
 {\color{green}~\textbullet}\algo{RF}~&~{\color{color0}16.8}~~&~{\color{color0}10.2}~~&~{\color{color0}6.47}~~&~{\color{color0}3.93}~~&~{\color{color0}2.31}~~&~{\color{color4}1.78}~~&~{\color{color0}2.17}~~&~{\color{color0}1.91}~~&~{\color{color0}2.39}~~&~{\color{color0}4.19}~~\\
 {\color{green}~\textbullet}\algo{BOM}~&~{\color{color0}26.5}~~&~{\color{color0}18.4}~~&~{\color{color0}14.3}~~&~{\color{color0}9.93}~~&~{\color{color0}6.36}~~&~{\color{color0}3.92}~~&~{\color{color0}3.22}~~&~{\color{color0}1.84}~~&~{\color{color13}1.06}~~&~{\color{color16}0.73}~~\\
 {\color{green}~\textbullet}\algo{BOM2}~&~{\color{color0}17.9}~~&~{\color{color0}10.8}~~&~{\color{color0}6.84}~~&~{\color{color0}4.12}~~&~{\color{color0}2.38}~~&~{\color{color6}1.74}~~&~{\color{color0}2.12}~~&~{\color{color18}1.29}~~&~{\color{color14}1.02}~~&~{\color{color0}2.22}~~\\
 {\color{green}~\textbullet}\algo{EBOM}~&~{\color{color2}12.1}~~&~{\color{color12}4.71}~~&~\colorbox{best}{\color{color20}2.59}~~&~\colorbox{best}{\color{color20}1.91}~~&~{\color{color12}1.75}~~&~{\color{color9}1.70}~~&~{\color{color4}1.95}~~&~{\color{color16}1.34}~~&~{\color{color11}1.20}~~&~{\color{color0}2.57}~~\\
 {\color{green}~\textbullet}\algo{FBOM}~&~{\color{color8}8.64}~~&~{\color{color10}5.36}~~&~{\color{color13}3.40}~~&~{\color{color13}2.29}~~&~{\color{color4}1.92}~~&~{\color{color0}1.85}~~&~{\color{color0}2.21}~~&~{\color{color14}1.41}~~&~{\color{color10}1.24}~~&~{\color{color0}2.59}~~\\
 {\color{green}~\textbullet}\algo{SEBOM}~&~{\color{color0}12.8}~~&~{\color{color11}5.01}~~&~{\color{color19}2.77}~~&~{\color{color18}2.03}~~&~{\color{color6}1.89}~~&~{\color{color2}1.81}~~&~{\color{color0}2.07}~~&~{\color{color13}1.46}~~&~{\color{color9}1.28}~~&~{\color{color0}2.61}~~\\
 {\color{green}~\textbullet}\algo{SFBOM}~&~{\color{color8}8.68}~~&~{\color{color10}5.42}~~&~{\color{color13}3.44}~~&~{\color{color12}2.36}~~&~{\color{color1}1.99}~~&~{\color{color0}1.93}~~&~{\color{color0}2.27}~~&~{\color{color12}1.48}~~&~{\color{color9}1.29}~~&~{\color{color0}2.63}~~\\
 \hline
 {\color{blu}~\textbullet}\algo{SO}~&~{\color{color4}10.8}~~&~{\color{color0}10.8}~~&~{\color{color0}10.8}~~&~{\color{color0}10.9}~~&~{\color{color0}10.8}~~&~{\color{color0}14.2}~~&~{\color{color0}14.2}~~&~{\color{color0}14.2}~~&~{\color{color0}14.2}~~&~{\color{color0}14.2}~~\\
 {\color{blu}~\textbullet}\algo{SA}~&~{\color{color5}10.7}~~&~{\color{color0}10.7}~~&~{\color{color0}10.7}~~&~{\color{color0}10.7}~~&~{\color{color0}10.7}~~&~{\color{color0}12.3}~~&~{\color{color0}12.3}~~&~{\color{color0}12.3}~~&~{\color{color0}12.3}~~&~{\color{color0}12.3}~~\\
 {\color{blu}~\textbullet}\algo{SBNDM}~&~{\color{color0}31.2}~~&~{\color{color0}10.9}~~&~{\color{color0}5.02}~~&~{\color{color5}2.73}~~&~{\color{color14}1.71}~~&~{\color{color8}1.72}~~&~{\color{color13}1.73}~~&~{\color{color4}1.72}~~&~{\color{color1}1.73}~~&~{\color{color3}1.72}~~\\
 {\color{blu}~\textbullet}\algo{TNDM}~&~{\color{color0}16.1}~~&~{\color{color0}9.55}~~&~{\color{color0}5.93}~~&~{\color{color0}3.49}~~&~{\color{color3}1.95}~~&~{\color{color0}1.94}~~&~{\color{color5}1.93}~~&~{\color{color0}1.94}~~&~{\color{color0}1.94}~~&~{\color{color0}1.95}~~\\
 {\color{blu}~\textbullet}\algo{TNDMa}~&~{\color{color0}14.4}~~&~{\color{color0}8.67}~~&~{\color{color0}5.79}~~&~{\color{color0}3.66}~~&~{\color{color0}2.00}~~&~{\color{color0}1.95}~~&~{\color{color4}1.95}~~&~{\color{color0}1.96}~~&~{\color{color0}1.94}~~&~{\color{color0}1.95}~~\\
 {\color{blu}~\textbullet}\algo{LBNDM}~&~{\color{color0}22.0}~~&~{\color{color0}12.2}~~&~{\color{color0}7.08}~~&~{\color{color0}4.08}~~&~{\color{color0}2.34}~~&~{\color{color9}1.70}~~&~{\color{color0}2.28}~~&~{\color{color10}1.53}~~&~{\color{color7}1.40}~~&~{\color{color6}1.49}~~\\
 {\color{blu}~\textbullet}\algo{SVM1}~&~{\color{color2}11.9}~~&~{\color{color0}8.87}~~&~{\color{color0}10.2}~~&~{\color{color0}7.67}~~&~{\color{color0}5.93}~~&~{\color{color0}13.5}~~&~{\color{color0}13.5}~~&~{\color{color0}13.5}~~&~{\color{color0}13.5}~~&~{\color{color0}13.5}~~\\
 {\color{blu}~\textbullet}\algo{SBNDM2}~&~{\color{color0}22.8}~~&~{\color{color0}8.14}~~&~{\color{color8}3.95}~~&~{\color{color14}2.25}~~&~{\color{color16}1.67}~~&~{\color{color15}1.60}~~&~{\color{color18}1.60}~~&~{\color{color8}1.59}~~&~{\color{color3}1.60}~~&~{\color{color5}1.60}~~\\
 {\color{blu}~\textbullet}\algo{SBNDM-BMH}~&~{\color{color5}10.4}~~&~{\color{color7}6.30}~~&~{\color{color6}4.17}~~&~{\color{color6}2.66}~~&~{\color{color13}1.73}~~&~{\color{color8}1.72}~~&~{\color{color13}1.72}~~&~{\color{color4}1.73}~~&~{\color{color1}1.72}~~&~{\color{color3}1.73}~~\\
 {\color{blu}~\textbullet}\algo{BMH-SBNDM}~&~{\color{color6}10.1}~~&~{\color{color10}5.43}~~&~{\color{color15}3.18}~~&~{\color{color17}2.11}~~&~{\color{color13}1.72}~~&~{\color{color4}1.77}~~&~{\color{color11}1.78}~~&~{\color{color2}1.80}~~&~{\color{color0}1.78}~~&~{\color{color3}1.78}~~\\
 {\color{blu}~\textbullet}\algo{FAOSOq2}~&~{\color{color0}23.1}~~&~{\color{color1}8.04}~~&~{\color{color0}7.02}~~&~{\color{color0}7.01}~~&~{\color{color0}6.64}~~&~{\color{color0}6.66}~~&~{\color{color0}6.65}~~&~{\color{color0}6.65}~~&~{\color{color0}6.65}~~&~{\color{color0}6.64}~~\\
 {\color{blu}~\textbullet}\algo{AOSO2}~ &~-~&~{\color{color4}7.24}~~&~{\color{color0}6.32}~~&~{\color{color0}6.32}~~&~{\color{color0}6.32}~~&~{\color{color0}5.56}~~&~{\color{color0}5.57}~~&~{\color{color0}5.55}~~&~{\color{color0}5.56}~~&~{\color{color0}5.57}~~\\
 {\color{blu}~\textbullet}\algo{AOSO4}~ &~-~ &~-~&~{\color{color5}4.23}~~&~{\color{color0}3.31}~~&~{\color{color0}3.31}~~&~{\color{color0}2.97}~~&~{\color{color0}2.97}~~&~{\color{color0}2.95}~~&~{\color{color0}2.96}~~&~{\color{color0}2.97}~~\\
 {\color{blu}~\textbullet}\algo{AOSO6}~&~\colorbox{best}{\color{color20}2.33}~~&~\colorbox{best}{\color{color20}2.32}~~&~{\color{color0}16.1}~~&~{\color{color0}3.29}~~&~{\color{color0}2.33}~~&~{\color{color0}2.15}~~&~{\color{color0}2.15}~~&~{\color{color0}2.16}~~&~{\color{color0}2.15}~~&~{\color{color0}2.16}~~\\
 {\color{blu}~\textbullet}\algo{FSBNDM}~&~{\color{color0}15.3}~~&~{\color{color2}7.77}~~&~{\color{color6}4.15}~~&~{\color{color11}2.42}~~&~\colorbox{best}{\color{color20}1.56}~~&~{\color{color18}1.56}~~&~\colorbox{best}{\color{color20}1.55}~~&~{\color{color10}1.54}~~&~{\color{color4}1.56}~~&~{\color{color5}1.56}~~\\
 {\color{blu}~\textbullet}\algo{BNDMq2}~&~{\color{color0}21.4}~~&~{\color{color2}7.61}~~&~{\color{color11}3.63}~~&~{\color{color18}2.05}~~&~{\color{color18}1.62}~~&~{\color{color6}1.74}~~&~{\color{color13}1.73}~~&~{\color{color3}1.75}~~&~{\color{color1}1.73}~~&~{\color{color3}1.73}~~\\
 {\color{blu}~\textbullet}\algo{BNDMq4}~ &~-~&~{\color{color0}31.5}~~&~{\color{color0}6.80}~~&~{\color{color0}2.99}~~&~{\color{color15}1.68}~~&~{\color{color0}2.06}~~&~{\color{color0}2.05}~~&~{\color{color0}2.06}~~&~{\color{color0}2.07}~~&~{\color{color0}2.06}~~\\
 {\color{blu}~\textbullet}\algo{SBNDMq2}~&~{\color{color0}21.1}~~&~{\color{color2}7.58}~~&~{\color{color10}3.70}~~&~{\color{color16}2.16}~~&~{\color{color17}1.63}~~&~{\color{color16}1.59}~~&~{\color{color18}1.60}~~&~{\color{color8}1.59}~~&~{\color{color3}1.61}~~&~{\color{color5}1.60}~~\\
 {\color{blu}~\textbullet}\algo{SBNDMq4}~ &~-~&~{\color{color0}29.7}~~&~{\color{color0}6.44}~~&~{\color{color2}2.85}~~&~{\color{color16}1.66}~~&~{\color{color15}1.61}~~&~{\color{color18}1.60}~~&~{\color{color8}1.60}~~&~{\color{color3}1.60}~~&~{\color{color5}1.61}~~\\
 {\color{blu}~\textbullet}\algo{UFNDMq2}~&~{\color{color0}19.9}~~&~{\color{color0}10.4}~~&~{\color{color0}5.59}~~&~{\color{color0}3.18}~~&~{\color{color0}2.01}~~&~{\color{color0}2.01}~~&~{\color{color1}2.01}~~&~{\color{color0}2.02}~~&~{\color{color0}2.02}~~&~{\color{color0}2.02}~~\\
\hline\end{tabular*}\\
\end{center}
\end{scriptsize}

In the case of very short patterns the \textsf{SBNDM-BMH} algorithm obtains the best performance. 
Other very good algorithms are \textsf{SSABS}, \textsf{TVSBS} and \textsf{FJS}.
For short patterns the algorithms based on bit-parallelism achieves better results, in particular \textsf{SBNDM2}, \textsf{FSBNDM}, \textsf{BNDMq2}, \textsf{SBNDMq2}. 
The \textsf{EBOM} and \textsf{SEBOM} algorithms maintain also good performance.
In the case of long patterns the algorithms \textsf{EBOM} and \textsf{SEBOM} are good choices. 
Very good results are obtained also by the \textsf{HASH$q$} algorithms, \textsf{LBNDM} and \textsf{TVSBS}. 
For very long patterns the best results are obtained by the \textsf{SSEF}, \textsf{ZT}, \textsf{BR}, \textsf{HASH$q$} and \textsf{TVSBS} algorithms. 
Among the algorithms based on automata the best results are obtained by the \textsf{BOM} algorithm.
Evaluating the overall performance the algorithms \textsf{GRASPm}, \textsf{BR} and \textsf{TVSBS} maintains good performance for all patterns.

\vfill

\subsection{Experimental Results on Bible}\label{sec:exp:bible}
In this section we present experimental results on a natural language text with 63
 different characters.

\begin{scriptsize}
\begin{center}
\begin{tabular*}{\textwidth}{@{\extracolsep{\fill}}|l|cc|ccc|ccc|cc|}
\hline
$m$ & 2 & 4 & 8 & 16 & 32 & 64 & 128 & 256 & 512 & 1024 \\
\hline

 {\color{red}~\textbullet}\algo{BM}~&~{\color{color0}19.0}~~&~{\color{color2}10.8}~~&~{\color{color0}6.39}~~&~{\color{color0}4.17}~~&~{\color{color0}2.95}~~&~{\color{color0}2.46}~~&~{\color{color2}2.59}~~&~{\color{color0}2.29}~~&~{\color{color2}1.90}~~&~{\color{color7}1.59}~~\\
 {\color{red}~\textbullet}\algo{HOR}~&~{\color{color0}23.4}~~&~{\color{color0}13.1}~~&~{\color{color0}7.87}~~&~{\color{color0}4.87}~~&~{\color{color0}3.45}~~&~{\color{color0}2.75}~~&~{\color{color0}3.05}~~&~{\color{color0}2.58}~~&~{\color{color0}2.20}~~&~{\color{color4}1.91}~~\\
 {\color{red}~\textbullet}\algo{KR}~&~{\color{color8}14.6}~~&~{\color{color0}13.3}~~&~{\color{color0}13.1}~~&~{\color{color0}13.1}~~&~{\color{color0}13.1}~~&~{\color{color0}13.1}~~&~{\color{color0}13.1}~~&~{\color{color0}13.1}~~&~{\color{color0}13.1}~~&~{\color{color0}13.1}~~\\
 {\color{red}~\textbullet}\algo{ZT}~&~{\color{color0}26.4}~~&~{\color{color0}13.8}~~&~{\color{color0}7.52}~~&~{\color{color0}4.44}~~&~{\color{color0}2.87}~~&~{\color{color7}2.26}~~&~{\color{color3}2.55}~~&~{\color{color12}1.77}~~&~{\color{color13}1.27}~~&~{\color{color14}0.98}~~\\
 {\color{red}~\textbullet}\algo{QS}~&~{\color{color12}13.2}~~&~{\color{color10}8.90}~~&~{\color{color7}5.56}~~&~{\color{color4}3.66}~~&~{\color{color2}2.73}~~&~{\color{color5}2.32}~~&~{\color{color0}2.75}~~&~{\color{color0}2.30}~~&~{\color{color1}1.97}~~&~{\color{color6}1.70}~~\\
 {\color{red}~\textbullet}\algo{TunBM}~&~{\color{color8}14.8}~~&~{\color{color12}8.41}~~&~{\color{color11}5.11}~~&~{\color{color10}3.35}~~&~{\color{color5}2.59}~~&~{\color{color5}2.30}~~&~{\color{color3}2.55}~~&~{\color{color0}2.25}~~&~{\color{color1}1.95}~~&~{\color{color7}1.65}~~\\
 {\color{red}~\textbullet}\algo{NSN}~&~{\color{color9}14.5}~~&~{\color{color0}15.8}~~&~{\color{color0}15.2}~~&~{\color{color0}15.7}~~&~{\color{color0}15.5}~~&~{\color{color0}15.8}~~&~{\color{color0}16.1}~~&~{\color{color0}15.9}~~&~{\color{color0}16.4}~~&~{\color{color0}16.0}~~\\
 {\color{red}~\textbullet}\algo{Smith}~&~{\color{color0}24.9}~~&~{\color{color0}15.5}~~&~{\color{color0}9.29}~~&~{\color{color0}5.79}~~&~{\color{color0}3.97}~~&~{\color{color0}3.03}~~&~{\color{color0}3.38}~~&~{\color{color0}2.81}~~&~{\color{color0}2.39}~~&~{\color{color2}2.08}~~\\
 {\color{red}~\textbullet}\algo{Raita}~&~{\color{color9}14.4}~~&~{\color{color15}7.90}~~&~{\color{color13}4.79}~~&~{\color{color13}3.18}~~&~{\color{color9}2.43}~~&~{\color{color9}2.18}~~&~{\color{color6}2.46}~~&~{\color{color2}2.18}~~&~{\color{color3}1.86}~~&~{\color{color7}1.59}~~\\
 {\color{red}~\textbullet}\algo{RCol}~&~{\color{color13}12.9}~~&~{\color{color17}7.39}~~&~{\color{color15}4.59}~~&~{\color{color16}3.04}~~&~{\color{color10}2.38}~~&~{\color{color11}2.13}~~&~{\color{color10}2.36}~~&~{\color{color5}2.07}~~&~{\color{color4}1.75}~~&~{\color{color9}1.47}~~\\
 {\color{red}~\textbullet}\algo{BR}~&~{\color{color17}11.3}~~&~{\color{color13}8.17}~~&~{\color{color9}5.28}~~&~{\color{color9}3.41}~~&~{\color{color10}2.40}~~&~{\color{color10}2.17}~~&~{\color{color0}2.63}~~&~{\color{color12}1.78}~~&~{\color{color12}1.30}~~&~{\color{color14}1.01}~~\\
 {\color{red}~\textbullet}\algo{FS}~&~{\color{color13}12.8}~~&~{\color{color17}7.46}~~&~{\color{color15}4.58}~~&~{\color{color15}3.09}~~&~{\color{color11}2.36}~~&~{\color{color11}2.13}~~&~{\color{color9}2.37}~~&~{\color{color4}2.09}~~&~{\color{color4}1.76}~~&~{\color{color9}1.47}~~\\
 {\color{red}~\textbullet}\algo{FFS}~&~{\color{color13}13.0}~~&~{\color{color16}7.57}~~&~{\color{color15}4.60}~~&~{\color{color16}3.02}~~&~{\color{color11}2.33}~~&~{\color{color11}2.12}~~&~{\color{color9}2.39}~~&~{\color{color3}2.13}~~&~{\color{color2}1.92}~~&~{\color{color3}1.96}~~\\
 {\color{red}~\textbullet}\algo{BFS}~&~{\color{color13}13.1}~~&~{\color{color16}7.62}~~&~{\color{color15}4.52}~~&~{\color{color16}3.01}~~&~{\color{color12}2.29}~~&~{\color{color12}2.08}~~&~{\color{color8}2.40}~~&~{\color{color3}2.14}~~&~{\color{color1}1.94}~~&~{\color{color3}1.99}~~\\
 {\color{red}~\textbullet}\algo{TS}~&~{\color{color14}12.7}~~&~{\color{color0}11.2}~~&~{\color{color0}8.97}~~&~{\color{color0}7.40}~~&~{\color{color0}6.22}~~&~{\color{color0}5.68}~~&~{\color{color0}5.51}~~&~{\color{color0}5.70}~~&~{\color{color0}5.68}~~&~{\color{color0}5.68}~~\\
 {\color{red}~\textbullet}\algo{SSABS}~&~{\color{color19}10.8}~~&~{\color{color18}7.17}~~&~{\color{color12}4.88}~~&~{\color{color8}3.43}~~&~{\color{color3}2.66}~~&~{\color{color4}2.33}~~&~{\color{color4}2.52}~~&~{\color{color0}2.25}~~&~{\color{color2}1.91}~~&~{\color{color7}1.65}~~\\
 {\color{red}~\textbullet}\algo{TVSBS}~&~\colorbox{best}{\color{color20}10.1}~~&~{\color{color18}7.16}~~&~{\color{color15}4.60}~~&~{\color{color14}3.12}~~&~{\color{color11}2.35}~~&~{\color{color10}2.15}~~&~{\color{color8}2.42}~~&~{\color{color14}1.71}~~&~{\color{color14}1.21}~~&~{\color{color15}0.94}~~\\
 {\color{red}~\textbullet}\algo{FJS}~&~{\color{color18}11.1}~~&~{\color{color16}7.53}~~&~{\color{color11}5.05}~~&~{\color{color6}3.53}~~&~{\color{color0}2.77}~~&~{\color{color0}2.45}~~&~{\color{color1}2.62}~~&~{\color{color0}2.32}~~&~{\color{color0}1.98}~~&~{\color{color6}1.69}~~\\
 {\color{red}~\textbullet}\algo{HASH3}~ &~-~&~{\color{color0}14.6}~~&~{\color{color8}5.42}~~&~\colorbox{best}{\color{color20}2.79}~~&~\colorbox{best}{\color{color20}1.97}~~&~\colorbox{best}{\color{color20}1.84}~~&~{\color{color19}2.09}~~&~{\color{color20}1.45}~~&~{\color{color16}1.06}~~&~{\color{color16}0.85}~~\\
 {\color{red}~\textbullet}\algo{HASH5}~ &~-~ &~-~&~{\color{color0}9.68}~~&~{\color{color1}3.81}~~&~{\color{color16}2.14}~~&~{\color{color19}1.90}~~&~{\color{color15}2.21}~~&~{\color{color19}1.49}~~&~{\color{color16}1.07}~~&~{\color{color15}0.89}~~\\
 {\color{red}~\textbullet}\algo{HASH8}~ &~-~ &~-~ &~-~&~{\color{color0}6.08}~~&~{\color{color0}2.77}~~&~{\color{color16}1.97}~~&~{\color{color13}2.28}~~&~{\color{color17}1.58}~~&~{\color{color15}1.15}~~&~{\color{color14}0.98}~~\\
 {\color{red}~\textbullet}\algo{TSW}~&~{\color{color13}13.1}~~&~{\color{color9}9.32}~~&~{\color{color2}6.21}~~&~{\color{color0}4.12}~~&~{\color{color0}2.91}~~&~{\color{color4}2.33}~~&~{\color{color0}3.14}~~&~{\color{color2}2.17}~~&~{\color{color6}1.63}~~&~{\color{color11}1.30}~~\\
 {\color{red}~\textbullet}\algo{GRASPm}~&~{\color{color8}14.7}~~&~{\color{color13}8.31}~~&~{\color{color12}4.97}~~&~{\color{color11}3.31}~~&~{\color{color8}2.46}~~&~{\color{color10}2.16}~~&~{\color{color4}2.52}~~&~{\color{color4}2.12}~~&~{\color{color5}1.69}~~&~{\color{color10}1.32}~~\\
 {\color{red}~\textbullet}\algo{SSEF}~ &~-~ &~-~ &~-~ &~-~&~{\color{color0}4.36}~~&~{\color{color0}2.59}~~&~{\color{color0}2.72}~~&~\colorbox{best}{\color{color20}1.44}~~&~\colorbox{best}{\color{color20}0.80}~~&~\colorbox{best}{\color{color20}0.45}~~\\
 \hline
 {\color{green}~\textbullet}\algo{BOM}~&~{\color{color0}34.6}~~&~{\color{color0}25.2}~~&~{\color{color0}19.9}~~&~{\color{color0}13.9}~~&~{\color{color0}9.36}~~&~{\color{color0}6.03}~~&~{\color{color0}4.90}~~&~{\color{color0}2.90}~~&~{\color{color5}1.72}~~&~{\color{color13}1.10}~~\\
 {\color{green}~\textbullet}\algo{BOM2}~&~{\color{color0}23.6}~~&~{\color{color0}15.4}~~&~{\color{color0}9.71}~~&~{\color{color0}5.83}~~&~{\color{color0}3.51}~~&~{\color{color2}2.40}~~&~{\color{color0}2.83}~~&~{\color{color14}1.70}~~&~{\color{color13}1.22}~~&~{\color{color0}2.35}~~\\
 {\color{green}~\textbullet}\algo{EBOM}~&~{\color{color7}15.3}~~&~\colorbox{best}{\color{color20}6.53}~~&~\colorbox{best}{\color{color20}3.87}~~&~{\color{color18}2.91}~~&~{\color{color8}2.47}~~&~{\color{color8}2.21}~~&~{\color{color3}2.55}~~&~{\color{color15}1.68}~~&~{\color{color10}1.41}~~&~{\color{color0}2.67}~~\\
 {\color{green}~\textbullet}\algo{FBOM}~&~{\color{color16}11.8}~~&~{\color{color16}7.51}~~&~{\color{color12}4.89}~~&~{\color{color7}3.51}~~&~{\color{color0}2.80}~~&~{\color{color0}2.44}~~&~{\color{color0}2.85}~~&~{\color{color12}1.77}~~&~{\color{color10}1.44}~~&~{\color{color0}2.69}~~\\
 {\color{green}~\textbullet}\algo{SEBOM}~&~{\color{color4}16.2}~~&~{\color{color19}6.84}~~&~{\color{color19}4.10}~~&~{\color{color15}3.07}~~&~{\color{color5}2.61}~~&~{\color{color4}2.33}~~&~{\color{color0}2.67}~~&~{\color{color12}1.80}~~&~{\color{color9}1.48}~~&~{\color{color0}2.71}~~\\
 {\color{green}~\textbullet}\algo{SFBOM}~&~{\color{color17}11.6}~~&~{\color{color17}7.46}~~&~{\color{color12}4.94}~~&~{\color{color5}3.59}~~&~{\color{color0}2.83}~~&~{\color{color0}2.49}~~&~{\color{color0}2.93}~~&~{\color{color11}1.84}~~&~{\color{color9}1.48}~~&~{\color{color0}2.73}~~\\
 {\color{green}~\textbullet}\algo{SBDM}~&~{\color{color0}23.5}~~&~{\color{color0}13.4}~~&~{\color{color0}7.74}~~&~{\color{color0}4.89}~~&~{\color{color0}3.44}~~&~{\color{color0}2.77}~~&~{\color{color0}3.06}~~&~{\color{color0}2.58}~~&~{\color{color0}2.21}~~&~{\color{color4}1.91}~~\\
 \hline
 {\color{blu}~\textbullet}\algo{SO}~&~{\color{color12}13.3}~~&~{\color{color0}13.3}~~&~{\color{color0}13.3}~~&~{\color{color0}13.3}~~&~{\color{color0}13.3}~~&~{\color{color0}17.6}~~&~{\color{color0}17.6}~~&~{\color{color0}17.6}~~&~{\color{color0}17.6}~~&~{\color{color0}17.6}~~\\
 {\color{blu}~\textbullet}\algo{SA}~&~{\color{color13}13.1}~~&~{\color{color0}13.1}~~&~{\color{color0}13.1}~~&~{\color{color0}13.1}~~&~{\color{color0}13.1}~~&~{\color{color0}15.3}~~&~{\color{color0}15.3}~~&~{\color{color0}15.3}~~&~{\color{color0}15.3}~~&~{\color{color0}15.3}~~\\
 {\color{blu}~\textbullet}\algo{SBNDM}~&~{\color{color0}38.3}~~&~{\color{color0}14.0}~~&~{\color{color0}6.77}~~&~{\color{color0}3.94}~~&~{\color{color6}2.55}~~&~{\color{color0}2.58}~~&~{\color{color3}2.57}~~&~{\color{color0}2.56}~~&~{\color{color0}2.56}~~&~{\color{color0}2.56}~~\\
 {\color{blu}~\textbullet}\algo{LBNDM}~&~{\color{color0}28.3}~~&~{\color{color0}16.4}~~&~{\color{color0}9.59}~~&~{\color{color0}5.57}~~&~{\color{color0}3.32}~~&~{\color{color9}2.19}~~&~{\color{color0}2.90}~~&~{\color{color8}1.94}~~&~{\color{color6}1.65}~~&~{\color{color9}1.48}~~\\
 {\color{blu}~\textbullet}\algo{SVM1}~&~{\color{color4}16.1}~~&~{\color{color0}12.3}~~&~{\color{color0}13.2}~~&~{\color{color0}9.77}~~&~{\color{color0}7.46}~~&~{\color{color0}16.9}~~&~{\color{color0}16.9}~~&~{\color{color0}16.9}~~&~{\color{color0}16.9}~~&~{\color{color0}16.9}~~\\
 {\color{blu}~\textbullet}\algo{SBNDM2}~&~{\color{color0}28.4}~~&~{\color{color3}10.5}~~&~{\color{color9}5.36}~~&~{\color{color11}3.30}~~&~{\color{color10}2.39}~~&~{\color{color7}2.26}~~&~{\color{color14}2.23}~~&~{\color{color0}2.25}~~&~{\color{color0}2.23}~~&~{\color{color0}2.24}~~\\
 {\color{blu}~\textbullet}\algo{SBNDM-BMH}~&~{\color{color11}13.8}~~&~{\color{color11}8.80}~~&~{\color{color3}6.02}~~&~{\color{color0}3.84}~~&~{\color{color6}2.54}~~&~{\color{color0}2.54}~~&~{\color{color3}2.56}~~&~{\color{color0}2.57}~~&~{\color{color0}2.58}~~&~{\color{color0}2.57}~~\\
 {\color{blu}~\textbullet}\algo{BMH-SBNDM}~&~{\color{color12}13.2}~~&~{\color{color17}7.41}~~&~{\color{color16}4.46}~~&~{\color{color18}2.94}~~&~{\color{color13}2.28}~~&~{\color{color5}2.31}~~&~{\color{color12}2.31}~~&~{\color{color0}2.31}~~&~{\color{color0}2.30}~~&~{\color{color0}2.29}~~\\
 {\color{blu}~\textbullet}\algo{AOSO2}~&~{\color{color0}30.8}~~&~{\color{color5}10.2}~~&~{\color{color0}7.86}~~&~{\color{color0}7.76}~~&~{\color{color0}7.76}~~&~{\color{color0}6.82}~~&~{\color{color0}6.81}~~&~{\color{color0}6.82}~~&~{\color{color0}6.81}~~&~{\color{color0}6.81}~~\\
 {\color{blu}~\textbullet}\algo{AOSO4}~ &~-~&~{\color{color0}26.5}~~&~{\color{color4}5.93}~~&~{\color{color0}4.09}~~&~{\color{color0}4.07}~~&~{\color{color0}3.62}~~&~{\color{color0}3.62}~~&~{\color{color0}3.62}~~&~{\color{color0}3.63}~~&~{\color{color0}3.64}~~\\
 {\color{blu}~\textbullet}\algo{FSBNDM}~&~{\color{color0}19.8}~~&~{\color{color4}10.3}~~&~{\color{color6}5.74}~~&~{\color{color6}3.56}~~&~{\color{color15}2.20}~~&~{\color{color9}2.18}~~&~{\color{color15}2.20}~~&~{\color{color2}2.18}~~&~{\color{color0}2.18}~~&~{\color{color0}2.19}~~\\
 {\color{blu}~\textbullet}\algo{BNDMq2}~&~{\color{color0}26.8}~~&~{\color{color5}10.1}~~&~{\color{color11}5.06}~~&~{\color{color14}3.15}~~&~{\color{color12}2.30}~~&~{\color{color0}2.67}~~&~{\color{color0}2.70}~~&~{\color{color0}2.71}~~&~{\color{color0}2.70}~~&~{\color{color0}2.69}~~\\
 {\color{blu}~\textbullet}\algo{BNDMq4}~ &~-~&~{\color{color0}38.8}~~&~{\color{color0}8.48}~~&~{\color{color1}3.79}~~&~{\color{color16}2.16}~~&~{\color{color0}2.66}~~&~{\color{color0}2.67}~~&~{\color{color0}2.68}~~&~{\color{color0}2.67}~~&~{\color{color0}2.67}~~\\
 {\color{blu}~\textbullet}\algo{BNDMq6}~ &~-~ &~-~&~{\color{color0}19.2}~~&~{\color{color0}5.79}~~&~{\color{color0}2.83}~~&~{\color{color0}2.58}~~&~{\color{color2}2.59}~~&~{\color{color0}2.57}~~&~{\color{color0}2.58}~~&~{\color{color0}2.58}~~\\
 {\color{blu}~\textbullet}\algo{SBNDMq2}~&~{\color{color0}26.3}~~&~{\color{color6}9.87}~~&~{\color{color10}5.14}~~&~{\color{color12}3.21}~~&~{\color{color10}2.37}~~&~{\color{color7}2.24}~~&~{\color{color14}2.23}~~&~{\color{color1}2.21}~~&~{\color{color0}2.24}~~&~{\color{color0}2.25}~~\\
 {\color{blu}~\textbullet}\algo{SBNDMq4}~ &~-~&~{\color{color0}36.5}~~&~{\color{color0}8.02}~~&~{\color{color5}3.61}~~&~{\color{color16}2.14}~~&~{\color{color14}2.04}~~&~\colorbox{best}{\color{color20}2.05}~~&~{\color{color5}2.05}~~&~{\color{color0}2.05}~~&~{\color{color2}2.05}~~\\
 {\color{blu}~\textbullet}\algo{SBNDMq6}~ &~-~ &~-~&~{\color{color0}21.9}~~&~{\color{color0}6.45}~~&~{\color{color0}3.02}~~&~{\color{color0}2.59}~~&~{\color{color2}2.58}~~&~{\color{color0}2.59}~~&~{\color{color0}2.59}~~&~{\color{color0}2.59}~~\\
\hline\end{tabular*}\\
\end{center}
\end{scriptsize}

In the case of very short patterns the best results are obtained by the \textsf{TVSBS} and \textsf{EBOM} algorithms.
For short patterns the \textsf{EBOM} algorithm obtains the  best results for patterns of length 8, while in the other cases the 
best results are obtained by the \textsf{HASH$q$} algorithm.
In the case of long patterns the algorithms in the \textsf{HASH$q$} family are good choices. 
Very good results are obtained also by the \textsf{SSEF} and \textsf{SBNDMq4} algorithms. 
For very long patterns the best results are obtained by the \textsf{SSEF} algorithm. 
Evaluating the overall performance the algorithm  \textsf{TVSBS} maintains good performance for all patterns. 

\vfill

\subsection{Experimental Results on World192}\label{sec:exp:world}
In this section we present experimental results on a natural language text with 94
 different characters.

\begin{scriptsize}
\begin{center}
\begin{tabular*}{\textwidth}{@{\extracolsep{\fill}}|l|cc|ccc|ccc|cc|}
\hline
$m$ & 2 & 4 & 8 & 16 & 32 & 64 & 128 & 256 & 512 & 1024 \\
\hline

 {\color{red}~\textbullet}\algo{BM}~&~{\color{color0}11.1}~~&~{\color{color2}6.08}~~&~{\color{color0}3.57}~~&~{\color{color0}2.22}~~&~{\color{color2}1.56}~~&~{\color{color9}1.31}~~&~{\color{color6}1.50}~~&~{\color{color4}1.21}~~&~{\color{color8}0.93}~~&~{\color{color11}0.77}~~\\
 {\color{red}~\textbullet}\algo{HOR}~&~{\color{color0}13.7}~~&~{\color{color0}7.41}~~&~{\color{color0}4.26}~~&~{\color{color0}2.59}~~&~{\color{color0}1.76}~~&~{\color{color0}1.44}~~&~{\color{color0}1.72}~~&~{\color{color0}1.33}~~&~{\color{color4}1.08}~~&~{\color{color8}0.89}~~\\
 {\color{red}~\textbullet}\algo{KR}~&~{\color{color9}8.59}~~&~{\color{color0}8.13}~~&~{\color{color0}8.01}~~&~{\color{color0}7.99}~~&~{\color{color0}7.98}~~&~{\color{color0}8.00}~~&~{\color{color0}8.01}~~&~{\color{color0}8.02}~~&~{\color{color0}8.02}~~&~{\color{color0}8.01}~~\\
 {\color{red}~\textbullet}\algo{ZT}~&~{\color{color0}15.6}~~&~{\color{color0}8.13}~~&~{\color{color0}4.46}~~&~{\color{color0}2.68}~~&~{\color{color0}1.77}~~&~{\color{color0}1.44}~~&~{\color{color1}1.60}~~&~{\color{color11}1.08}~~&~{\color{color13}0.76}~~&~{\color{color15}0.56}~~\\
 {\color{red}~\textbullet}\algo{OM}~&~{\color{color0}10.9}~~&~{\color{color0}7.00}~~&~{\color{color0}4.50}~~&~{\color{color0}2.98}~~&~{\color{color0}2.06}~~&~{\color{color0}1.59}~~&~{\color{color0}1.68}~~&~{\color{color0}1.41}~~&~{\color{color1}1.17}~~&~{\color{color3}1.15}~~\\
 {\color{red}~\textbullet}\algo{MS}~&~{\color{color0}10.6}~~&~{\color{color0}6.92}~~&~{\color{color0}4.39}~~&~{\color{color0}2.79}~~&~{\color{color0}1.91}~~&~{\color{color0}1.48}~~&~{\color{color3}1.57}~~&~{\color{color0}1.33}~~&~{\color{color0}1.23}~~&~{\color{color0}1.60}~~\\
 {\color{red}~\textbullet}\algo{QS}~&~{\color{color14}7.38}~~&~{\color{color13}4.76}~~&~{\color{color9}2.99}~~&~{\color{color9}1.96}~~&~{\color{color8}1.45}~~&~{\color{color13}1.26}~~&~{\color{color0}1.62}~~&~{\color{color3}1.23}~~&~{\color{color6}0.99}~~&~{\color{color9}0.84}~~\\
 {\color{red}~\textbullet}\algo{TunBM}~&~{\color{color11}8.18}~~&~{\color{color14}4.63}~~&~{\color{color13}2.72}~~&~{\color{color15}1.75}~~&~{\color{color11}1.39}~~&~{\color{color12}1.27}~~&~{\color{color9}1.44}~~&~{\color{color7}1.16}~~&~{\color{color8}0.94}~~&~{\color{color11}0.76}~~\\
 {\color{red}~\textbullet}\algo{NSN}~&~{\color{color12}7.96}~~&~{\color{color0}8.41}~~&~{\color{color0}8.41}~~&~{\color{color0}8.45}~~&~{\color{color0}8.35}~~&~{\color{color0}8.24}~~&~{\color{color0}8.43}~~&~{\color{color0}8.40}~~&~{\color{color0}8.39}~~&~{\color{color0}8.34}~~\\
 {\color{red}~\textbullet}\algo{Smith}~&~{\color{color0}15.0}~~&~{\color{color0}9.31}~~&~{\color{color0}5.57}~~&~{\color{color0}3.39}~~&~{\color{color0}2.22}~~&~{\color{color0}1.62}~~&~{\color{color0}1.97}~~&~{\color{color0}1.50}~~&~{\color{color0}1.20}~~&~{\color{color6}1.01}~~\\
 {\color{red}~\textbullet}\algo{Raita}~&~{\color{color10}8.39}~~&~{\color{color14}4.57}~~&~{\color{color13}2.73}~~&~{\color{color15}1.75}~~&~{\color{color13}1.36}~~&~{\color{color16}1.22}~~&~{\color{color8}1.46}~~&~{\color{color6}1.18}~~&~{\color{color8}0.91}~~&~{\color{color11}0.75}~~\\
 {\color{red}~\textbullet}\algo{RCol}~&~{\color{color14}7.47}~~&~{\color{color18}4.10}~~&~{\color{color16}2.50}~~&~{\color{color19}1.62}~~&~{\color{color15}1.31}~~&~{\color{color15}1.23}~~&~{\color{color12}1.40}~~&~{\color{color9}1.12}~~&~{\color{color10}0.87}~~&~{\color{color11}0.74}~~\\
 {\color{red}~\textbullet}\algo{BR}~&~{\color{color18}6.60}~~&~{\color{color14}4.69}~~&~{\color{color7}3.13}~~&~{\color{color5}2.09}~~&~{\color{color2}1.55}~~&~{\color{color2}1.42}~~&~{\color{color0}1.65}~~&~{\color{color10}1.10}~~&~{\color{color12}0.78}~~&~{\color{color14}0.61}~~\\
 {\color{red}~\textbullet}\algo{FS}~&~{\color{color14}7.39}~~&~{\color{color18}4.10}~~&~{\color{color16}2.48}~~&~{\color{color19}1.64}~~&~{\color{color15}1.32}~~&~{\color{color17}1.20}~~&~{\color{color12}1.40}~~&~{\color{color8}1.13}~~&~{\color{color9}0.89}~~&~{\color{color12}0.71}~~\\
 {\color{red}~\textbullet}\algo{FFS}~&~{\color{color14}7.44}~~&~{\color{color17}4.26}~~&~{\color{color16}2.50}~~&~{\color{color19}1.64}~~&~{\color{color14}1.33}~~&~{\color{color14}1.25}~~&~{\color{color9}1.45}~~&~{\color{color1}1.28}~~&~{\color{color0}1.19}~~&~{\color{color0}1.39}~~\\
 {\color{red}~\textbullet}\algo{BFS}~&~{\color{color14}7.43}~~&~{\color{color18}4.18}~~&~{\color{color16}2.48}~~&~{\color{color19}1.63}~~&~{\color{color15}1.32}~~&~{\color{color15}1.23}~~&~{\color{color8}1.46}~~&~{\color{color0}1.29}~~&~{\color{color0}1.23}~~&~{\color{color0}1.46}~~\\
 {\color{red}~\textbullet}\algo{TS}~&~{\color{color16}7.02}~~&~{\color{color0}6.44}~~&~{\color{color0}5.50}~~&~{\color{color0}4.71}~~&~{\color{color0}4.09}~~&~{\color{color0}3.72}~~&~{\color{color0}3.66}~~&~{\color{color0}3.62}~~&~{\color{color0}3.15}~~&~{\color{color0}2.87}~~\\
 {\color{red}~\textbullet}\algo{SSABS}~&~{\color{color20}6.04}~~&~{\color{color18}4.13}~~&~{\color{color14}2.65}~~&~{\color{color13}1.83}~~&~{\color{color8}1.44}~~&~{\color{color12}1.28}~~&~{\color{color8}1.46}~~&~{\color{color6}1.18}~~&~{\color{color8}0.93}~~&~{\color{color11}0.77}~~\\
 {\color{red}~\textbullet}\algo{TVSBS}~&~\colorbox{best}{\color{color20}5.91}~~&~{\color{color17}4.21}~~&~{\color{color11}2.88}~~&~{\color{color8}1.98}~~&~{\color{color3}1.54}~~&~{\color{color1}1.43}~~&~{\color{color3}1.56}~~&~{\color{color12}1.05}~~&~{\color{color13}0.74}~~&~{\color{color15}0.54}~~\\
 {\color{red}~\textbullet}\algo{PBMH}~&~{\color{color0}12.1}~~&~{\color{color0}6.67}~~&~{\color{color0}3.80}~~&~{\color{color0}2.31}~~&~{\color{color0}1.58}~~&~{\color{color6}1.36}~~&~{\color{color0}1.78}~~&~{\color{color0}1.55}~~&~{\color{color0}2.07}~~&~{\color{color0}5.05}~~\\
 {\color{red}~\textbullet}\algo{FJS}~&~{\color{color20}6.07}~~&~{\color{color19}3.98}~~&~{\color{color13}2.70}~~&~{\color{color12}1.86}~~&~{\color{color6}1.48}~~&~{\color{color9}1.32}~~&~{\color{color6}1.51}~~&~{\color{color5}1.20}~~&~{\color{color7}0.96}~~&~{\color{color10}0.82}~~\\
 {\color{red}~\textbullet}\algo{HASH3}~ &~-~&~{\color{color0}8.85}~~&~{\color{color4}3.31}~~&~{\color{color16}1.72}~~&~\colorbox{best}{\color{color20}1.22}~~&~\colorbox{best}{\color{color20}1.15}~~&~{\color{color17}1.29}~~&~\colorbox{best}{\color{color20}0.88}~~&~{\color{color16}0.63}~~&~{\color{color16}0.50}~~\\
 {\color{red}~\textbullet}\algo{HASH5}~ &~-~ &~-~&~{\color{color0}5.94}~~&~{\color{color0}2.32}~~&~{\color{color15}1.32}~~&~{\color{color19}1.17}~~&~{\color{color14}1.36}~~&~{\color{color18}0.93}~~&~{\color{color15}0.68}~~&~{\color{color15}0.57}~~\\
 {\color{red}~\textbullet}\algo{HASH8}~ &~-~ &~-~ &~-~&~{\color{color0}3.74}~~&~{\color{color0}1.70}~~&~{\color{color15}1.23}~~&~{\color{color10}1.43}~~&~{\color{color15}0.99}~~&~{\color{color14}0.72}~~&~{\color{color14}0.63}~~\\
 {\color{red}~\textbullet}\algo{TSW}~&~{\color{color12}7.92}~~&~{\color{color5}5.68}~~&~{\color{color0}3.83}~~&~{\color{color0}2.59}~~&~{\color{color0}1.94}~~&~{\color{color0}1.61}~~&~{\color{color0}2.05}~~&~{\color{color0}1.41}~~&~{\color{color5}1.04}~~&~{\color{color9}0.85}~~\\
 {\color{red}~\textbullet}\algo{GRASPm}~&~{\color{color10}8.39}~~&~{\color{color14}4.65}~~&~{\color{color13}2.72}~~&~{\color{color15}1.75}~~&~{\color{color13}1.35}~~&~{\color{color15}1.23}~~&~{\color{color9}1.45}~~&~{\color{color7}1.16}~~&~{\color{color10}0.87}~~&~{\color{color12}0.70}~~\\
 {\color{red}~\textbullet}\algo{SSEF}~ &~-~ &~-~ &~-~ &~-~&~{\color{color0}2.74}~~&~{\color{color0}1.69}~~&~{\color{color0}1.73}~~&~{\color{color18}0.93}~~&~\colorbox{best}{\color{color20}0.48}~~&~\colorbox{best}{\color{color20}0.29}~~\\
 \hline
 {\color{green}~\textbullet}\algo{BOM}~&~{\color{color0}19.9}~~&~{\color{color0}13.9}~~&~{\color{color0}11.7}~~&~{\color{color0}8.39}~~&~{\color{color0}5.80}~~&~{\color{color0}3.92}~~&~{\color{color0}3.15}~~&~{\color{color0}1.92}~~&~{\color{color0}1.20}~~&~{\color{color9}0.84}~~\\
 {\color{green}~\textbullet}\algo{BOM2}~&~{\color{color0}13.2}~~&~{\color{color0}7.89}~~&~{\color{color0}5.04}~~&~{\color{color0}3.13}~~&~{\color{color0}1.99}~~&~{\color{color0}1.45}~~&~{\color{color0}1.73}~~&~{\color{color10}1.09}~~&~{\color{color8}0.91}~~&~{\color{color0}2.18}~~\\
 {\color{green}~\textbullet}\algo{EBOM}~&~{\color{color6}9.36}~~&~\colorbox{best}{\color{color20}3.84}~~&~\colorbox{best}{\color{color20}2.20}~~&~{\color{color19}1.64}~~&~{\color{color6}1.48}~~&~{\color{color0}1.44}~~&~{\color{color0}1.62}~~&~{\color{color6}1.18}~~&~{\color{color3}1.11}~~&~{\color{color0}2.49}~~\\
 {\color{green}~\textbullet}\algo{FBOM}~&~{\color{color17}6.84}~~&~{\color{color16}4.40}~~&~{\color{color12}2.81}~~&~{\color{color7}2.01}~~&~{\color{color0}1.68}~~&~{\color{color0}1.61}~~&~{\color{color0}1.86}~~&~{\color{color3}1.24}~~&~{\color{color1}1.16}~~&~{\color{color0}2.51}~~\\
 {\color{green}~\textbullet}\algo{SEBOM}~&~{\color{color4}9.90}~~&~{\color{color18}4.10}~~&~{\color{color18}2.35}~~&~{\color{color14}1.80}~~&~{\color{color0}1.62}~~&~{\color{color0}1.56}~~&~{\color{color0}1.75}~~&~{\color{color1}1.28}~~&~{\color{color0}1.19}~~&~{\color{color0}2.54}~~\\
 {\color{green}~\textbullet}\algo{SFBOM}~&~{\color{color16}6.86}~~&~{\color{color16}4.38}~~&~{\color{color11}2.87}~~&~{\color{color5}2.08}~~&~{\color{color0}1.75}~~&~{\color{color0}1.68}~~&~{\color{color0}1.93}~~&~{\color{color0}1.33}~~&~{\color{color0}1.23}~~&~{\color{color0}2.57}~~\\
 {\color{green}~\textbullet}\algo{SBDM}~&~{\color{color0}13.7}~~&~{\color{color0}7.59}~~&~{\color{color0}4.24}~~&~{\color{color0}2.62}~~&~{\color{color0}1.76}~~&~{\color{color1}1.43}~~&~{\color{color0}1.74}~~&~{\color{color0}1.32}~~&~{\color{color5}1.04}~~&~{\color{color9}0.88}~~\\
 \hline
 {\color{blu}~\textbullet}\algo{SO}~&~{\color{color11}8.12}~~&~{\color{color0}8.10}~~&~{\color{color0}8.12}~~&~{\color{color0}8.11}~~&~{\color{color0}8.10}~~&~{\color{color0}10.7}~~&~{\color{color0}10.8}~~&~{\color{color0}10.7}~~&~{\color{color0}10.8}~~&~{\color{color0}10.8}~~\\
 {\color{blu}~\textbullet}\algo{SA}~&~{\color{color12}8.02}~~&~{\color{color0}8.01}~~&~{\color{color0}8.02}~~&~{\color{color0}8.02}~~&~{\color{color0}8.04}~~&~{\color{color0}9.39}~~&~{\color{color0}9.40}~~&~{\color{color0}9.38}~~&~{\color{color0}9.38}~~&~{\color{color0}9.39}~~\\
 {\color{blu}~\textbullet}\algo{SBNDM}~&~{\color{color0}23.5}~~&~{\color{color0}8.32}~~&~{\color{color0}3.87}~~&~{\color{color1}2.19}~~&~{\color{color11}1.39}~~&~{\color{color0}1.44}~~&~{\color{color8}1.47}~~&~{\color{color0}1.45}~~&~{\color{color0}1.46}~~&~{\color{color0}1.46}~~\\
 {\color{blu}~\textbullet}\algo{LBNDM}~&~{\color{color0}16.5}~~&~{\color{color0}9.14}~~&~{\color{color0}5.28}~~&~{\color{color0}3.15}~~&~{\color{color0}1.90}~~&~{\color{color10}1.30}~~&~{\color{color0}1.66}~~&~{\color{color12}1.05}~~&~{\color{color13}0.76}~~&~{\color{color13}0.64}~~\\
 {\color{blu}~\textbullet}\algo{SVM1}~&~{\color{color8}8.80}~~&~{\color{color0}6.60}~~&~{\color{color0}7.73}~~&~{\color{color0}5.70}~~&~{\color{color0}4.38}~~&~{\color{color0}10.3}~~&~{\color{color0}10.3}~~&~{\color{color0}10.3}~~&~{\color{color0}10.3}~~&~{\color{color0}10.3}~~\\
 {\color{blu}~\textbullet}\algo{SBNDM2}~&~{\color{color0}17.2}~~&~{\color{color0}6.27}~~&~{\color{color8}3.07}~~&~{\color{color13}1.81}~~&~{\color{color13}1.35}~~&~{\color{color11}1.29}~~&~{\color{color18}1.27}~~&~{\color{color0}1.29}~~&~{\color{color0}1.30}~~&~{\color{color0}1.28}~~\\
 {\color{blu}~\textbullet}\algo{SBNDM-BMH}~&~{\color{color13}7.72}~~&~{\color{color14}4.62}~~&~{\color{color8}3.03}~~&~{\color{color7}2.01}~~&~{\color{color11}1.40}~~&~{\color{color0}1.47}~~&~{\color{color9}1.45}~~&~{\color{color0}1.47}~~&~{\color{color0}1.46}~~&~{\color{color0}1.48}~~\\
 {\color{blu}~\textbullet}\algo{BMH-SBNDM}~&~{\color{color13}7.60}~~&~{\color{color18}4.18}~~&~{\color{color17}2.41}~~&~\colorbox{best}{\color{color20}1.58}~~&~{\color{color17}1.28}~~&~{\color{color10}1.30}~~&~{\color{color17}1.30}~~&~{\color{color0}1.29}~~&~{\color{color0}1.32}~~&~{\color{color0}1.32}~~\\
 {\color{blu}~\textbullet}\algo{FAOSOq2}~&~{\color{color0}15.7}~~&~{\color{color2}6.13}~~&~{\color{color0}5.34}~~&~{\color{color0}5.27}~~&~{\color{color0}4.97}~~&~{\color{color0}4.99}~~&~{\color{color0}4.99}~~&~{\color{color0}4.99}~~&~{\color{color0}4.99}~~&~{\color{color0}5.00}~~\\
 {\color{blu}~\textbullet}\algo{FAOSOq4}~ &~-~&~{\color{color0}13.2}~~&~{\color{color2}3.50}~~&~{\color{color0}2.86}~~&~{\color{color0}2.69}~~&~{\color{color0}2.67}~~&~{\color{color0}2.69}~~&~{\color{color0}2.67}~~&~{\color{color0}2.68}~~&~{\color{color0}2.68}~~\\
 {\color{blu}~\textbullet}\algo{AOSO2}~&~{\color{color0}14.3}~~&~{\color{color6}5.60}~~&~{\color{color0}4.81}~~&~{\color{color0}4.79}~~&~{\color{color0}4.76}~~&~{\color{color0}4.18}~~&~{\color{color0}4.19}~~&~{\color{color0}4.19}~~&~{\color{color0}4.18}~~&~{\color{color0}4.20}~~\\
 {\color{blu}~\textbullet}\algo{AOSO4}~ &~-~&~{\color{color0}11.9}~~&~{\color{color8}3.08}~~&~{\color{color0}2.52}~~&~{\color{color0}2.50}~~&~{\color{color0}2.23}~~&~{\color{color0}2.25}~~&~{\color{color0}2.25}~~&~{\color{color0}2.25}~~&~{\color{color0}2.25}~~\\
 {\color{blu}~\textbullet}\algo{FSBNDM}~&~{\color{color0}11.4}~~&~{\color{color2}6.04}~~&~{\color{color5}3.27}~~&~{\color{color9}1.95}~~&~{\color{color18}1.26}~~&~{\color{color14}1.25}~~&~\colorbox{best}{\color{color20}1.23}~~&~{\color{color3}1.24}~~&~{\color{color0}1.24}~~&~{\color{color1}1.25}~~\\
 {\color{blu}~\textbullet}\algo{BNDMq2}~&~{\color{color0}16.2}~~&~{\color{color4}5.87}~~&~{\color{color10}2.89}~~&~{\color{color18}1.67}~~&~{\color{color15}1.32}~~&~{\color{color0}1.47}~~&~{\color{color8}1.46}~~&~{\color{color0}1.45}~~&~{\color{color0}1.47}~~&~{\color{color0}1.47}~~\\
 {\color{blu}~\textbullet}\algo{BNDMq4}~ &~-~&~{\color{color0}23.7}~~&~{\color{color0}5.19}~~&~{\color{color0}2.31}~~&~{\color{color14}1.33}~~&~{\color{color0}1.64}~~&~{\color{color0}1.63}~~&~{\color{color0}1.63}~~&~{\color{color0}1.64}~~&~{\color{color0}1.64}~~\\
 {\color{blu}~\textbullet}\algo{BNDMq6}~ &~-~ &~-~&~{\color{color0}11.8}~~&~{\color{color0}3.57}~~&~{\color{color0}1.76}~~&~{\color{color0}1.59}~~&~{\color{color1}1.60}~~&~{\color{color0}1.60}~~&~{\color{color0}1.60}~~&~{\color{color0}1.61}~~\\
 {\color{blu}~\textbullet}\algo{SBNDMq2}~&~{\color{color0}15.9}~~&~{\color{color4}5.84}~~&~{\color{color10}2.92}~~&~{\color{color15}1.75}~~&~{\color{color14}1.34}~~&~{\color{color11}1.29}~~&~{\color{color18}1.27}~~&~{\color{color0}1.30}~~&~{\color{color0}1.28}~~&~{\color{color0}1.29}~~\\
 {\color{blu}~\textbullet}\algo{SBNDMq4}~ &~-~&~{\color{color0}22.3}~~&~{\color{color0}4.89}~~&~{\color{color0}2.22}~~&~{\color{color15}1.32}~~&~{\color{color13}1.26}~~&~{\color{color19}1.26}~~&~{\color{color2}1.26}~~&~{\color{color0}1.24}~~&~{\color{color1}1.27}~~\\
\hline\end{tabular*}\\
\end{center}
\end{scriptsize}

In the case of very short patterns the \textsf{TVSBS} and the \textsf{EBOM} algorithms obtain the best performance. In particular the 
\textsf{TVSBS} algorithm is the fastest for patterns of length 2 while the \textsf{EBOM} algorithm obtains the best results for
pattern of length 8.
For short patterns the algorithms \textsf{EBOM}, \textsf{SBNDM-BMH} and \textsf{HASH$q$} obtain good performance for patterns of
length 8, 16 and 32, respectively.
In the case of long patterns the algorithms \textsf{HASH$q$} are the best algorithms. 
However sporadically they are outperformed by the \textsf{FSBNDM} algorithm.
For very long patterns the best results are obtained by the \textsf{SSEF} algorithm.
Evaluating the overall performance the algorithms \textsf{TVSBS}, \textsf{SSABS} and \textsf{FS} maintain good performance for all patterns.

\vfill

\section{Overall Discussion}

\begin{figure}[t]
\begin{center}
\includegraphics[width=\textwidth]{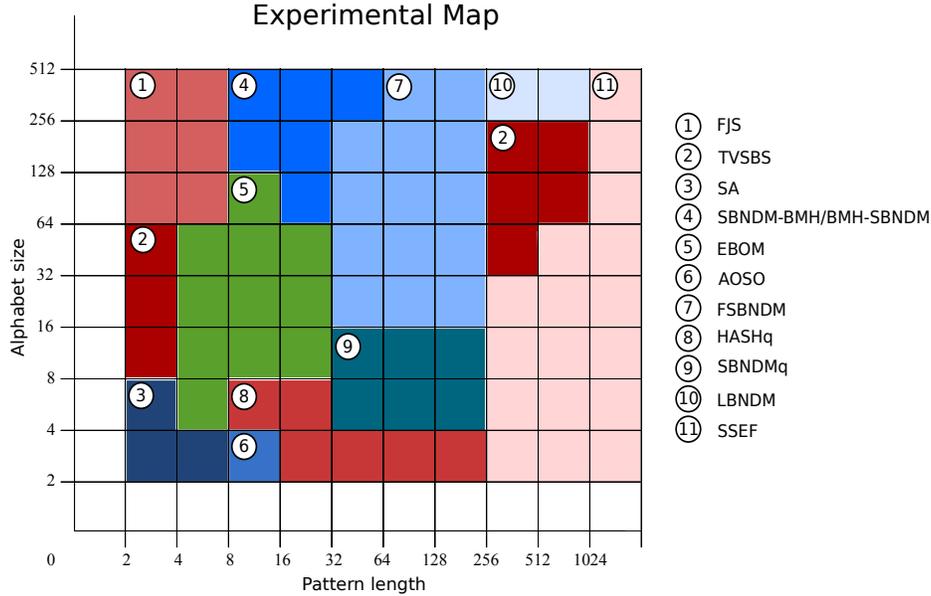}
\end{center}
\caption{\label{fig:map}Experimental map of the best results obtained in our evaluation. 
Comparison based algorithms are presented in red gradations, automata based algorithms are presented in green gradations and
bit parallel algorithms are presented in blu gradations.}
\end{figure}

We performed comparisons between 85 exact string matching algorithms with 12 text of
 different types.
We divide the patterns into four classes according to their length $m$: very short ($m\leq 4$),
 short ($4< m\leq 32$), long ($32< m \leq 256$) and very long ($m>256$).
We proceed in the same way for the alphabet according to their size $\sigma$:
 very small ($\sigma < 4$), small ($4 \le \sigma < 32$), large ($32 \le \sigma < 128$)
 and very large ($\sigma > 128$).
According to our experimental results, we conclude that the following algorithms
 are the most efficient in the following situations (see \figurename~\ref{fig:map}):
\begin{itemize}
 \item \textsf{SA}: very short patterns and very small alphabets.
 \item \textsf{TVSBS}: very short patterns and small alphabets, and long patterns and large alphabets.
 \item \textsf{FJS}: very short patterns and large and very large alphabets.
 \item \textsf{EBOM}: short patterns and large and very large alphabets.
 \item \textsf{SBNDM-BMH} and \textsf{BMH-SBNDM}: short patterns and very large alphabets.
 \item \textsf{HASH$q$}: short and large patterns and small alphabets.
 \item \textsf{FSBNDM}: long patterns and large and very larghe alphabets.
 \item \textsf{SBNDMq}: long pattern and small alphabets.
 \item \textsf{LBNDM}: very long patterns and very large alphabets.
 \item \textsf{SSEF}: very long patterns.
\end{itemize}
Among these algorithms all but one (the \textsf{SA} algorithm) have been designed during the last decade,
four of them are based on comparison of characters, one of them is based on automata (the \textsf{EBOM} algorithm) while six of them are bit-parallel algorithms.
\newpage

\bibliographystyle{alpha}
\newcommand{\etalchar}[1]{$^{#1}$}

\end{document}